\documentclass[a4paper,11pt]{article}
\pdfoutput=1 

\usepackage{jheppub} 

\usepackage[T1]{fontenc} 

\usepackage{physics}
\usepackage{slashed}
\usepackage[toc,page]{appendix}
\usepackage{float}
\usepackage[usenames,dvipsnames,svgnames]{xcolor}
\usepackage{multirow}
\usepackage{bm}
\usepackage[caption=false]{subfig}
\usepackage[shortlabels,inline]{enumitem}
\setlist[enumerate,1]{label=\textbf{\textrm{\arabic*)}}}
\usepackage[normalem]{ulem}
\usepackage[capitalise]{cleveref}
\usepackage{tikz-feynman}
\usepackage{tikz}
\usepackage{amsmath}
\DeclareMathAlphabet{\mathpzc}{OT1}{pzc}{m}{it}
\usepackage{threeparttable, tablefootnote}
\PassOptionsToPackage{demo}{graphicx}

\newcommand{\bea}{\begin{eqnarray}}
\newcommand{\eea}{\end{eqnarray}}
\newcommand{\beq}{\begin{equation}}
\newcommand{\eeq}{\end{equation}}
\newcommand{\ec}{\end{center}}
\newcommand{\bc}{\begin{center}}

\newcommand{\pdir}{p\kern -5.2pt\raise 0.2ex\hbox {/}}

\newcommand{\vdir}{v\kern -5.75pt\raise 0.15ex\hbox {/}}
\newcommand{\kdir}{k\kern -5.75pt\raise 0.15ex\hbox {/}}
\newcommand{\epsdir}{\epsilon\kern -5.0pt\raise 0.15ex\hbox {/}}
\newcommand{\bvdir}{\bar{v}\kern -5.75pt\raise 0.15ex\hbox {/}}
\newcommand{\Ddir}{D\kern -7.75pt\raise 0.20ex\hbox {/}}
\newcommand{\Adir}{A\kern -7.75pt\raise 0.20ex\hbox {/}}
\newcommand{\ldir}{l\kern -5.0pt\raise 0.2ex\hbox{/}}
\newcommand{\varepsdir}{\varepsilon\kern -5.5pt\raise 0.15ex\hbox{/}}

\makeatother
\pagebreak

\def\tew{T_{\rm{EW}}}

\def\be{\begin{equation}}
\def\ee{\end{equation}}

\def\s{\bm{s}}
\def\Tew{T_{\rm EW}}

\def\sc{\color{red}}

\newif\ifstartedinmathmode
\newcommand\encircled[1]{%
  \relax\ifmmode\startedinmathmodetrue\else\startedinmathmodefalse\fi%
  \tikz[baseline,anchor=base]{%
  \node[draw,circle,outer sep=0pt,inner sep=.2ex]
    {\ifstartedinmathmode$#1$\else#1\fi};}%
}

\usepackage{framed}
\colorlet{shadecolor}{blue!20}

\title{\boldmath Electroweak Symmetry Breaking and WIMP-FIMP Dark Matter}
\author[a]{Subhaditya Bhattacharya,}
\author[b]{Sreemanti Chakraborti}
\author[a]{and Dipankar Pradhan}
\affiliation[a]{Department of Physics, Indian Institute of Technology Guwahati, Assam 781039, India.}
\affiliation[b]{LAPTh, Univ. Grenoble Alpes, USMB,CNRS, F-74940 Annecy, France}

\emailAdd{subhab@iitg.ac.in}
\emailAdd{chakraborti@lapth.cnrs.fr}
\emailAdd{d.pradhan@iitg.ac.in}

\abstract{Electroweak Symmetry Breaking (EWSB) is known to produce a massive universe that we live in.  However, it may also provide an important boundary for freeze-in or freeze-out of dark matter (DM) connected to Standard Model via Higgs portal as processes contributing to DM relic differ across the boundary. We explore such possibilities in a two-component DM framework, where a massive $U(1)_X$ gauge boson DM freezes-in and a scalar singlet DM freezes-out, that inherits the effect of EWSB for both the cases in a correlated way. Amongst different possibilities, we study two sample cases; first when one DM component freezes in and the other freezes out from thermal bath both necessarily {\it before} EWSB and the second, when both freeze-in and freeze-out occur {\it after} EWSB. We find some prominent distinctive features in the available parameter space of the model for these two cases, after addressing relic density and the recent most direct search constraints from XENON1T, some of which can be borrowed in a model independent way.
}
\begin{document}

\begin{flushright}
LAPTH-039/21
\end{flushright}

\maketitle
\flushbottom

\section{Introduction}
\label{sec:intro}

Electroweak Symmetry Breaking (EWSB) is one of the most important phenomena that fundamental particle physics has taught us. 
The discovery of the Higgs-like boson with mass $125$ GeV in 2012 at the LHC \cite{ATLAS:2012yve,CMS:2012qbp}, has established 
EWSB as a law of nature and Standard Model (SM) of particle physics as the most appropriate theory to describe electromagnetic, weak and 
strong interactions amongst fundamental particles. Weak gauge boson ($W$ and $Z$) masses provide the scale of EWSB to be $\sim$ 246 
GeV, (equivalent to a temperature of $\sim 160$ GeV) when the phase transition occurs. Albeit the plethora of knowledge accumulated for EWSB, 
there are several unanswered questions like whether the Higgs boson responsible for EWSB is SM like, how to stabilize the {\it metastable} vacua \cite{Isidori:2001bm,Markkanen:2018pdo} that we live in, or how to solve the gauge hierarchy problem \cite{Khoury:2021zao}, together with other experimental 
observations like tiny but non-zero neutrino mass \cite{Cheng:1980qt,RevModPhys.59.671,Schechter:1980gr}, dark matter, baryon asymmetry of the universe \cite{Shaposhnikov:1987tw,Morrissey:2012db}, that leave ample scope to study physics beyond the SM (BSM). 

One of the most important hints for BSM physics arises from the evidences of dark matter (DM) obtained from astrophysical 
observations \cite{Zwicky:1933gu,Zwicky:1937zza,Sofue:2000jx}. A particle realisation of DM is highly motivated, although the 
discovery is still awaited. The most well known parameter for DM physics comes from DM relic density to constitute $\sim$23\% 
of the energy budget of the universe. This is measured from the anisotropies in cosmic microwave background radiation (CMBR) 
\cite{Bullock:2017xww} in experiments like WMAP \cite{WMAP:2012nax} and PLANCK \cite{Planck:2018vyg}, often expressed 
in terms of $\Omega h^2 \simeq0.12$ \cite{Planck:2018vyg}, where $\Omega$ is 
the cosmological density and $h$ is today's reduced Hubble constant in the units of 100 km/s/MPc. 

   
The major classification for particle DM comes from the initial condition. If the DM was in chemical and 
thermal equilibrium with the Standard Model (SM) in early universe and {\it freezes out} as universe expands 
\cite{Gondolo:1990dk,Kolb:1990vq} then the DM annihilation cross-section needs to be of the order of weak 
interaction strength ($\sim 10^{-10} ~\rm{GeV}^{-2}$) to be compatible with the observed relic density. Therefore such kind 
of DM is dubbed as Weakly Interacting Massive Particle (WIMP) \cite{Bertone:2004pz, Feng:2010gw, Bergstrom:2012fi}. 
On the contrary, DM may remain out of equilibrium due to very feeble interaction with the SM and gets produced from 
the decay or annihilation of particles in thermal bath to {\it freeze-in} when the temperature drops below the DM mass \cite{Hall:2009bx}. 
For freeze-in production of DM, the relic density is proportional to the decay width or the cross-section through which it is produced.
For renormalizable interactions\footnote{For non-renormalizable effective interactions, the relevant parameters are the new 
physics scale ($\Lambda$), DM mass and reheating temperature ($T_{RH}<\Lambda$), often providing a quick DM saturation, 
is called ultra-violet (UV) freeze in \cite{Elahi:2014fsa,Heeba:2018wtf}. There are also studies on mixed UV-IR freeze in scenarios 
\cite{Biswas:2019iqm}.}, the DM abundance builds up slowly with coupling strength as low as $\lesssim 10^{-10}$, is known as Infrared (IR) 
freeze-in and the DM is justifiably called Feebly Interacting Massive Particle (FIMP) (for a review on possible models, see \cite{Bernal:2017kxu}). 
WIMP paradigm turns phenomenologically more appealing, particularly for the prospect of discovery at present direct search experiments such 
as XENON1T \cite{XENON:2018voc} as well as upcoming experiments like XENONnT \cite{ XENON:2020kmp}, PANDAX-II \cite{PandaX-II:2020oim} 
and LUX-ZEPLIN (LZ) \cite{LUX-ZEPLIN:2018poe} and also in collider searches, for example, at the Large Hadron Collider (LHC) 
\cite{Nath:2010zj,Kahlhoefer:2017dnp,Roszkowski:2017nbc}. FIMP models on the other hand, owing to its feeble coupling 
remains mostly undetectable, although possibilities of producing the `decaying' particle at collider and seeing a charge track or a displaced vertex have 
been considered as a signal for such cases \cite{Belanger:2018sti,Chakraborti:2019ohe, Aboubrahim:2019qpc, Ghosh2017}.

Our aim of this analysis is to study the effect of EWSB as a boundary for DM freeze-in and freeze-out. EWSB can provide an important 
boundary, mainly because of two reasons: first SM particles become massive and second, additional channels open up for DM production 
or annihilation after EWSB, particularly for DM that connects to SM via Higgs portal, both of which alter the yield. Specifically, we are interested in exploring 
the difference between the resulting relic density and direct search allowed parameter spaces of the model, if the DM freezes in (or freezes out) before 
EWSB (bEWSB) to that when it freezes-in (or freezes out) after EWSB (aEWSB).  Although the phenomena is well understood, the authors are not aware of any systematic 
comparative analysis that distinguishes these two possibilities in details. It is worthy to point out that freeze-in production of a light (KeV-MeV) scalar has been studied 
\cite{Heeba:2018wtf} in five steps with emphasis on finite temperature effects and quantum statistics around EWSB scale to show that the production magnifies 
around that scale, although the results do not apply to our case, as heavy DMs are considered. To study the effect in both freeze-in and freeze-out context, we choose a two component DM 
setup with one WIMP and one FIMP like DM. 

Multi-particle dark sector constituted of different kinds of DM particles is motivated from several reasons. WIMP-WIMP combination 
is studied widely in different contexts \cite{Cao:2007fy,Zurek:2008qg,Profumo:2009tb,Bhattacharya:2013hva,Biswas:2013nn,Bian:2013wna,Bhattacharya:2016ysw, DiFranzo:2016uzc, Ahmed:2017dbb, Bhattacharya:2017fid, Barman:2018esi, Chakraborti:2018lso, Chakraborti:2018aae, Bhattacharya:2018cgx, Elahi:2019jeo, Bhattacharya:2019fgs, Biswas:2019ygr, Bhattacharya:2019tqq,Betancur:2020fdl, Nam:2020twn, Belanger:2021lwd}; 
but WIMP and FIMP together has not been studied exhaustively excepting for a few cases like \cite{DuttaBanik:2016jzv,Borah:2019epq}. 
Having a WIMP and FIMP together in the dark sector has some important phenomenological implications, particularly concerning the interaction 
between the DM particles, which we highlight upon, has not been elaborated so far. There exists studies on FIMP-FIMP combinations as well, see for example, \cite{PeymanZakeri:2018zaa,Pandey:2017quk}.

 We choose an abelian vector boson DM (VBDM) in an $U(1)_X$ gauge extension of SM 
\cite{Duch:2017khv, Barman:2020ifq, Delaunay:2020vdb,Choi:2020kch,Barman:2021lot} to constitute a FIMP like DM. 
A scalar singlet on the other hand, is considered as WIMP (such DM is perhaps the most popular, amongst many studies, see 
\cite{McDonald:1993ex,Guo:2010hq,Cline:2013gha,Steele:2013fka,Silveira:1985rk,Burgess:2000yq}). 
VBDM has also been studied extensively as single component DM, both in the context of WIMP 
\cite{Hambye:2008bq,Hambye:2009fg,Bhattacharya:2011tr,Farzan:2012hh,Hu:2021pln,Baek:2012se, Ko:2014gha,Duch:2015jta,Duch:2017nbe,YaserAyazi:2019caf,Choi:2020dec,Lebedev:2011iq,Davoudiasl:2013jma} 
and FIMP \cite{Duch:2017khv, Barman:2020ifq, Delaunay:2020vdb,Choi:2020kch,Barman:2021lot}.
The stability of both DM components is ensured by added $\mathcal{Z}_2 \times \mathcal{Z}_2^{'}$ symmetry 
under which they transform non-trivially. While the model serves as an example of a two component 
WIMP-FIMP set up where the effect of EWSB is studied, VBDM freeze-in provides an additional scale via $U(1)_X$ 
breaking, which helps achieving a rich phenomenology both before and after EWSB as we illustrated. Apart from that, the interplay of the scalar fields to 
address the correct Higgs mass and bounds can also be adopted in other WIMP-FIMP frameworks having 
extended scalar sector and Higgs portal interaction. 

The paper is organised as follows: first we discuss the model in section \ref{sec:model}, features of the model with VBDM as FIMP 
and scalar singlet as WIMP is discussed next in section \ref{sec:possibilities}; FIMP freeze-in and WIMP freeze-out before EWSB is 
discussed in section \ref{sec:bEWSB}, while the case when freeze-in and freeze-out both occur after EWSB is elaborated in section 
\ref{sec:aEWSB}, we finally conclude in section \ref{sec:conclusions}.  Appendices \ref{sec:bEWSB-details}, \ref{sec:aEWSB-details}, 
\ref{sec:invisible}, \ref{sec:directsearch} provide some necessary details omitted in the main text.

\section{Model}
\label{sec:model}

The model consists of two DM components: (i) an abelian vector boson $X$ (VBDM) 
arising from an $U(1)_X$ gauge extension of SM and (ii) a real scalar singlet ($\phi$) having 
Higgs portal interaction with the SM. The scalar doublet $H$ is responsible for spontaneous EWSB.
Both DM candidates are rendered electromagnetic charge neutral by having zero SM hypercharge. 
$U(1)_X$ symmetry spontaneously breaks (to no remnant symmetry) via non-zero vacuum expectation 
value (vev) of a complex scalar singlet ($S$) transforming under $U(1)_X$, \footnote{The $U(1)_X$ charge of $S$ remains indetermined in absence of any term containing a single $S$ field to cater to $\mathcal{Z}_2$ invariance.} to yield $X$ massive. 
A stabilising symmetry (we choose the simplest possibility $\mathcal{Z}_2$) is further imposed under which $X \to -X$ 
to make it a stable VBDM. The real scalar singlet ($\phi$) also needs to be stabilised for becoming the second 
DM component of the universe and the simplest possibility is yet again to consider an additional symmetry 
$\mathcal{Z}_2^{'}: \phi \to -\phi$, different from $\mathcal{Z}_2$\footnote{Two different symmetries $\mathcal{Z}_2\times \mathcal{Z}_2^{'}$ 
are required to stabilise two DM, as the lightest particle under a symmetry is stable, while the heavier ones transforming under the same 
symmetry, decay to the lightest.}. However, $X$ does not have a direct renormalizable coupling to $\phi$; $X$ couples to complex scalar 
$S$, which has portal interactions to both $\phi$ and SM Higgs ($H$). Therefore, $\phi$ is {\it apparently stable} even if it transforms 
under the same $\mathcal{Z}_2$ symmetry as of $X$, absent a direct interaction with each other. However, 
an effective dimension five operator involving $U(1)_X$ gauge field strength tensor $X^{\mu\nu}$, 
SM hypercharge field strength tensor $B_{\mu\nu}$ and $\phi$ can be written as:
\bea
\mathcal{L}_{\text {dim}~5} \supset \frac{1}{\Lambda}X^{\mu\nu}B_{\mu\nu}\phi\,;
\label{eq:EFT}
\eea
invariant under $\text {SM} \times U(1)_X \times \mathcal{Z}_2$ symmetry. This will in turn allow the heavier between 
$\phi$ and $X$ to decay into the other and provides a single component DM model. 
The phenomenology of such higher dimensional operator to study DM production of $X$ 
in context of both freeze-in (see \cite{Barman:2020ifq,Bhattacharya:2021edh}) and freeze out limit \cite{Fitzpatrick:2012ix,Fortuna:2020wwx,Matsumoto:2014rxa,Bell:2015sza,DeSimone:2016fbz,Cao:2009uw,Cheung:2012gi,Busoni:2013lha,Duch:2014xda} has been studied. Therefore having two different symmetries for two DM components is necessary, which prohibits an 
operator like in Eq.~\ref{eq:EFT} and renders both DM components stable. The charges of the fields under 
$\mathcal{Z}_2 \times \mathcal{Z}^{'}_2 \times U(1)_X$ are mentioned in Table \ref{tab:charges}. Note that none of the SM fields 
possess any charge under the dark symmetry and none of the additional fields has SM charges. 

\begin{table}[htb]
\begin{center}
\begin{tabular}{|c |c |c |}
\hline
 Particles & $\mathcal{Z}_2$ & $\mathcal{Z}^{'}_2$ \\
\hline
\hline
$U(1)_X$ Gauge Boson $X$ & $-X$ & $+X$ \\
\hline
Complex scalar $S$ & $S^*$ & $S$ \\ 
\hline
Real scalar $\phi$ & $\phi$  & $-\phi$  \\
\hline
Complex scalar doublet $H$ & $H$  & $H$ \\
\hline
\hline
\end{tabular}
\end{center} 
\caption{Fields beyond the SM together with SM Higgs doublet ($H$) and their 
charges under the symmetry $\mathcal{Z}_2 \times \mathcal{Z}^{'}_2 $.} 
\label{tab:charges}
\end{table}

The Lagrangian for the model having field content and charges as in 
Table \ref{tab:charges} is: 

\begin{eqnarray}
\mathcal{L}=\mathcal{L}_{SM}+\frac{1}{2}|\partial_{\mu}\phi|^2+|D_{\mu}S|^2+\frac{1}{4} X^{\mu\nu}{X_{\mu\nu}}-V(H, \phi,S)\,;
\end{eqnarray}
where,$$ D_{\mu}=\partial_{\mu}+ig_X X_{\mu}; ~~X^{\mu\nu}=\partial^\mu X^\nu- \partial^\nu X^\mu\,; ~ {\rm and}
$$
\begin{eqnarray}
V(H,\phi,S) &=&\mu_{H}^2 ( H^{\dagger}H)+\lambda_{H}( H^{\dagger}H)^2 +\frac{1}{2}\mu_{\phi}^2\phi^2+
\frac{1}{4!}\lambda_{\phi}\phi^4+\mu_{S}^2(S^{*}S)+\lambda_S(S^{*}S)^2 
\nonumber \\ 
& & + \frac{1}{2}\lambda_{\phi H}\phi^2 (H^{\dagger}H)+\lambda_{H S}(H^{\dagger}H) (S^{*}S)+ 
\frac{1}{2}\lambda_{\phi S}\phi^2 (S^{*}S)\,.
\label{eq:potential}
\end{eqnarray}

In the scalar potential $V(H,\phi, S)$ above (in Eq.~\ref{eq:potential}), we choose $\lambda_{H},\lambda_{S},\mu_\phi^2>0$ and $\mu_{H}^2,\mu_{S}^2<0$
so that it provides a minimum with the following vacuum:
\bea
H=\begin{pmatrix}
\phi^+ & \\
\frac{v+h+i\phi_0}{\sqrt{2}} &\\
\end{pmatrix}  \to 
\langle H \rangle =\begin{pmatrix}
0 & \\
\frac{v}{\sqrt{2}} &\\
\end{pmatrix}\,; 
~ S =\frac{1}{\sqrt{2}}(v_s+s+iA) \to \langle S \rangle =\frac{1}{\sqrt{2}}v_s; 
~\langle \phi \rangle=0\,. 
\eea 
Therefore, two scalar fields acquire non-zero vev: $\langle S \rangle=v_s/\sqrt{2}$, which breaks 
$SM\times U(1)_X \to SM$ and  $\langle H \rangle = v/\sqrt{2}$, which causes spontaneous EWSB: 
$SU(2)_L\times U(1)_Y \to U(1)_{EM}$. In above, $\phi^{\pm,0},A$ denote Nambu-Goldstone Bosons 
which disappear in the unitary gauge after EWSB. We draw the reader's attention here to a notation 
used further in the draft, where $S$ is referred to the complex scalar singlet field, while $s$ refers to the 
physical scalar particle after $U(1)_X$ breaking. Note also that $v_s$ renders the $U(1)_X$ gauge boson massive via: 
\bea
m_X=g_X v_s\,,
\label{mrel}
\eea
where $g_X$ denotes $U(1)_X$ gauge coupling constant. The value of $v_s$ denotes the scale of $U(1)_X$ symmetry breaking and
is crucially governed by the condition whether $X$ is FIMP or WIMP. EWSB scale ($v$) is known from 
SM gauge boson masses to be $v=246$ GeV. Physical particles that arise in the model, 
depends on the scale (before or after EWSB) and will be elaborated in the respective regimes.

\begin{figure}[htb]
$$
\includegraphics[scale=0.35]{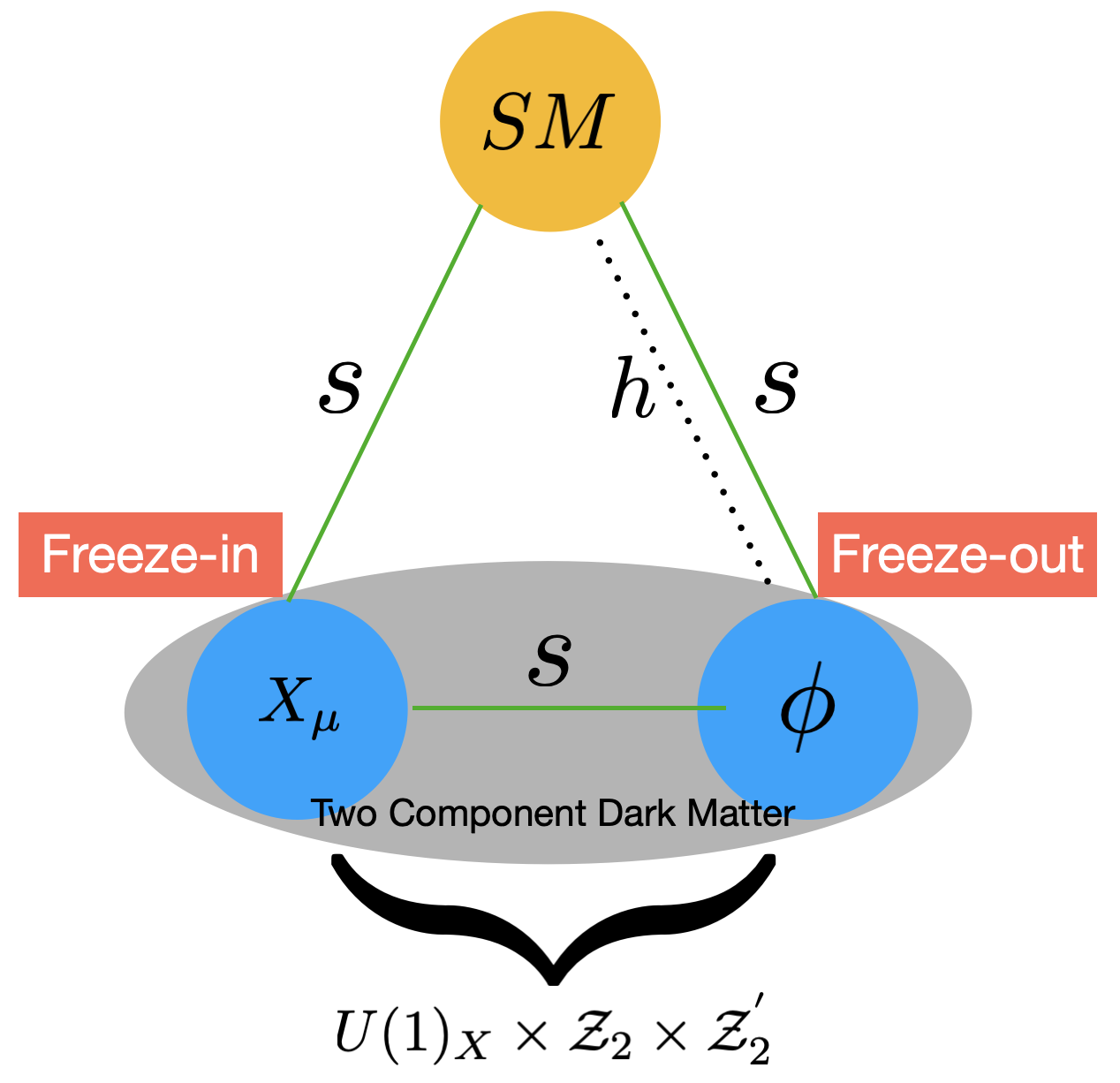}
\includegraphics[scale=0.5]{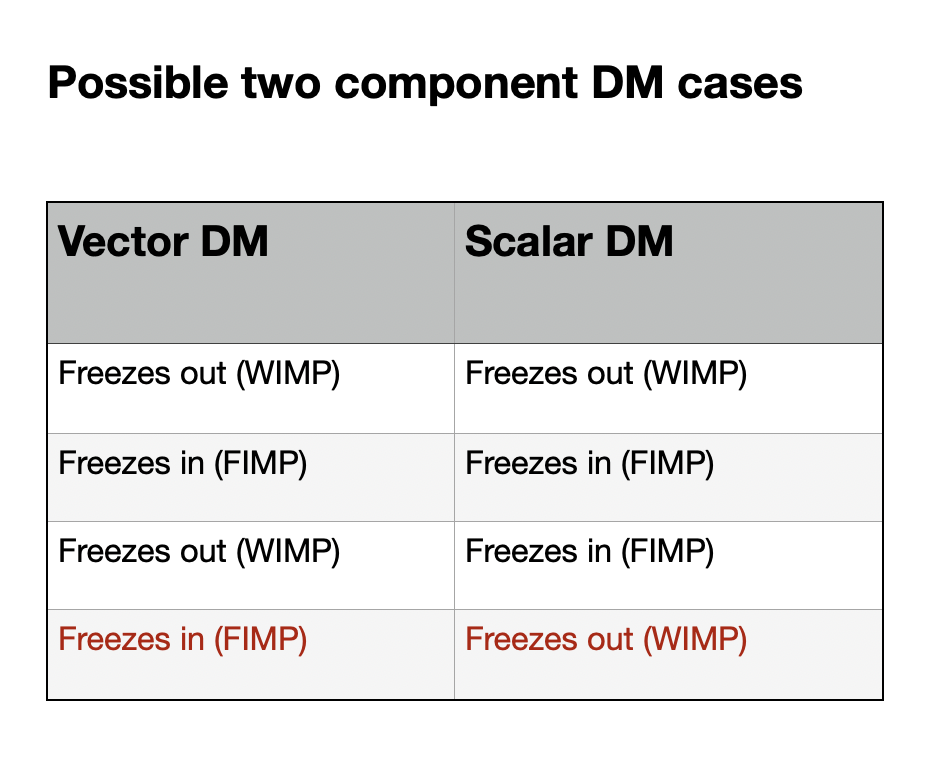}
$$
\caption{Left: A cartoon depicting the DM components in the model and their interactions to visible sector and amongst them (see text for notation); Right: Possible 
scenarios in the two component DM set-up, the one in colour is considered in this analysis.}
\label{fig:cartoon}
\end{figure}

We further note that $X$ being odd under $\mathcal{Z}_2$, requires: 
\begin{equation}
{\mathcal{Z}}_2: X \to -X,~~\implies {\mathcal{Z}}_2: S \to S^{*}.
\end{equation}
The last relation follows from straightforward calculation. First of all,
\begin{eqnarray}
|{D}^\mu S|^2 &&= \left[(\partial^\mu+ig_XX^\mu) S\right]^{*}\left[(\partial^\mu+ig_{X}X^\mu) S\right], \nonumber\\
&&= (\partial^\mu-ig_{X}X^\mu)S^{*}(\partial^\mu+ig_XX^\mu)S\,.
\label{eq:z21}
\end{eqnarray}
The transformation of the kinetic piece $|{D}^\mu S|^2$ under ${\mathcal{Z}}_2$ goes as,
\begin{eqnarray}
{\mathcal{Z}}_2: |{D}^\mu S|^2 &&= (\partial^\mu-ig_XX^\mu)S^{*}(\partial^\mu+ig_XX^\mu)S \nonumber \\ &&\rightarrow  
(\partial^\mu+ig_XX^\mu)S^{*}(\partial^\mu-ig_XX^\mu)S\,.
\label{eq:z22}
\end{eqnarray}
Comparing Eq.~\ref{eq:z22} with Eq.~\ref{eq:z21}, we get ${\mathcal{Z}}_2: S \to S^{*}$.

To summarise, the model inherits two DM components, a VBDM $X$ and a scalar singlet $\phi$ in 
$U(1)_X \times \mathcal{Z}_2 \times \mathcal{Z}^{'}_2$ extension of SM. Both of them interact with each other 
and with SM via scalar particle $s$\footnote{$s-H$ mixing after EWSB also connects VBDM to SM via SM Higgs.}, 
while $\phi$ also interacts via SM Higgs ($H$) portal. The dark sector particles and their interactions are sketched in a cartoon 
in the left panel of Fig.~\ref{fig:cartoon}. Four different phenomenological situations emerge depending on which DM freezes 
in (FIMP) and which freezes out (WIMP), as shown in the right panel of Fig.~\ref{fig:cartoon}. We explore the possibility when 
$X$ is a FIMP and $\phi$ is an WIMP like DM.

\section{Possibilities with $X$ freezing-in and $\phi$ freezing-out }
\label{sec:possibilities}

\begin{figure}[htb]
$$
\includegraphics[scale=0.43]{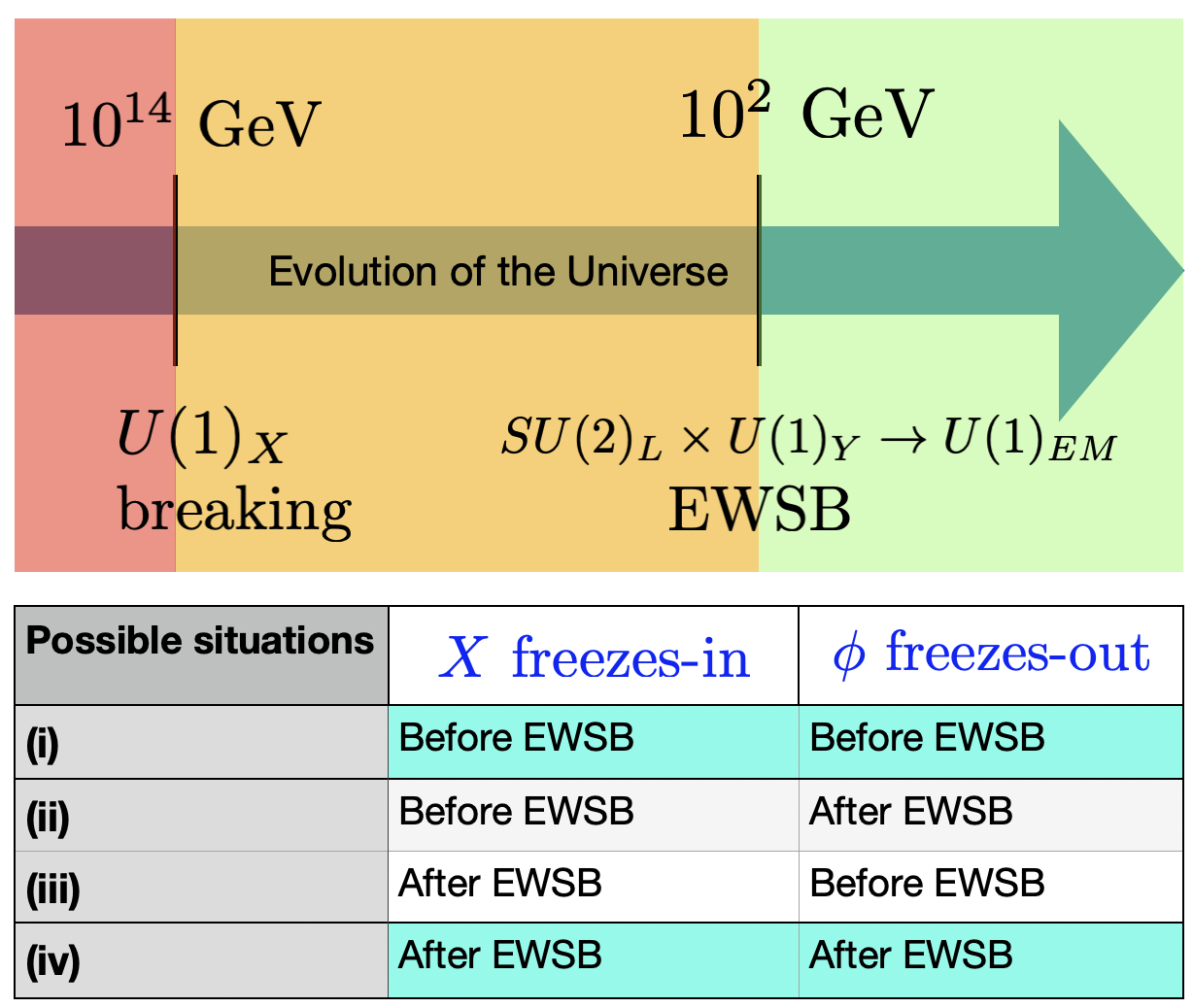}
$$
\caption{The relevant symmetry breaking scales of the model : $U(1)_X$ breaking and EWSB (top panel) and 
different phenomenological situations for $X$ to freeze-in and $\phi$ to freeze-out (bottom panel).}
\label{fig:cartoon2}
\end{figure}

The first noteworthy feature of the model is the presence of two widely different symmetry breaking scales: (i) $SM\times U(1)_X \to SM$, 
and (ii) EWSB: $SU(2)_L\times U(1)_Y \to U(1)_{EM}$, which are pictorially depicted in Fig.~\ref{fig:cartoon2}. 
While EWSB scale is known, $U(1)_X$ breaking scale crucially depends on whether $X$ freezes in or freezes out. 
It is explained in a moment. $X$ as FIMP and $\phi$ as WIMP inherit yet another set of phenomenological possibilities that 
the model offers and are noted in the bottom panel of Fig.~\ref{fig:cartoon2}. 

\begin{itemize}[wide, labelwidth=!, labelindent=0pt]
  \item \textbf{Freeze-in of $X$ and $U(1)_X$ breaking scale}:
The VBDM $X$ to be a cold DM dictates the scale for $U(1)_X$ breaking. The $U(1)_X$ gauge coupling 
($g_X$) which provides DM-SM interaction, is required to be feeble (roughly 
$g_X \sim 10^{-11}$) to keep it out of equilibrium. The DM yield that generates correct relic is proportional to 
the production cross section (or decay width). For $m_X \sim$ TeV, so that it behaves as a cold dark matter (CDM),
the $U(1)_X$ breaking scale turns out to be $v_s\sim 10^{14}$ GeV (following Eq.~\ref{mrel}). 
On the other hand, EWSB occurs at $\Tew \sim$ 160 GeV \cite{Carena:2004ha,Baker:2018vos}, 
corresponding to $v_{\rm EW}\sim v$ = 246 GeV. Therefore, the hierarchy $v_s \gg v$ implies $T_{U(1)} \gg \Tew$ 
(see Fig.~\ref{fig:cartoon2}), which further aids to the distinction between freeze-in before EWSB (bEWSB) and after EWSB (aEWSB).
   
   \begin{enumerate}[(a)]
      \item \underline{Freeze-in bEWSB ($\Tew < m_X < T_{U(1)}$)}:  
      When the DM freezes in completely bEWSB, the DM production saturates before $\Tew$; 
      then characteristic freeze-in scale is depicted by $T_{\rm{FI}}$ or $ x_{\rm{FI}}=\frac{m_X}{T_{\rm{FI}}}$ requires
      \bea
      T_{\rm{FI}}>T_{\rm{EW}}; ~\implies x_{\rm{FI}}<x_{\rm {EW}}\,.
      \eea
      Note also, that the characteristic freeze-in temperature is correlated to the DM mass, $T_{\rm{FI}}\sim m_X$.
     Obviously, $x_{\rm {EW}}=\frac{m_X}{\Tew}$. In this regime, only the singlet scalar {$S$} acquires a vev ($v_{\rm s}$) to give mass to $X$. 
     Other scalars (Higgs and $\phi$) are also massive due to bare mass term, while all the SM fields are massless. 
     In such a situation $X$ has no connection to SM and the production occurs via the interaction with the physical scalar $s$, which 
     is assumed to be in the thermal bath due to sizeable portal coupling with SM Higgs ($H$). The details of the production processes 
     will be discussed in section \ref{sec:bEWSB} when we elaborate such a scenario\footnote{ We shall also note that freeze-in bEWSB do not 
     include the possibility of $T_{\rm{FI}}>T_{U(1)}$, as $X$ is massless in that regime.}.

      \item \underline{Freeze-in aEWSB ($m_X \lesssim \Tew $): }
      When DM production from thermal bath continues even aEWSB, we have,
      \bea
      T_{\rm{FI}}\sim m_X<T_{\rm{EW}}; ~\implies  x_{\rm{FI}}> x_{\rm {EW}}\,.
      \eea
      
      After EWSB, the Higgs doublet ($H$) acquires a vev ($v$); $H$ and $s$ mix to yield physical scalars 
      $h_1$ and $h_2$, where $h_1$ is assumed to be SM Higgs (dominantly doublet), the one observed at LHC with 
      $m_{h_1}\sim$ 125 GeV, while $h_2$ is dominantly a singlet, heavier or lighter than the SM Higgs following the existing bounds. 
      Naturally, DM can be additionally produced from the SM particles in thermal bath aEWSB, providing a different allowed parameter space. 
      The detailed discussion is taken up in section \ref{sec:aEWSB}. 

      A cartoon of freeze-in bEWSB (in blue) and aEWSB (in red) is shown in the left panel of Fig.~\ref{fig:cartoon3} 
      in $Y-x$ plane, where $Y=\frac{n}{\s}$ refers to DM yield with $\s$ denoting to entropy density and $x=\frac{m}{T}$, 
      with $m$ denoting DM mass and $T$ denoting temperature of the bath (details in subsection \ref{sec:bEWSB-cBEQ}). 
      From left to right along x-axis, $x$ becomes larger with $T$ dropping. In Fig.~\ref{fig:cartoon3} (left panel) we consider 
      two DM species denoted by $\encircled{1}$ and $\encircled{2}$ with $m_2>m_1$. Following $T_{\rm{FI}}\sim m_X$, we have $x_{\rm{FI}}\sim 1$ 
      for both cases. Then following, $x_{\rm{FI}}<x_{\rm {EW}}$ for freeze-in bEWSB, we need $m>T_{\rm{EW}}$, while for freeze-in aEWSB, 
      $x_{\rm{FI}}> x_{\rm {EW}}$ requires $m<T_{\rm{EW}}$. Therefore, we can have $m_2>T_{\rm{EW}}>m_1$ where $\encircled{2}$ freezes in 
      bEWSB and $\encircled{1}$ freezes in aEWSB. The red vertical dotted line shows $x_{\rm{FI}}\sim 1$, while pink and blue dotted vertical lines 
      indicate $(x_1)_{\rm{EW}}$ and $(x_2)_{\rm{EW}}$ with $(x_2)_{\rm{EW}}>(x_1)_{\rm{EW}}$. The relative abundance shown here depends 
      on DM-SM interaction and has no implication unless discussed in context of the model.
    
    \end{enumerate}
    \item \textbf{Freeze-out of $\phi$ and EWSB}: The real scalar singlet $\phi$ is assumed to be in thermal bath via non-suppressed portal 
    couplings. It freezes out through the dominant $2\to 2$ annihilation to SM and also to other DM candidate (if kinematically allowed). The freeze-out 
    do not crucially dictate any scale in the model unlike freeze-in. Here also, two possibilities emerge:
    \begin{enumerate}[(a)]
    \item \underline{Freeze-out bEWSB} : For $\phi$ to freeze out bEWSB, one requires the characteristic freeze-out temperature ($T_{\rm{FO}}$) 
    to follow,
    \bea
    T_{\rm{FO}}>\Tew \implies x_{\rm{FO}}<x_{\rm{EW}}\, ;
    \eea
    where $x_{\rm{FO}}=\frac{m_\phi}{T_{\rm{FO}}}$.
    The processes that are responsible for freeze out of $\phi$ bEWSB are only through the coupling with the physical scalar 
    $s$ and will be elaborated in section \ref{sec:bEWSB}.
    \item \underline{Freeze-out aEWSB} : Freeze out of $\phi$ aEWSB implies:
    \bea
    T_{\rm{FO}}<\Tew \implies x_{\rm{FO}}>x_{\rm{EW}}\,.
    \eea
    In this regime, the interaction between $\phi$ and SM arises through $h-s$ mixing and occurs via both
    the physical scalars $h_1$ and $h_2$. Therefore, new channels contribute to DM number depletion, 
    we discuss them in details in section \ref{sec:aEWSB}.
    \end{enumerate}
    
    \begin{figure}[htb]
$$
\includegraphics[scale=0.43]{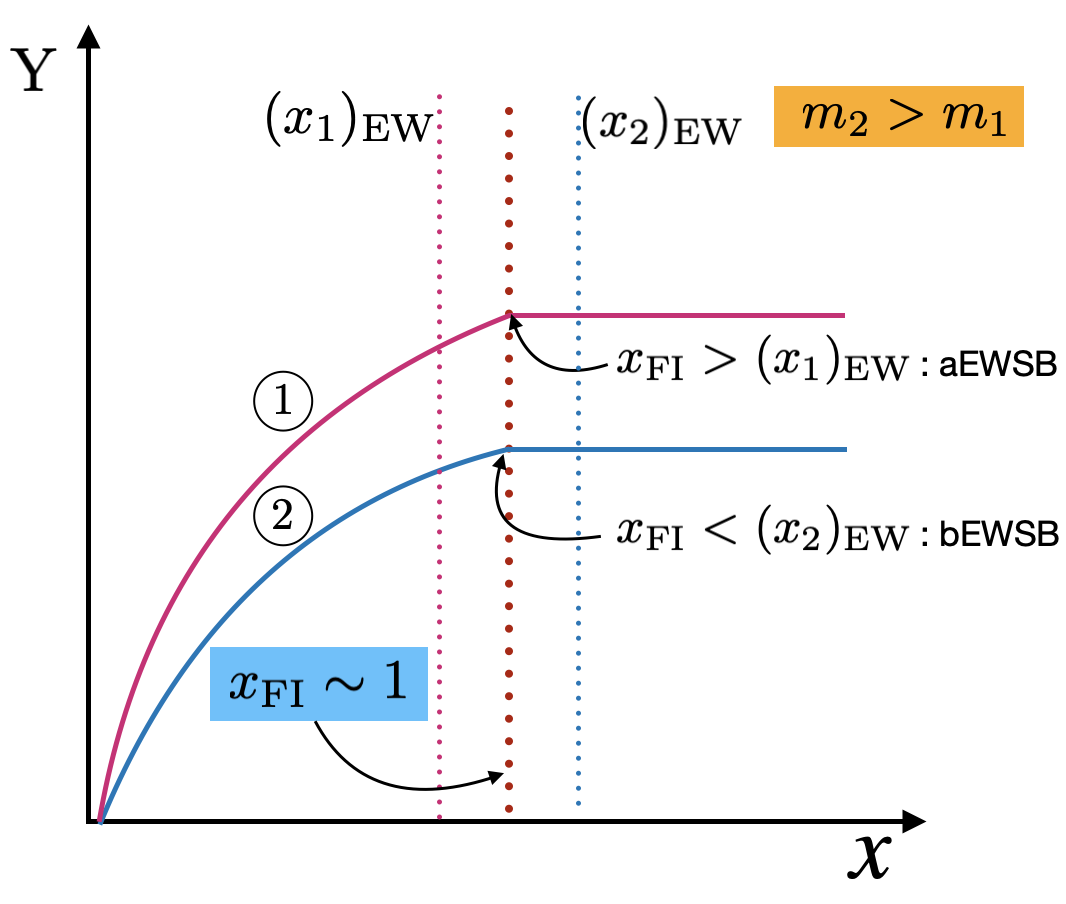}
\includegraphics[scale=0.42]{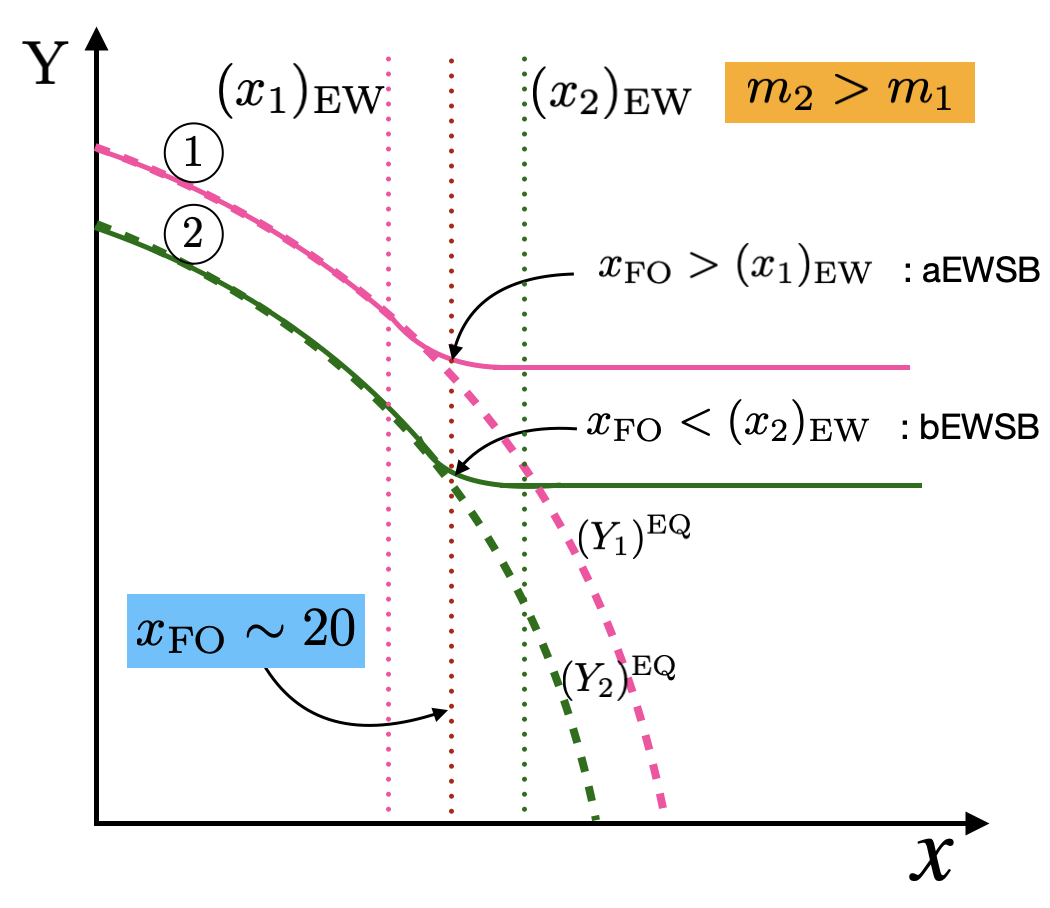}
$$
\caption{Left: A cartoon freeze-in in the $Y-x$ plane for two sample DMs with $m_2>m_1$ 
which shows freeze-in aEWSB (in red) and bEWSB (in blue). Right: Cartoon freeze-out which 
shows freeze-out aEWSB (in pink) and bEWSB (in green) from the respective equilibrium distributions 
(in dashed curves) shown in the right panel. }
\label{fig:cartoon3}
\end{figure}

    Unlike freeze-in, freeze-out of DM does not directly constrain the DM mass. 
    However, for freeze-out to render correct relic density, $x_{\rm{FO}}$ remains in the ballpark 
    $x_{\rm{FO}}\sim 20-25$. Inevitably, freeze-out bEWSB or aEWSB can be 
    realized for different DM masses ($m$), which changes $x_{\rm{EW}}=\frac{m}{\Tew}$ to lie above or below $x_{\rm{FO}}$, 
    as depicted in the right panel of Fig.~\ref{fig:cartoon3}. We again consider two DM species denoted by $\encircled{1}$ and $\encircled{2}$ 
    with $m_2>m_1$ (shown by green and pink lines); so that $(x_2)_{\rm {EW}}>(x_1)_{\rm {EW}}$ (vertical dotted lines) depicts freeze-out bEWSB ($x_{\rm{FO}}<(x_2)_{\rm {EW}}$) and aEWSB ($x_{\rm{FO}}>(x_1)_{\rm {EW}}$) respectively from their equilibrium distributions. 
    The distributions and the relative yields in this figure have not been sketched for a particular model or interaction, so the relative strengths have no 
    implications, this is just for the illustration purpose. 

\end{itemize}

Out of different freeze-in epochs of $X$ and freeze-out of $\phi$, as noted in the bottom panel of Fig.~\ref{fig:cartoon2}, we explore 
two sample cases here with (i) both freeze-in and freeze-out occurring bEWSB and (ii) both occurring aEWSB, which capture the most interesting 
distinctions of allowed parameter space of the model. 

\section{Dark Matter phenomenology bEWSB}
\label{sec:bEWSB}

Here we address in details the freeze-in of $X$ and freeze-out of $\phi$ both occurring bEWSB (option (i) of the bottom 
panel in Fig.~\ref{fig:cartoon2}.). To be specific, the temperatures around which freeze-in ($T_{\rm{FI}}$) and 
freeze-out ($T_{\rm{FO}}$) occur, lie between $U(1)_X$ breaking and EWSB, i.e.
\bea
T_{U(1)}>T_{\rm FI}>\Tew; ~~T_{U(1)}>T_{\rm FO}>\Tew\,.
\eea
In the following subsections, we discuss the physical particles and interactions in this regime for DM freeze-in and freeze-out via 
coupled BEQ, relic density and direct search allowed parameter space of the model.

\subsection{Physical states and parameters}
\label{sec:bEWSB-parameters}

As the regime is dictated by interactions after spontaneous $U(1)_X$ breaking and bEWSB, in unitary gauge we have,
\bea
S =\frac{1}{\sqrt{2}}(v_s+s) \to \langle S \rangle =\frac{1}{\sqrt{2}}v_s, ~~\langle H \rangle=0,~~\langle \phi\rangle=0\,.
\eea
The scalar potential in this limit reads:
\begin{eqnarray}
 V_{\rm scalar}=\mu_{H}^2(H^{\dagger}H)+\lambda_H(H^{\dagger}H)^2+\frac{1}{2}\mu_{\phi}^2\phi^2+\frac{1}{4!}\lambda_{\phi}\phi^4 +\frac{1}{2}\mu_{S}^2(v_s+s)^2+\frac{1}{4}\lambda_{S}(v_s+s)^4  \nonumber \\
+\frac{1}{2}\lambda_{\phi H}(H^{\dagger}H)\phi^2+\frac{\lambda_{H S}}{2}(H^{\dagger}H)(v_s+s)^2+\frac{\lambda_{\phi S}}{4}\phi^2(v_s+s)^2\,. 
\label{eq:potBEWSB}
\end{eqnarray} 
The physical scalars can be identified from extremization of the potential, which provides following relations between the neutral physical 
scalars and parameters of the model:

\begin{align}\begin{split}
&\frac{\partial V_{\rm scalar}}{\partial s}=0 \to \mu_S^2=-\lambda_Sv_s^2 \,,\\
& \frac{\partial ^2V_{\rm scalar}}{\partial H^{\dagger}\partial H}=m_H^2 \to \mu_H^2=m_{H}^2-\frac{\lambda_{H S}}{2}v_s^2 \,,  \\
& \frac{\partial^2 V_{\rm scalar}}{\partial \phi^2}=m_{\phi}^2 \to \mu_{\phi}^2=m_{\phi}^2-\frac{\lambda_{\phi S}}{2}v_s^2 \,, \\
& \frac{\partial^2 V_{\rm scalar}}{\partial s^2}=m_{s}^2 \to \mu_{S}^2=m_{s}^2-3\lambda_{S}v_s^2 \,;
\end{split}\label{mrel_3}
\end{align}

where $m_H$ is the mass of the SM like Higgs bEWSB and $m_{\phi}$ is the mass of real scalar DM $(\phi)$ bEWSB. 
We must note here that $H$ bEWSB represents {\it four} massive scalar degrees of freedom (d.o.f) \cite{Biswas:2019iqm,Heeba:2018wtf,
DeRomeri:2020wng,Chianese:2018dsz}  being part of the complex isodoublet, while $\phi$ has only one d.o.f being a real scalar singlet. Using  Eq.~\ref{eq:potBEWSB} and Eq.~\ref{mrel_3}, it is easy to show that the mass of the $U(1)_X$ complex 
scalar turns out to be $m_s^2=2\lambda_S v_s^2$. Although $m_H$ and $m_s$ can be treated as free parameters, they must reproduce
correct Higgs mass and mixing, see discussions in the next subsection \ref{sec:constraints} and appendix \ref{sec:aEWSB-details}.

This allows us further to identify the parameters of the model that are relevant for the analysis. All the physical masses and the couplings 
controlling the relic density of DM are chosen as the external parameters. Quartic self couplings like $\lambda_{H}$ 
and $\lambda_{\phi}$ are fixed at values within the limit of DM self scattering ($\sim 0.1$) and plays minimal role in DM-SM interaction. 
All the other parameters are considered as internal parameters which are defined by the relations described in Eqns. \ref{mrel} and \ref{mrel_3}. 
Table \ref{tab:constraints} summarises the parameters of the model bEWSB. The parameters of the model are subject to further constraints as 
explained in the next subsection. 
\begin{table}[htb]
\centering
\begin{tabular}{|c |c |} \hline
External parameters & Internal parameters \\ \hline \hline
$m_{H}$, $m_s$, $m_{\phi}$, $m_X$, $g_X$,
$\lambda_H$, $\lambda_{\phi}$, $\lambda_{H S}$, $\lambda_{\phi S}$, $\lambda_{\phi H}$ 
& $\mu_H$, $\mu_{\phi}$, $\mu_{S}$, $v_s$, $\lambda_S$ \\ \hline
\end{tabular}
\caption{External and internal parameters of the model bEWSB.}
\label{tab:constraints}
\end{table}

\subsection{Constraints and bounds}
\label{sec:constraints}

In this section, we discuss the possible theoretical and experimental constraints on parameters of the model relevant for our analysis.

\begin{itemize} 

 \item {\bf Stability:\\} In order to get the potential bounded from below, the quadratic couplings of the potential $ V_{\rm scalar}$ must satisfy the following co-positivity conditions as \cite{Elias-Miro:2012eoi,Kannike:2016fmd,Chakrabortty:2013mha},
\bea
\begin{split}
&\lambda_H\geq 0,\lambda_{\phi}\geq0,\lambda_S\geq0,\\&
\lambda_{\phi H}+2\sqrt{\lambda_H\lambda_{\phi}}\geq0,\lambda_{HS}+2\sqrt{\lambda_H\lambda_S}\geq0,\lambda_{\phi S}+2\sqrt{\lambda_{\phi}\lambda_S}\geq0 \,.
\end{split}
\eea
In this analysis, we choose all the couplings positive, which satisfy the above conditions trivially.

\item{\bf Perturbativity:\\}
In oder to maintain perturbativity of the theory, the quartic couplings of the
scalar potential $ V_{\rm scalar}$ and the gauge couplings obey \cite{Bhattacharyya:2015nca,Bhattacharya:2019fgs},
\bea
\begin{split}
&\lambda_H<4\pi,\hspace{2cm}\lambda_S<4\pi,\hspace{2cm}\lambda_{\phi}<4\pi,\hspace{2cm}g_X<\sqrt{4\pi},\\&
\lambda_{HS}<4\pi,\hspace{1.86cm}\lambda_{\phi S}<4\pi,\hspace{1.82cm}\lambda_{\phi H}<4\pi\,.
\end{split}
\eea
 \item{\bf Tree level unitarity:\\}
 Tree level unitarity of the theory, coming from all possible $2\to 2$ scattering amplitude, can be ensured \cite{Horejsi:2005da,Bhattacharyya:2015nca,Bhattacharya:2019fgs} via following constraints
\bea
|\lambda_{\phi}|<8\pi,~|\lambda_{S}|<4\pi,~|\lambda_{H}|<4\pi,~|\lambda_{\phi S}|<8\pi,~|\lambda_{ H S}|<8\pi,~|\lambda_{\phi H}|<8\pi\,.
\eea

\item {\bf Constraints on DM mass bEWSB:\\} For freeze-in of $X$ to complete bEWSB, it is  required that the freeze-in scale ($T_{{\rm FI}}\sim m_X$) 
has to be larger than $\tew\sim$ 160 GeV; then, $m_X \gtrsim$ 160 GeV. Similarly, freeze-out 
of $\phi$ bEWSB forces the freeze-out temperature $T_{\rm FO}$ to be larger than $\tew$, i.e, $T_{\rm FO} \gtrsim$ 160 GeV. 
This condition automatically implies that $x_{\rm FO}= m_{\phi}/{T_{\rm FO}}$, which is typically $\sim\ $25 for WIMP freeze-out, 
requires the following condition on WIMP and FIMP masses:
\begin{equation}
{\rm WIMP}:~m_{\phi} \gtrsim 4\ {\rm TeV}; ~~{\rm FIMP}:~ m_X \gtrsim 160 \ {\rm GeV}.
\label{m_fi_2}
\end{equation}
 
\item {\bf Relic density}: One of the most important constraints on the parameters of the model comes from the observed relic abundance of DM. 
The latest observations from anisotropies in CMBR in experiments like WMAP \cite{WMAP:2012nax}, and PLANCK \cite{Planck:2018vyg} indicate
\bea
\Omega_{\rm DM}h^2=\Omega_{X}h^2+\Omega_{\phi}h^2=0.1200\pm 0.0012 \,,
\label{eq:totrelic}
\eea
where $\Omega=\frac{\rho}{\rho_c}$ refers to cosmological density, with $\rho_c$ indicating critical density, $h$ is the present Hubble parameter 
scaled in units of 100 km/s/Mpc. In the two component WIMP-FIMP set up that we explore here, the individual relic densities of $X$ and $\phi$ 
shall add to the total observed relic, where each of the individual components will be under-abundant, {\it ie}, $\Omega_{X,\phi}h^2\lesssim 0.12$.
We elaborate on the relic density of the WIMP and FIMP components of the model in the next section.

\item {\bf Direct detection (DD) constraints}: WIMP ($\phi$) has a direct search prospect, while the FIMP ($X$) does not
\footnote{ The interaction of $X$ with SM occurs via physical scalars $h_{1,2}$. Due to tiny $\lambda_{HS}$ coupling 
$(\sim10^{-12})$ and heavy mass of the mediators $\rm{\sim \mathcal{O} (100)GeV}$ as considered here, FIMP $X$ does not have any direct search prospect. 
But, the smallness of the dominantly singlet scalar mass $m_{h_2}\sim\rm{MeV}$ may bring $X$ under the DD scanner \cite{Duch:2017khv}.}. 
In this regime, although freeze out of $\phi$ occurs bEWSB, after EWSB $\phi$ couples to SM, yielding a possibility of direct search of $\phi$. 
 Non observation of DM from ongoing experiments like XENON1T \cite{XENON:2018voc} sets an stringent upper limit on WIMP-nucleon 
spin-independent elastic scattering cross-section at 90\% C.L,
\bea\begin{split}
\sigma_{\rm{SI}}\sim\begin{cases}4.1\times 10^{-47}~\rm{cm}^2~~\rm{~~~~at~30~GeV/c^2~~(XENON1T)}\\1.4\times 10^{-48}~\rm{cm}^2~~\rm{~~~~at~50~GeV/c^2~~(XENONnT)}\end{cases}
\end{split}\label{dd_xenon1T-nT}\eea
 Here, we also mention the projected XENONnT \cite{XENON:2020kmp} sensitivity. The relevant couplings $g_X$ and 
$\lambda_{\phi S}$,\footnote{Also, as $v_s\sim 10^{14}$ GeV, the requirement of keeping $X$ out-of-equilibrium demands that the coupling $\lambda_{\phi S}$ should be as small as $\sim 10^{-12}$.} are constrained to be $\sim 10^{-12}$ both from the freeze-in requirements and direct search bounds.
In addition, $\lambda_{\phi H} \sim 10^{-3}$ keeps the DD cross section safely below the experimental direct search bounds. 
We provide a detailed account for direct search of the model in the Appendix \ref{sec:directsearch}.

 \item {\bf Higgs mass and mixing}: It is important to note that independent of DM freeze-out or freeze-in to occur bEWSB or aEWSB, there is a mixing of Higgs ($h$) with $s$ after EWSB to render two physical states: $h_1$, assumed to be SM like Higgs 
with $m_{h_1}=125.1$ GeV and a heavy or light $h_2$, dominantly a singlet. $h-s$ mixing angle ($\theta$) 
is restricted by LHC data \cite{Chalons:2016lyk} within:
\bea
|\sin\theta|\lesssim 0.3 \, .
\eea
The requirement of correct Higgs mass as well as mixing puts limit on parameters $m_H,m_s,\lambda_{HS}$ etc. For details of the mass eigenstates, mixing 
and relations with model parameters, refer to the discussions in appendix \ref{sec:aEWSB-details}. We note here one important exception, if 
$m_s\gtrsim 2m_X$ then dominant FIMP production occurs from $s$ decay, and additionally if the FIMP production from late decay of $s$ saturates bEWSB, then there is no 
$s$ which remains in the bath to mix with $h$ and consequently no $h_2$ state to appear aEWSB. Then Higgs mass bEWSB and aEWSB are related by 
$m_H^2=\frac{m_{h_1}^2}{2}$. Note further, collider bound on scalar singlet WIMP DM mass $m_\phi$ is mild 
\cite{Barger:2007im,Fuchs:2020cmm}, while no significant bound on FIMP mass can be obtained.

 \item{\bf Invisible Decay of Higgs :} SM Higgs ($h_1$) can decay into pairs of DM ($X$ and $\phi$) as well as to pairs of $h_2$ 
in our model if kinematically allowed. Since these decays contribute to invisible decay of Higgs, corresponding $h_1\phi\phi,~h_1XX,~h_1h_2h_2$ 
couplings get severely constrained by the experimental data. They can be traced from expressions of Higgs decay width to 
$\phi,X,h_2$ as provided in Appendix \ref{sec:invisible}. As per the latest experimental data given by ATLAS 
(for 139 $fb^{-1}$ luminosity at $\sqrt{s}=13$ TeV), the strictest upper limit on $\mathcal{B}_{h_1\to\text{inv}}$ can be set to $\sim$\, 0.13 at 
95\% CL \cite{Milosevic:2020xup,ATLAS:2020kdi,Okawa:2020jea}. A comparison of the invisible Higgs decay bound from latest 
ATLAS and CMS data is given by:

\bea\begin{split}
&\mathcal{B}_{h_1\to\rm{inv}}< \begin{cases}0.13~~~\rm( ATLAS)\\0.19\rm~~~( CMS)\end{cases}\\&
\Gamma_{h_1\to\rm{inv}}<  \begin{cases}0.61\rm{~MeV}~~~\rm( ATLAS)\\0.95\rm{~MeV}~~~( CMS)\end{cases}
\end{split}\label{higgs_invisible_decay}\eea\par

For simplicity, we do not scan the region of parameter space where Higgs invisible decay to 
DM is possible, with $m_X>m_{h_1}/2$, and $m_\phi>m_{h_1}/2$. 
However, $h_2\to XX$ is considered for the analysis, where there is no bound.
\end{itemize}

\subsection{Processes contributing to DM Relic}
\label{sec:bEWSB-process}

\begin{figure}[htb!]
	\centering
  \begin{tikzpicture}[baseline={(current bounding box.center)}]
  \begin{feynman}
  \vertex (a);
  \vertex[above left=1cm and 1cm of a] (a1){\(\phi\)}; 
  \vertex[above right=1cm and 1cm of a] (a3){\(s\)}; 
  \vertex[below left=1cm and 1cm of a] (a2){\(\phi\)}; 
  \vertex[below right=1cm and 1cm of a] (a4){\(s\)}; 
  \diagram* {
  	(a1) -- [scalar] (a) -- [scalar] (a3),
  	(a2) -- [scalar] (a) -- [scalar] (a4)
  };\end{feynman}
\end{tikzpicture}\hspace{0.5cm}	
  \begin{tikzpicture}[baseline={(current bounding box.center)}]
  \begin{feynman}
  \vertex (a);
  \vertex[right=1cm of a] (b);
  \vertex[above left=1cm and 1cm of a] (a1){\(\phi\)}; 
  \vertex[below left=1cm and 1cm of a] (a2){\(\phi\)}; 
  \vertex[above right=1cm and 1cm of b] (a3){\(s\)}; 
  \vertex[below right=1cm and 1cm of b] (a4){\(s\)}; 
  \diagram* {
  	(a) -- [scalar] (a1),(a2) -- [scalar] (a),
  	(a3) -- [scalar]	(b) -- [scalar] (a4),(a)  --[scalar,edge label={\(s\)}] (b)
  };\end{feynman}
\end{tikzpicture}\hspace{0.5cm}	
  \begin{tikzpicture}[baseline={(current bounding box.center)}]
  \begin{feynman}
  \vertex (a);
  \vertex[below=2cm of a] (b);
  \vertex[ left=1cm and 1cm of a] (a1){\(\phi\)}; 
  \vertex[ right=1cm and 1cm of a] (a2){\(s\)}; 
  \vertex[ left=1cm and 1cm of b] (a3){\(\phi\)}; 
  \vertex[ right=1cm and 1cm of b] (a4){\(s\)}; 
  \diagram* {
  	(a) -- [scalar] (a1),(a2) -- [scalar] (a),
  	(a3) -- [scalar] (b) -- [scalar] (a4),(a)  --[scalar,edge label={\(\phi\)}] (b)
  };\end{feynman}
  \end{tikzpicture}
  
  \begin{tikzpicture}[baseline={(current bounding box.center)}]
  \begin{feynman}
  \vertex (a);
  \vertex[below=2cm of a] (b);
  \vertex[ left=1cm and 1cm of a] (a1){\(\phi\)}; 
  \vertex[ right=1cm and 1cm of a] (a2){\(s\)}; 
  \vertex[ left=1cm and 1cm of b] (a3){\(\phi\)}; 
  \vertex[ right=1cm and 1cm of b] (a4){\(s\)}; 
  \diagram* {
  	(a) -- [scalar] (a1),(a4) -- [scalar] (a),
  	(a3) -- [scalar] (b) -- [scalar] (a2),(b)  --[scalar,edge label={\(\phi\)}] (a)
  };\end{feynman}
\end{tikzpicture}\hspace{0.5cm}	
  \begin{tikzpicture}[baseline={(current bounding box.center)}]
  \begin{feynman}
  \vertex (a);
  \vertex[right=1cm of a] (b);
  \vertex[above left=1cm and 1cm of a] (a1){\(\phi\)}; 
  \vertex[below left=1cm and 1cm of a] (a2){\(\phi\)}; 
  \vertex[above right=1cm and 1cm of b] (a3){\(H\)}; 
  \vertex[below right=1cm and 1cm of b] (a4){\(H\)}; 
  \diagram* {
  	(a) -- [scalar] (a1),(a2) -- [scalar] (a),
  	(b) -- [charged scalar]	(a3),(a4) -- [charged scalar] (b),(a)  --[scalar,edge label={\(s\)}] (b)
  };\end{feynman}
\end{tikzpicture}\hspace{0.5cm}	
  \begin{tikzpicture}[baseline={(current bounding box.center)}]
  \begin{feynman}
  \vertex (a);
  \vertex[above left=1cm and 1cm of a] (a1){\(\phi\)}; 
  \vertex[above right=1cm and 1cm of a] (a3){\(H\)}; 
  \vertex[below left=1cm and 1cm of a] (a2){\(\phi\)}; 
  \vertex[below right=1cm and 1cm of a] (a4){\(H\)}; 
  \diagram* {
  	(a1) -- [scalar] (a) -- [charged scalar] (a3),
  	(a) -- [scalar] (a2),(a4) -- [charged scalar] (a)
  };\end{feynman}
  \end{tikzpicture}
  
  \begin{tikzpicture}[baseline={(current bounding box.center)}]
  \begin{feynman}
  \vertex (a);
  \vertex[right=1cm of a] (b);
  \vertex[above left=1cm and 1cm of a] (a1){\(\phi\)}; 
  \vertex[below left=1cm and 1cm of a] (a2){\(\phi\)}; 
  \vertex[above right=1cm and 1cm of b] (a3){\(X\)}; 
  \vertex[below right=1cm and 1cm of b] (a4){\(X\)}; 
  \diagram* {
  	(a) -- [scalar] (a1),(a2) -- [scalar] (a),
  	(a3) -- [boson]	(b) -- [boson] (a4),(a)  --[scalar,edge label={\(s\)}] (b)
  };\end{feynman}
  \end{tikzpicture}
  \caption{Feynman diagrams contributing to $\phi$ freeze-out via $2 \to 2$ depletion processes bEWSB.}
  \label{fig:frzout-FD-EWSB}
\end{figure}
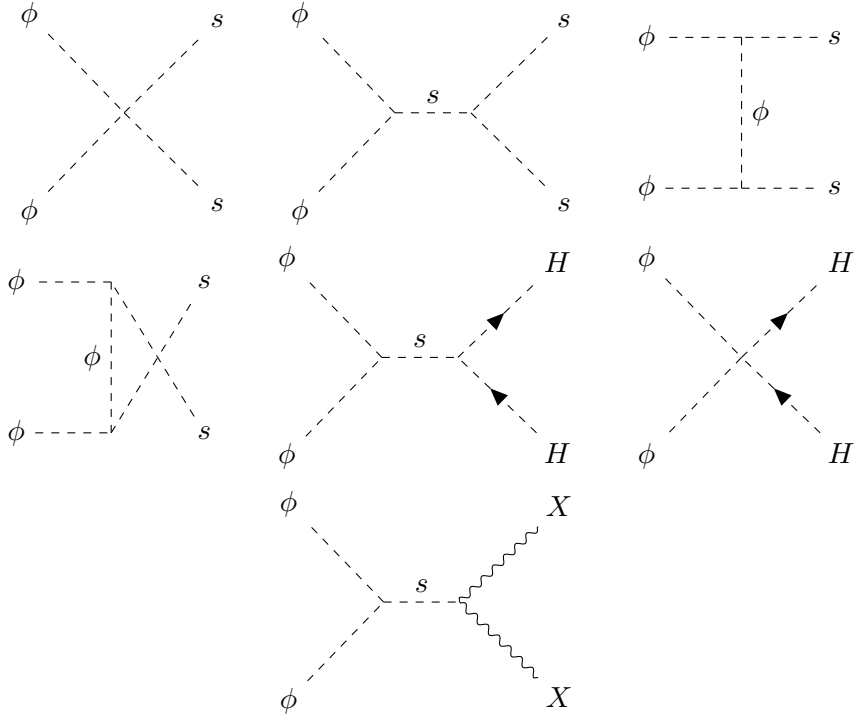

The processes that contribute to freeze-in of vector boson $X$ bEWSB are shown in Fig.~\ref{fig:FD-frzin-bEWSB}. The initial abundance of $X$ in the early universe is assumed negligible, while it builds up via production from the particles in thermal bath, namely $s,\phi$ and $H$. Decay of $s$ contributes the most, subject to the kinematic constraint $m_s \ge 2m_X$. Scattering processes $s s\to XX$ also contribute but it is suppressed compared to the decay. In addition,  $HH^{\dagger}\to XX$ and $\phi \phi\to XX$ also contribute, mediated by s-channel $s$. The process $\phi \phi \to XX$ is WIMP-FIMP conversion and it occurs as $\phi$ is assumed present in the thermal bath. We will discuss the effect of such conversion contributions in details. It is worth noting that unsuppressed $s s\to H^{\dagger} H,~\phi\phi\to H^{\dagger} H,~\phi\phi\to s s$ keep $\phi,~s,~$and $H$, all in equilibrium in early universe. The processes which contribute to freeze out of $\phi$ bEWSB are shown in Fig.~\ref{fig:frzout-FD-EWSB}. They are all known, and include $\phi\phi \to HH^{\dagger}$ via the quartic coupling as well as that mediated by $s$; additionally $\phi\phi \to s s$ occurs via quartic portal coupling, and s-channel mediation by $s$ and t-channel mediation by $\phi$. Finally, WIMP-FIMP conversion $\phi\phi \to XX$ via s-channel mediation of $s$ is also possible as shown in the bottom panel of Fig.~\ref{fig:frzout-FD-EWSB}, otherwise absent in single component case. Let us recall again that all the processes initiated by $H$ and those produce $H$ assume four massive scalar d.o.f.
\begin{figure}[htb!]
	\centering
\begin{tikzpicture}[baseline={(current bounding box.center)}]
\begin{feynman}
\vertex (a1){\(s\)};
\vertex[right=1cm of a1] (a2);
\vertex[above right=1cm and 1cm of a2] (a3){\(X\)}; 
\vertex[below right=1cm and 1cm of a2] (a4){\(X\)}; 
\diagram* {
	(a2) -- [scalar] (a1),(a2) -- [boson] (a3),
	(a4) -- [boson] (a2)
};\end{feynman}
\end{tikzpicture}\hspace{0.5cm}	
\begin{tikzpicture}[baseline={(current bounding box.center)}]
\begin{feynman}
\vertex (a);
\vertex[above left=1cm and 1cm of a] (a1){\(s\)}; 
\vertex[below left=1cm and 1cm of a] (a2){\(s\)}; 
\vertex[above right=1cm and 1cm of a] (a3){\(X\)}; 
\vertex[below right=1cm and 1cm of a] (a4){\(X\)}; 
\diagram* {
	(a) -- [scalar] (a1),(a2) -- [scalar] (a),
	(a3) -- [boson]	(a) -- [boson] (a4)
};\end{feynman}
\end{tikzpicture}\hspace{0.5cm}	
\begin{tikzpicture}[baseline={(current bounding box.center)}]
\begin{feynman}
\vertex (a);
\vertex[right=1cm of a] (b);
\vertex[above left=1cm and 1cm of a] (a1){\(s\)}; 
\vertex[below left=1cm and 1cm of a] (a2){\(s\)}; 
\vertex[above right=1cm and 1cm of b] (a3){\(X\)}; 
\vertex[below right=1cm and 1cm of b] (a4){\(X\)}; 
\diagram* {
	(a) -- [scalar] (a1),(a2) -- [scalar] (a),
	(a3) -- [boson]	(b) -- [boson] (a4),(a)  --[scalar,edge label={\(s\)}] (b)
};\end{feynman}
\end{tikzpicture}	

\begin{tikzpicture}[baseline={(current bounding box.center)}]
\begin{feynman}
\vertex (a);
\vertex[below=2cm of a] (b);
\vertex[left=1cm and 1cm of a] (a1){\(s\)}; 
\vertex[right=1cm and 1cm of a] (a2){\(X\)}; 
\vertex[left=1cm and 1cm of b] (a3){\(s\)}; 
\vertex[right=1cm and 1cm of b] (a4){\(X\)}; 
\diagram* {
	(a) -- [scalar] (a1),(a2) -- [boson] (a),
	(a3) -- [scalar] (b) -- [boson] (a4),(a)  --[boson,edge label={\(X\)}] (b)
};\end{feynman}
\end{tikzpicture}\hspace{0.5cm}	
\begin{tikzpicture}[baseline={(current bounding box.center)}]
\begin{feynman}
\vertex (a);
\vertex[right=1cm of a] (b);
\vertex[above left=1cm and 1cm of a] (a1){\(H\)}; 
\vertex[below left=1cm and 1cm of a] (a2){\(H\)}; 
\vertex[above right=1cm and 1cm of b] (a3){\(X\)}; 
\vertex[below right=1cm and 1cm of b] (a4){\(X\)}; 
\diagram* {
	(a1) -- [charged scalar] (a),(a) -- [charged scalar] (a2),
	(a3) -- [boson]	(b) -- [boson] (a4),(a)  --[scalar,edge label={\(s\)}] (b)
};\end{feynman}
\end{tikzpicture}\hspace{0.5cm}	
\begin{tikzpicture}[baseline={(current bounding box.center)}]
\begin{feynman}
\vertex (a);
\vertex[right=1cm of a] (b);
\vertex[above left=1cm and 1cm of a] (a1){\(\phi\)}; 
\vertex[below left=1cm and 1cm of a] (a2){\(\phi\)}; 
\vertex[above right=1cm and 1cm of b] (a3){\(X\)}; 
\vertex[below right=1cm and 1cm of b] (a4){\(X\)}; 
\diagram* {
	(a1) -- [scalar] (a),(a) -- [scalar] (a2),
	(a3) -- [boson]	(b) -- [boson] (a4),(a)  --[scalar,edge label={\(s\)}] (b)
};\end{feynman}
\end{tikzpicture}
\caption{Feynman diagrams showing non-thermal production channels of $X$ bEWSB.}
\label{fig:FD-frzin-bEWSB}
\end{figure}
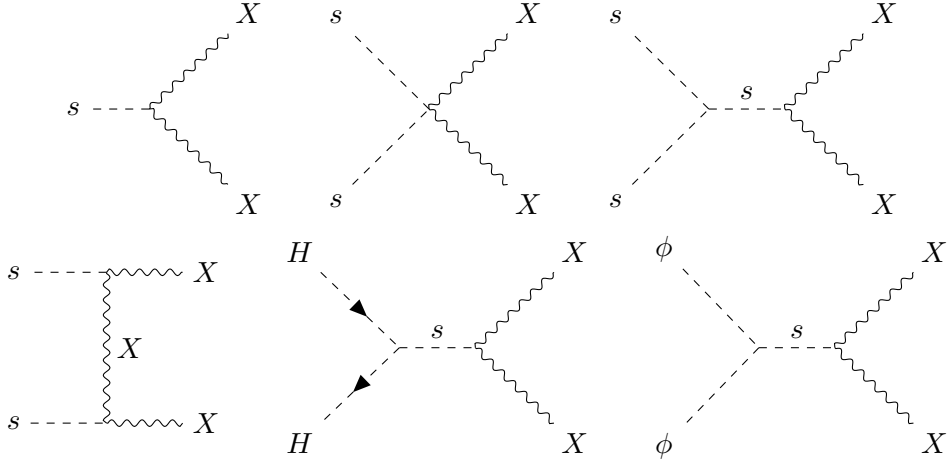

\subsection{Coupled Boltzmann Equations and conversion}
\label{sec:bEWSB-cBEQ}

The DM number density for the WIMP-FIMP scenario can be evaluated by the coupled Boltzmann equations (cBEQs):

\begin{align}
 \frac{dY_X}{dx}
 =&\,
  -\,\frac{2\,{\rm M_{Pl}}}{1.66}\frac{x}{m_X^2}\frac{\sqrt{g_*(x)}}{g_s^*(x)}\langle\Gamma_{s\to XX}\rangle\left(Y_X- Y_s^{eq}\right) \ \nonumber  \\
 -& \frac{4\pi^2{\rm M_{Pl}}\sqrt{g_*(x)}}{45\times 1.66}\frac{m_X}{x^2}
   \left[ \sum_{i=s,H}\langle\sigma \mathpzc{v} \rangle_{ii\to XX}\left( Y_X^2-(Y_{i}^{eq})^2 \right)
 - \langle \sigma \mathpzc{v} \rangle_{\phi\phi \to XX}\left(Y_{\phi}^2-\frac{(Y_{\phi}^{eq})^2}{(Y_X^{eq})^2}Y_X^2\right) \right],
 \label{BEQ-FIMP}\\
 \frac{dY_{\phi}}{dx}
 =&
 -\frac{2\pi^2\,{\rm M_{Pl}}\sqrt{g_*(x)}}{45\times 1.66}\frac{m_{\phi}}{x^2}\left[ \sum_{i=s,H} \langle\sigma \mathpzc{v} \rangle_{\phi\phi\to ii}\left(Y_{\phi}^2-(Y_{\phi}^{eq})^2\right)
 + \langle \sigma \mathpzc{v} \rangle_{\phi\phi \to XX}\left(Y_{\phi}^2-\frac{(Y_{\phi}^{eq})^2}{(Y_X^{eq})^2}Y_X^2\right) \right]\,.
 \label{BEQ-WIMP}
\end{align}

In the above, $x=m/T,~Y_{X,\phi}=\frac{n_{X,\phi}}{\s}$, where $n_{X,\phi}$ refers to the number of $X\text{ and }\phi$ 
respectively, and $\s$ represents entropy density given by $\s=\frac{2\pi^2}{45}g_*^s(T)T^3$. The equilibrium number density with respect 
to comoving volume for non-relativistic species ($X,\phi$) is given by Boltzmann distribution (assuming chemical potential to be zero)
$$Y^{eq}=\frac{n^{eq}}{\s}=0.145\frac{g}{g^s_{*}}x^{3/2}e^{-x}.$$
Further note
$\sqrt{g_*(T)}\simeq\frac{g^s_*(T)}{\sqrt{g^{\rho}_*(T)}}$, where $g^s_*$ and $g^{\rho}_*$ denote d.o.f corresponding to entropy 
and energy density of the Universe. Further, $\rm{M_{Pl}}=1.22 \times 10^{19}$ GeV denotes reduced Planck mass.
Thermal average of decay width $\langle\Gamma_{A \to B B}\rangle$ and annihilation cross-section times the velocity
$\langle\sigma \mathpzc{v}\rangle_{AA \to BB}$ are given by, 

\bea
\langle \Gamma_{A \to B B} \rangle &=& \Gamma_{A \to B B} \frac{K_1\left(m_B/T\right)}{K_2\left(m_B/T\right)}, \nonumber \\
\langle \sigma \mathpzc{v} \rangle_{AA\to BB}&=&\frac{1}{8m_A^4TK_2^2(m_A/T)}\int_{4m_B^2}^{\infty}d\mathtt{s}\ \sigma_{AA \to BB}(\mathtt{s})(\mathtt{s}-4m_A^2)\sqrt{\mathtt{s}}\,K_1\left(\sqrt{\mathtt{s}}/T\right).
\eea

$K_{1,2}$ are modified Bessel functions of first and second kind respectively and  $\mathpzc{v}$ refers to M$\ddot{o}$llar velocity defined by
$\mathpzc{v}=\frac{\sqrt{(p_1.p_2)^2-m_1^2m_2^2}}{E_1E_2}$. In $\langle \Gamma_{A \to B B} \rangle$, the particle ($A$) is decaying at 
rest and the thermal average do not involve an integration over the centre-of-mass energy $\sqrt{\mathtt{s}}$, while for annihilation cross-section 
$\langle \sigma \mathpzc{v} \rangle_{AA\to BB}$, a lower limit $\mathtt{s}=4m_B^2$ is required for the reaction to occur and it diminishes at high $\sqrt{\mathtt{s}}$, owing to the 
presence of $K_1\left(\sqrt{\mathtt{s}}/T\right)$ for a particular $T$.

One important point to note before we proceed further. The WIMP-FIMP conversion $\phi\phi \to XX$, which makes the BEQs 
(Eq.~\ref{BEQ-FIMP} and \ref{BEQ-WIMP}) coupled, requires to be of the order of freeze-in production cross-section, as otherwise
it will thermalize the FIMP, suppressing the non-thermal production (this exercise will be discussed in details elsewhere). 
This in turn, makes conversion process negligible compared to other annihilation cross-sections of $\phi$ (first term in Eq.~\ref{BEQ-WIMP}). 
However, this conversion can still be significant for non-thermal production of $X$. This feature is generic to any two-component 
WIMP-FIMP model, where the freeze-out of WIMP can be marked unaffected by the conversion to FIMP, while the FIMP production can be substantial 
due to WIMP. This feature importantly reduces the cBEQs as in Eq. \ref{BEQ-FIMP} and  Eq. \ref{BEQ-WIMP}
to two individual uncoupled BEQs, where the conversion can be dropped from Eq. \ref{BEQ-WIMP} to yield:

\begin{align}
\frac{dY_X}{dx}=&\frac{2\,{\rm M_{Pl}}}{1.66\sqrt{g^{\rho}_*(x)}}\frac{x}{m_X^2}\langle\Gamma_{s\to XX}\rangle Y_s^{eq}\nonumber \\
& +\frac{4\pi^2\, {\rm M_{Pl}}}{45\times 1.66}\frac{g^s_*(x)}{\sqrt{g^{\rho}_*(x)}}\frac{m_X}{x^2}\left(\sum_{i=s,H}\langle\sigma \mathpzc{v} \rangle_{ii\to XX}(Y_i^{eq})^2+\langle \sigma \mathpzc{v} \rangle_{\phi\phi \to XX}Y_{\phi}^2\right), \label{eq:split-FIMP} \\
\frac{dY_{\phi}}{dx}=& -\frac{2\pi^2\, {\rm M_{Pl}}}{45\times 1.66}\frac{g^s_*(x)}{\sqrt{g^{\rho}_*(x)}}\frac{m_{\phi}}{x^2}\sum_{i=s,H} \langle\sigma \mathpzc{v} \rangle_{\phi\phi\to ii}\left( Y_{\phi}^2-(Y_{\phi}^{eq})^2\right).
\label{split-coup-BEQ}
\end{align}

Eq.~\ref{eq:split-FIMP} and~\ref{split-coup-BEQ} allow us to treat the freeze-in of $X$ and freeze-out of $\phi$ separately as we do next. 
It also allows to treat $x=m_X/T$ in Eq. \ref{eq:split-FIMP} and $x=m_\phi/T$ in Eq.~\ref{split-coup-BEQ} as two separate variables
\footnote{Otherwise in cBEQ, one needs to define a common $x=\mu/T$, where $\mu=\frac{m_Xm_\phi}{m_X+m_\phi}$ (see \cite{Bhattacharya:2017fid}).}. 
 We further note that in view of small $Y_X$ and feeble interaction, we have dropped the terms $\propto Y_X$ and $Y_X^2$ ($X\, X \to s$ and $X\, X \to i\,i$) in Eq.~\ref{BEQ-FIMP}
to obtain Eq.~\ref{eq:split-FIMP}. 

\subsection{Freeze-in of X}
\label{sec:bEWSB-freeze-in}

Now let us discuss the freeze-in of $X$, bEWSB in details. The main point is that the initial abundance of $X$ in the early universe 
is negligible, builds up from the decay or scattering of the particles in thermal bath and saturates when the photon temperature falls 
below DM mass. One essentially then needs to solve BEQ. \ref{eq:split-FIMP} from $x\simeq 0$ to $x=m_X/\tew$, 
using non-thermal production of $X$, indicated in Fig. \ref{fig:FD-frzin-bEWSB}, and include:
\begin{enumerate*}
\item $s$ decays to $X$ pair while in thermal equilibrium and after $s$ freezes out,
\item $s$ and $H$ scattering to $X$ pairs,
\item $\phi$ pair annihilation to $X$ pairs.
\end{enumerate*} 
 
\begin{figure}[htb!]
	\centering
	\begin{tikzpicture}[baseline={(current bounding box.center)}]
	\begin{feynman}
	\vertex (a);
	\vertex[above left=1cm and 1cm of a] (a1){\(s\)}; 
	\vertex[below left=1cm and 1cm of a] (a2){\(s\)}; 
	\vertex[above right=1cm and 1cm of a] (a3){\(X\)}; 
	\vertex[below right=1cm and 1cm of a] (a4){\(X\)}; 
	\diagram* {
		(a) -- [scalar] (a1),(a2) -- [scalar] (a),
		(a3) -- [boson]	(a) -- [boson] (a4)
	};\end{feynman}
	\end{tikzpicture}\hspace{0.5cm}	
	\begin{tikzpicture}[baseline={(current bounding box.center)}]
	\begin{feynman}
	\vertex (a);
	\vertex[right=1cm of a] (b);
	\vertex[above left=1cm and 1cm of a] (a1){\(s\)}; 
	\vertex[below left=1cm and 1cm of a] (a2){\(s\)}; 
	\vertex[above right=1cm and 1cm of b] (a3){\(X\)}; 
	\vertex[below right=1cm and 1cm of b] (a4){\(X\)}; 
	\diagram* {
		(a) -- [scalar] (a1),(a2) -- [scalar] (a),
		(a3) -- [boson]	(b) -- [boson] (a4),(a)  --[scalar,edge label={\(s\)}] (b)
	};\end{feynman}
	\end{tikzpicture}\hspace{0.5cm}	
	\begin{tikzpicture}[baseline={(current bounding box.center)}]
	\begin{feynman}
	\vertex (a);
	\vertex[below=2cm of a] (b);
	\vertex[left=1cm and 1cm of a] (a1){\(s\)}; 
	\vertex[right=1cm and 1cm of a] (a2){\(X\)}; 
	\vertex[left=1cm and 1cm of b] (a3){\(s\)}; 
	\vertex[right=1cm and 1cm of b] (a4){\(X\)}; 
	\diagram* {
		(a) -- [scalar] (a1),(a2) -- [boson] (a),
		(a3) -- [scalar] (b) -- [boson] (a4),(a)  --[boson,edge label={\(X\)}] (b)
	};\end{feynman}
	\end{tikzpicture}
	
		\begin{tikzpicture}[baseline={(current bounding box.center)}]
	\begin{feynman}
	\vertex (a);
	\vertex[above left=1cm and 1cm of a] (a1){\(s\)}; 
	\vertex[below left=1cm and 1cm of a] (a2){\(s\)}; 
	\vertex[above right=1cm and 1cm of a] (a3){\(H\)}; 
	\vertex[below right=1cm and 1cm of a] (a4){\(H\)}; 
	\diagram* {
		(a) -- [scalar] (a1),(a2) -- [scalar] (a),
		(a) -- [charged scalar]	(a3) ,(a4)-- [charged scalar] (a)
	};\end{feynman}
	\end{tikzpicture}\hspace{0.5cm}		
	\begin{tikzpicture}[baseline={(current bounding box.center)}]
	\begin{feynman}
	\vertex (a);
	\vertex[right=1cm of a] (b);
	\vertex[above left=1cm and 1cm of a] (a1){\(s\)}; 
	\vertex[below left=1cm and 1cm of a] (a2){\(s\)}; 
	\vertex[above right=1cm and 1cm of b] (a3){\(H\)}; 
	\vertex[below right=1cm and 1cm of b] (a4){\(H\)}; 
	\diagram* {
		(a) -- [scalar] (a1),(a2) -- [scalar] (a),
		(a4) -- [charged scalar]	(b) -- [ charged scalar] (a3),(a)  --[scalar,edge label={\(s\)}] (b)
	};\end{feynman}
	\end{tikzpicture}\hspace{0.5cm}	
	\begin{tikzpicture}[baseline={(current bounding box.center)}]
	\begin{feynman}
	\vertex (a);
	\vertex[below=2cm of a] (b);
	\vertex[left=1cm and 1cm of a] (a1){\(s\)}; 
	\vertex[right=1cm and 1cm of a] (a2){\(H\)}; 
	\vertex[left=1cm and 1cm of b] (a3){\(s\)}; 
	\vertex[right=1cm and 1cm of b] (a4){\(H\)}; 
	\diagram* {
		(a) -- [scalar] (a1),(a) -- [charged scalar] (a2),
		(a4) -- [charged scalar] (b) -- [ scalar] (a3),(b)  --[ charged scalar,edge label={\(H\)}] (a)
	};\end{feynman}
	\end{tikzpicture}
		
\begin{tikzpicture}[baseline={(current bounding box.center)}]
\begin{feynman}
\vertex (a);
\vertex[above left=1cm and 1cm of a] (a1){\(s\)}; 
\vertex[below left=1cm and 1cm of a] (a2){\(s\)}; 
\vertex[above right=1cm and 1cm of a] (a3){\(\phi\)}; 
\vertex[below right=1cm and 1cm of a] (a4){\(\phi\)}; 
\diagram* {
	(a) -- [scalar] (a1),(a2) -- [scalar] (a),
	(a) -- [ scalar]	(a3) ,(a4)-- [ scalar] (a)
};\end{feynman}
\end{tikzpicture}\hspace{0.5cm}	
\begin{tikzpicture}[baseline={(current bounding box.center)}]
\begin{feynman}
\vertex (a);
\vertex[right=1cm of a] (b);
\vertex[above left=1cm and 1cm of a] (a1){\(s\)}; 
\vertex[below left=1cm and 1cm of a] (a2){\(s\)}; 
\vertex[above right=1cm and 1cm of b] (a3){\(\phi\)}; 
\vertex[below right=1cm and 1cm of b] (a4){\(\phi\)}; 
\diagram* {
	(a) -- [scalar] (a1),(a2) -- [scalar] (a),
	(a4) -- [ scalar]	(b) -- [ scalar] (a3),(a)  --[scalar,edge label={\(s\)}] (b)
};\end{feynman}
\end{tikzpicture}\hspace{0.5cm}	
\begin{tikzpicture}[baseline={(current bounding box.center)}]
\begin{feynman}
\vertex (a);
\vertex[below=2cm of a] (b);
\vertex[left=1cm and 1cm of a] (a1){\(s\)}; 
\vertex[right=1cm and 1cm of a] (a2){\(\phi\)}; 
\vertex[left=1cm and 1cm of b] (a3){\(s\)}; 
\vertex[right=1cm and 1cm of b] (a4){\(\phi\)}; 
\diagram* {
	(a) -- [scalar] (a1),(a2) -- [ scalar] (a),
	(b) -- [ scalar] (a4),(b) -- [ scalar] (a3),(a)  --[  scalar,edge label={\(\phi\)}] (b)
};\end{feynman}
\end{tikzpicture}
\caption{Feynman diagrams showing all possible annihilation channels of $s$, which causes $s$ to freeze out from thermal bath bEWSB.}
\label{S-annihilation}
\end{figure}
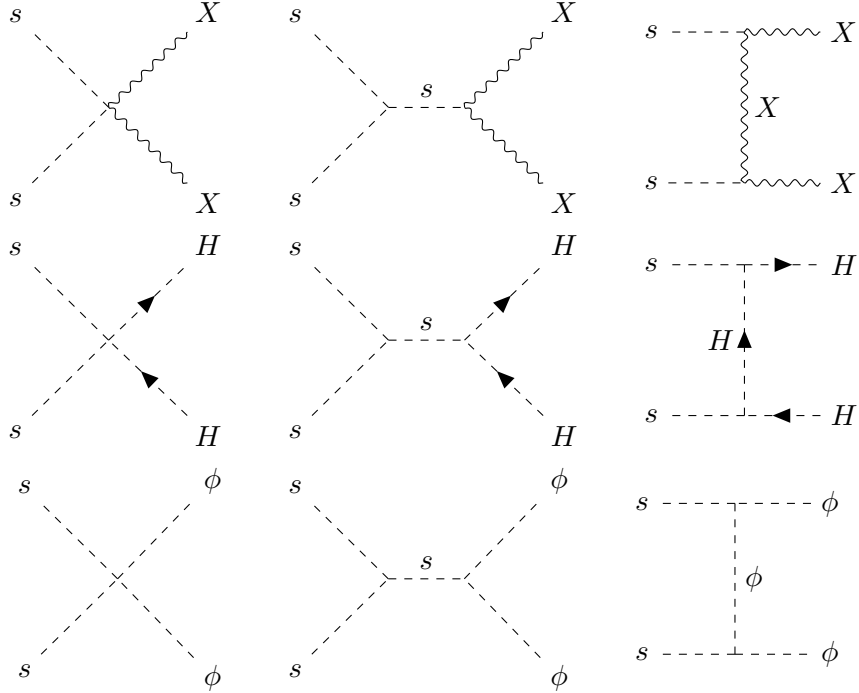

 However, there is a slight twist to the story. The decay contribution of $s$ to $X$ pairs as written in Eq.~\ref{eq:split-FIMP} 
 is only applicable when the decaying particle is in equilibrium with the thermal bath. The decay process however continues even after 
 $s$ freezes out from thermal bath, i.e. beyond $x\ge x_D$, where $x_D$ denotes freeze-out point of $s$. The decay contribution after 
 freeze-out of $s$ from thermal bath is often termed as `late decay' (LD) of $s$. The dynamics of such effect can be captured by yet another 
 coupled BEQ written together with the evolution of yield $Y_s$ (see Appendix \ref{sec:bEWSB-details}), where the freeze out of $s$ is 
 governed by its annihilation channels to SM, as shown in Feynman graph in Fig.~\ref{S-annihilation}. The coupled BEQ for this case can 
 be simplified to a single BEQ with an additional term to in-equilibrium decay (for derivation, see \cite{Barman:2019lvm,Buch:2016jjp}): 
 \begin{align}
\frac{dY_X}{dx}=&\frac{45}{3.32\pi^4}\frac{g_s\, M_{{\rm Pl}}\,m_s^2\, \Gamma_{s\to XX}}{m_X^4}\left(\frac{x^3 K_1\left[\frac{m_s}{m_X}x\right]}{g^s_*(x)\sqrt{g_*^{\rho}(x)}}\Theta\left(x_D-x\right)
\right. \nonumber \\
&\left.+ \ e^{-\frac{0.602\, M_{{\rm Pl}}\, \Gamma_{s \to XX}}{m_X^2\sqrt{g^{\rho}_*(x)}}(x^2-x_D^2)}\frac{ x^2x_D}{\eta(x,x_D)}K_1\left[\alpha(x,x_D)\frac{m_s}{m_X}\frac{x^2}{x_D}\right]e^{\frac{m_s}{m_X}\left(\alpha(x,x_D)\frac{x^2}{x_D}-x_D\right)} \Theta(x-x_D)\right) \nonumber \\
& + \frac{4\pi^2\, \rm{M_{ Pl}}}{45\times 1.66}\frac{g^s_*(x)}{\sqrt{g^{\rho}_*(x)}}\frac{m_X}{x^2}\left(\sum_{i=s,H}\langle\sigma \mathpzc{v} \rangle_{ii\to XX}(Y_i^{eq})^2 +\langle \sigma \mathpzc{v} \rangle_{\phi\phi \to XX}Y_{\phi}^2\right).
\label{BEQ_FIMP2}
\end{align}
In the above, the term proportional to $\Theta(x_D-x)$ in the first parenthesis indicates FIMP production from `in-equilibrium' decay of $s$ and
the second term in the first parenthesis captures the late decay contribution with $\Theta(x-x_D)$ denoting the Heaviside theta function. Also 
note that $g_s$ (internal d.o.f for $s$) is 1. The freeze out point of $s$ is denoted by $x_D$ which can be found out by following expression:
\begin{equation}
x_D={\rm ln}[\Lambda]-\frac{1}{2}{\rm ln[ln}[\Lambda]],\ \Lambda=0.038~\frac{g_s m_s \rm{M_{ Pl}}}{\sqrt{g^{\rho}_*}}\sum_{SM=H,\phi,X}\sigma^0_{ss\rightarrow \rm{SM~SM}},
\label{xD_def}
\end{equation}
where $\sigma^0_{ss\rightarrow \rm{SM~SM}}$ denotes annihilation cross-section of $s$ to SM particles at threshold ($\mathtt{s}=4\,m^2_{{\rm s}}$) and corresponding expressions are provided in the 
appendix \ref{sec:bEWSB-details}. Also note in Eq.~\ref{BEQ_FIMP2}, the factor $\eta(x,x_D)$ and $\alpha(x,x_D)$ are given by:
\bea
\eta(x,x_D)=\alpha(x,x_D)g_*^s(x)\sqrt{g_*^{\rho}(x)}, ~~\alpha(x,x_D)=\left[\frac{g_*^s(x_D)}{g_*^s(x)}\right]^{1/3}\ \left[\frac{g_*^{\rho}(x_D)}{g_*^{\rho}(x)}\right]^{1/4}.
\label{alpha_xd}
\eea

Late decay contribution provides significant contribution to DM yield. However, when we consider freeze-in to occur bEWSB, 
late decay contribution should also accumulate fully bEWSB. This evidently requires freeze-out of $s$ to occur bEWSB with 
$T_D>\tew $, with $x_D=\frac{m_s}{T_D}$ (see Eq. \ref{xD_def} for details) varying typically in the range of $\sim 20-25$. We can achieve 
this limit by having heavy $m_s$ for which $x_D \le m_s/\tew$, resulting a limit on $m_s$ as:
 \bea
 T_{\rm EW}\lesssim \frac{m_s}{25}\implies m_s \gtrsim 4 ~\rm{TeV}\,,
  \label{eq:masslimit}
 \eea
which is not surprisingly the same limit on WIMP mass to freeze-out bEWSB as in Eq.~\ref{m_fi_2}.

The yield of $X$ in the pre-EWSB regime is given by, 

\begin{align}
Y_X^{{\rm bEWSB}}&=\frac{45}{3.32 \pi^4 }\frac{{\rm M_{Pl}}\,m_s^2\,\Gamma_{s \to XX}}{m_X^4}\int_{0}^{m_X/\tew}\biggl[ \frac{x^3 K_1\left[\frac{m_s}{m_X}x\right]}{\sqrt{g_*^{\rho}(x)}g^s_*(x)} \Theta\left(x_D-x\right)\nonumber \\
&+ \ e^{-\frac{0.602\, {\rm M_{Pl}}\, \Gamma_{s \to XX}}{m_X^2\sqrt{g^{\rho}_*(x)}}(x^2-x_D^2)}\frac{ x^2x_D}{\eta(x,x_D)}K_1\left[\alpha(x,x_D)\frac{m_s}{m_X}\frac{x^2}{x_D}\right]e^{\frac{m_s}{m_X}\left(\alpha(x,x_D)\frac{x^2}{x_D}-x_D\right)} \Theta(x-x_D)\biggr]dx \nonumber \\
& + \frac{45\,{\rm M_{Pl}}}{4 \pi^6\times 1.66}\int_{0}^{m_X/\tew}\sum_{i=s,H}\langle \sigma \mathpzc{v} \rangle_{ii \to XX}~\frac{m_i^4}{m_X^3}\frac{x^2 K_2^2\left[\frac{m_i}{m_X}x\right]}{g_*^s(x)\sqrt{g_*^{\rho}(x)}}dx \nonumber \\
&+ \frac{4\pi^2\rm{M_{Pl}}m_X}{45\times 1.66} \int_0^{m_X/\tew}\langle \sigma \mathpzc{v} \rangle_{\phi\phi \to XX}~\frac{g_*^s(x)}{\sqrt{g_{\rho}^*(x)}} \frac{Y_{\phi}^2}{x^2}dx\,.
\label{frzin_int}
\end{align}

 We note that although the limit of $x$ integration above is taken upto EWSB scale ($x: 0 \to m_X/\tew$), the result does not alter if we extend the limit to 
smaller temperature or larger $x \to \infty$, as the parameters are chosen in a way that freeze-in occurs bEWSB. Freeze-in bEWSB is ensured by checking 
$Y_{x>x_{EW}}=Y_{x_{EW}}$.
Finally, the relic density for $X$ can be written in terms of $Y_X$ and we want to probe under abundant region, as $X$ 
constitutes a part of two component framework, then,
\begin{equation}
\Omega_X \rm h^2 \simeq 2.744 \times 10^8\ m_X Y_X^{\rm{bEWSB}}; ~~ \Omega_X \rm h^2\le 0.1212 \,.
\label{omg_X}
\end{equation}
 where the FIMP dark matter relic density is written in terms of the reduced Hubble parameter, $h$ in units of 100 km/s/Mpc.

\subsubsection{Phenomenology}
\label{sec:pheno-freeze-in-bEWSB}
 As argued before, FIMP production from decay is always dominant over the scattering processes in our model due to the presence of either feeble couplings at both vertices, 
a heavy mediator or heavy initial state particles for the latter. Therefore, in this study we can divide the FIMP parameter space into two purely separate mass regimes, where 
decay and scattering contributions to FIMP production are mutually exclusive. However, in cases where scattering can create a heavy mediator on-shell with unsuppressed 
production and decays subsequently to DM, there may arise a potential double counting when both decay and scattering processes are considered together \cite{Belanger:2018ccd}, 
which needs to be accounted. For us there is no such issue with the following segregation of kinematic regimes which yield different phenomenology:
\begin{itemize}
\item \textbf{Case-I} ($m_s \ge 2m_X$): $X$ is produced mainly from $s$ decay, annihilation processes are smaller and neglected.

\item \textbf{Case-II} ($m_s< 2m_X $): Decay channel ($s\to XX$) is forbidden, scattering processes contribute to $X$ production.
\end{itemize}

\paragraph{\textbf{\underline{Case-I}} $\left(m_s \ge 2m_X\right):$\\}

In this kinematical region, given that even late decay of $s$ occurs bEWSB, it leaves no trace of $s$ aEWSB. So, there is no 
$s-H$ mixing and dark sector remains detached from the SM. As mentioned previously, for this case, we need to choose 
$m_H^2=m_{h_1}^2 /2$ to get the correct Higgs mass aEWSB. Together we also demand that $\phi$ freezes-out bEWSB, then 
DM components do not have any {\it direct} coupling to SM, except the quartic interaction $\phi \phi \to H H$ proportional to 
$\lambda_{\phi H}$.  But as discussed, constraints from Direct search on $\lambda_{\phi H}$ makes this coupling weak, this 
particular kinematical region with freeze-in and freeze-out both occurring bEWSB, is difficult to probe by any experiment 
in the near future and is thus constrained very feebly by direct detection or collider search constraints. 

\begin{figure}[htb!]
	\centering
	\subfloat[]{\includegraphics[width=0.5\linewidth]{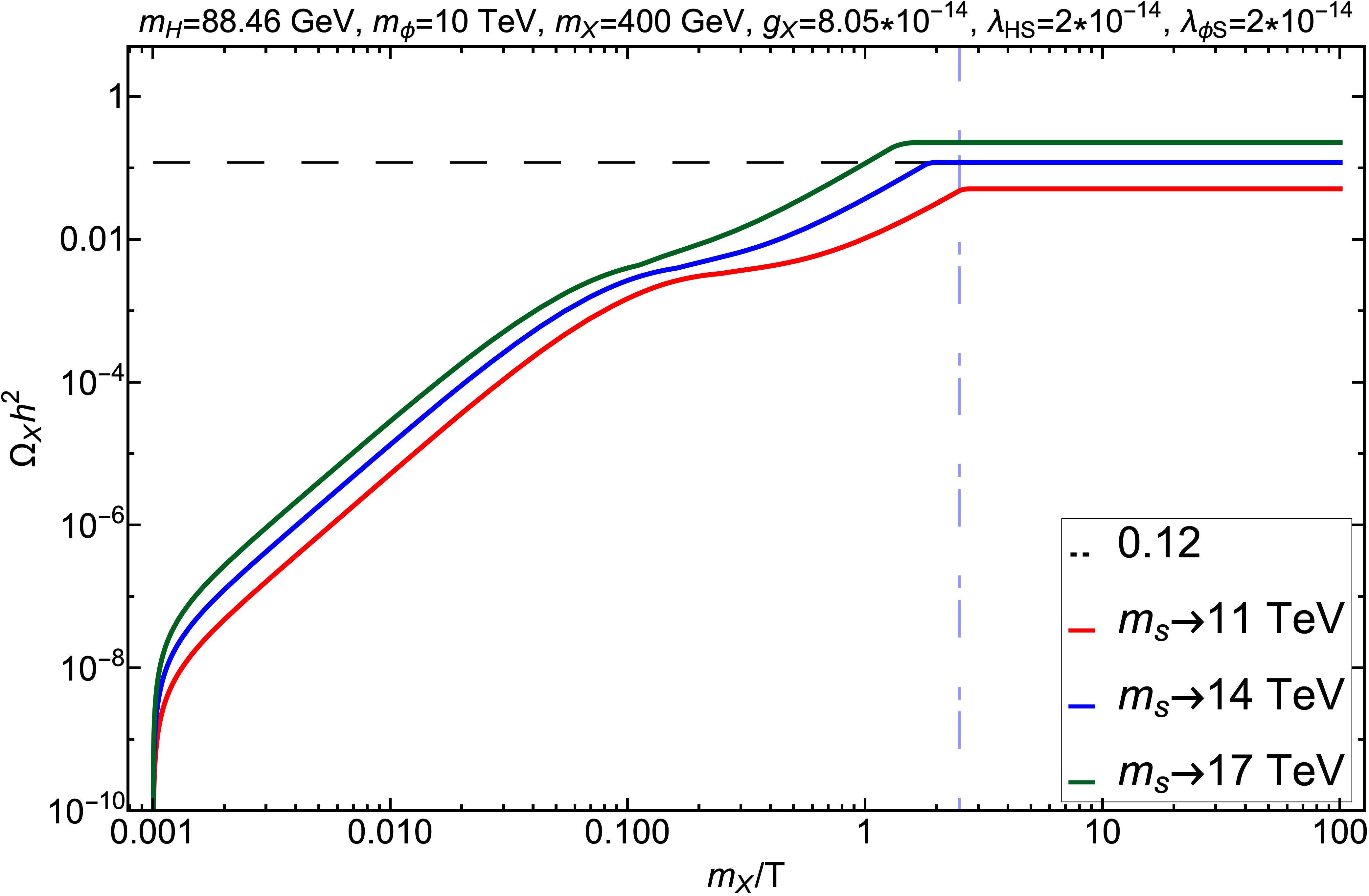}\label{frzin_fig1}}~~
	\subfloat[]{\includegraphics[width=0.5\linewidth]{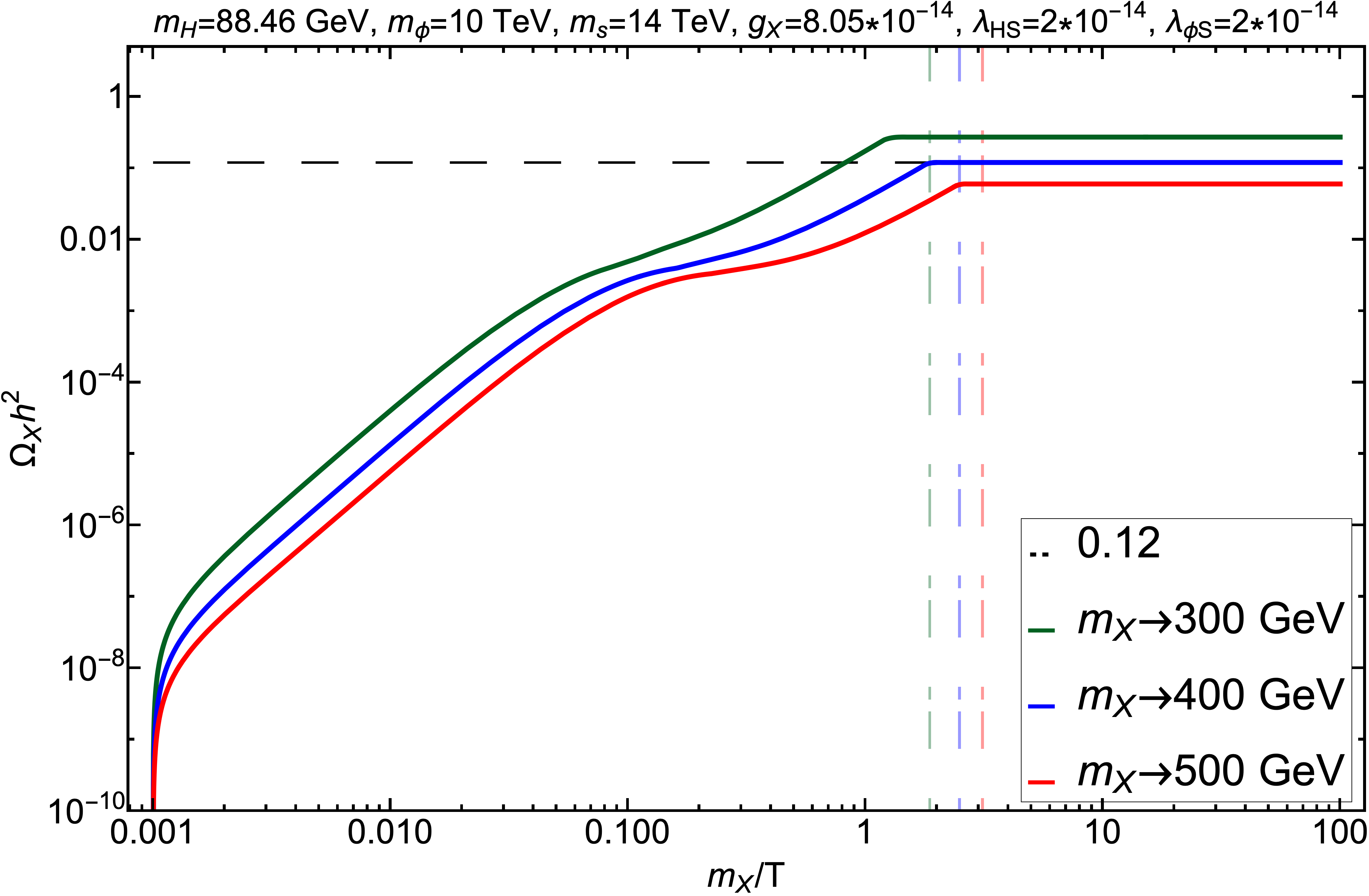}\label{frzin_fig2}}\\
	\subfloat[]{\includegraphics[width=0.5\linewidth]{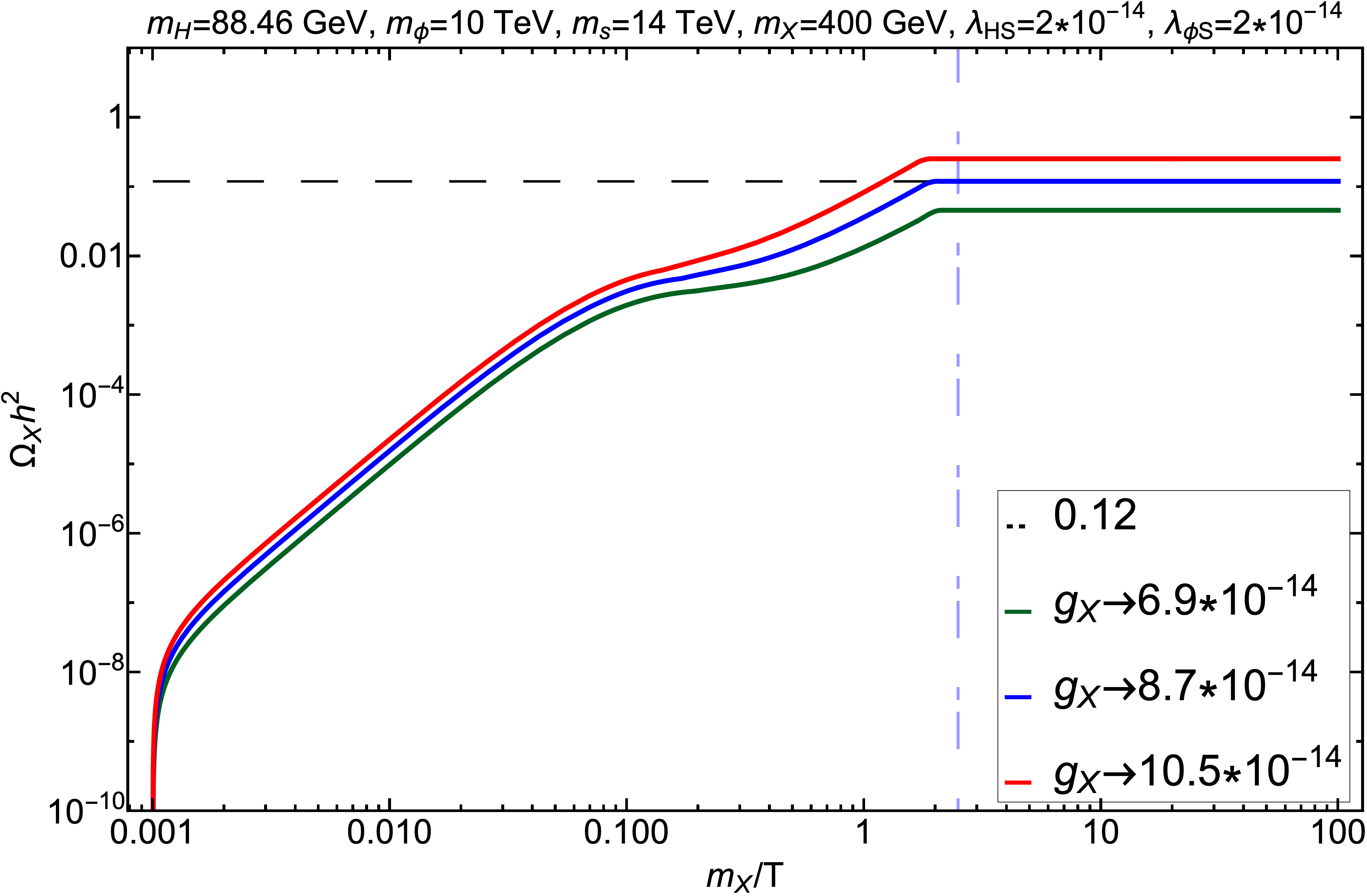}\label{frzin_fig3}}~~
	\subfloat[]{\includegraphics[width=0.5\linewidth]{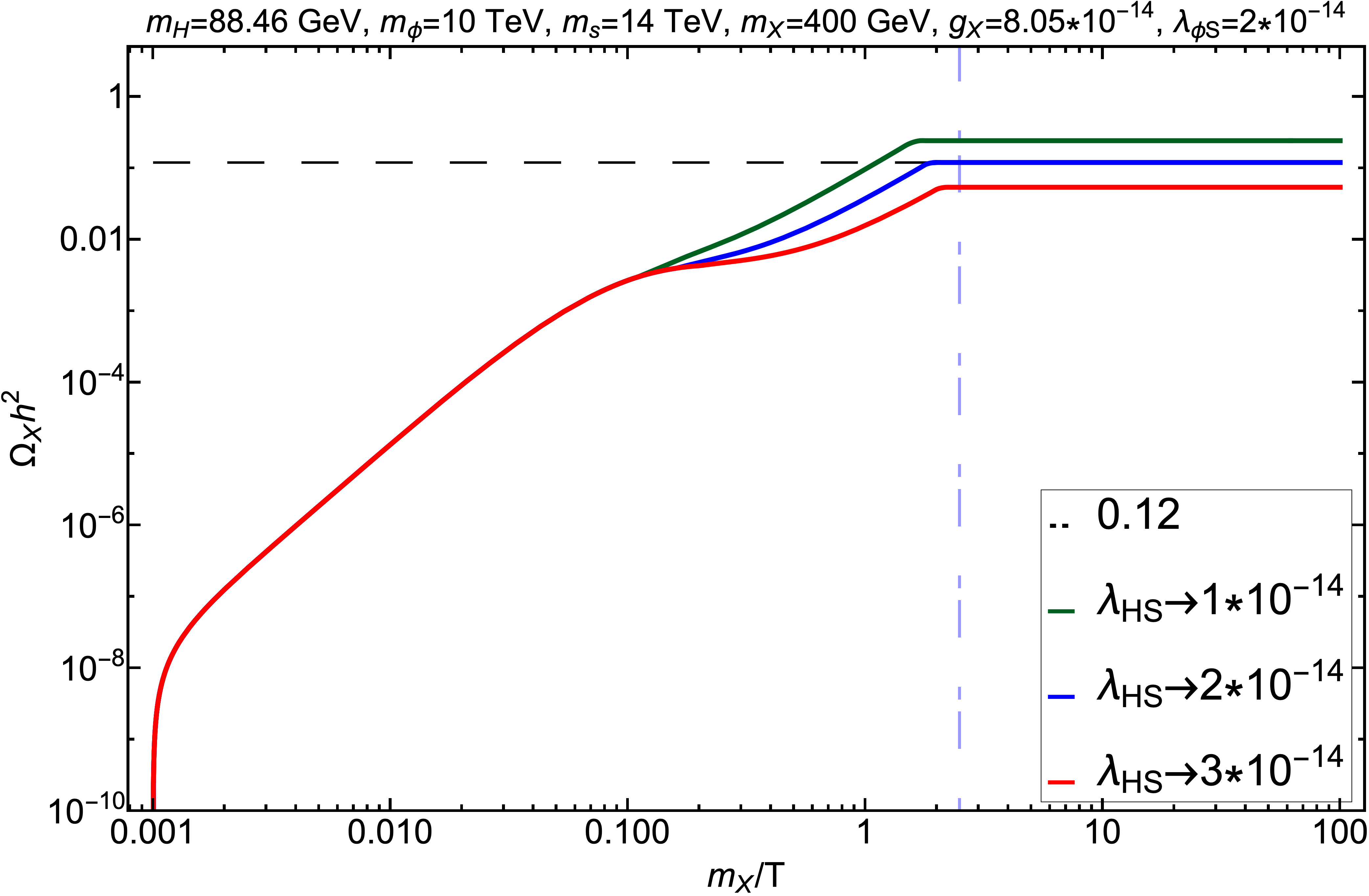}\label{frzin_fig4}}\\
	\subfloat[]{\includegraphics[width=0.5\linewidth]{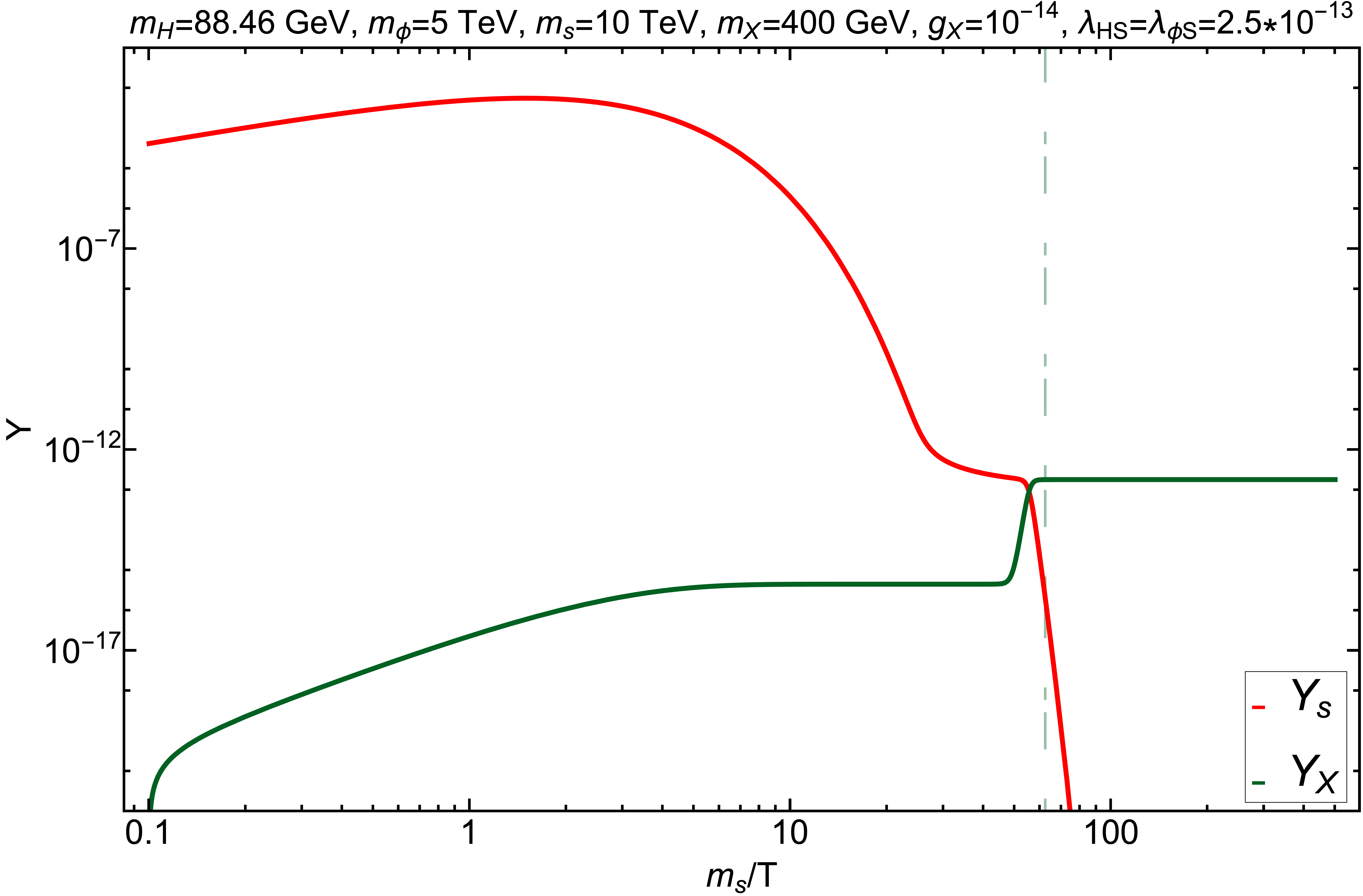}\label{frzin_fig5}}
	\caption{ Freeze-in for $X$ bEWSB in relic density ($\Omega_Xh^2$) versus $x=\frac{m_X}{T}$ plane obtained 
	by solving BEQ (Eq.~\ref{BEQ_FIMP2}) for the kinematic region $2m_X \leq m_s$. 
	Fig. \ref{frzin_fig1}, \ref{frzin_fig2}, \ref{frzin_fig3}, \ref{frzin_fig4} shows variation with respect to parameters 
	$m_s, m_X,g_X,\lambda_{HS}$, having three different values where one provides correct relic, 
	one under abundance and one over abundance. Parameters chosen for the plot are mentioned in figure inset and heading.  
	Horizontal black dashed line shows correct relic density. The vertical dot-dashed lines indicate EWSB. 
	Fig. \ref{frzin_fig5} shows late decay contribution of $s$ to freeze-in of $X$ bEWSB, obtained by solving cBEQ (Eq.~\ref{cbeq_sX}) using Mathematica.}
\label{frzin_plot_1}
\end{figure}

 The plots in Fig. \ref{frzin_plot_1} show change in $\Omega_Xh^2$ in terms of $x=m_X/T$ where freeze-in necessarily occurs bEWSB. All 
 the plots are generated by solving BEQ (Eq.~\ref{BEQ_FIMP2}) using Mathematica. In Fig. \ref{frzin_fig1}, three different coloured lines in red, 
 blue, green correspond to three different values of $m_{s}$ (mentioned in figure inset with other parameters kept fixed are mentioned in the figure heading), 
 so that $X$ freezes-in bEWSB. The vertical blue dotted line shows EWSB $(x_{\rm{EW}}=m_X/\tew$). As $X$ is produced from $s$ decay (and late decay 
 of $s$), where the decay width of $s$ is proportional to $m_s$, it is clear that with larger $m_s$, the $X$ abundance increases. We also see that 
 the entire freeze-in of $X$, takes place in two steps. Firstly, when $s$ is in equilibrium i.e., for $T>T_D$ ($T_D$ denotes freeze-out point of $s$), 
 then $X$ yield increases from zero and reaches the first plateau when $T< m_X$. Afterwards, when $T$ drops to $T\lesssim T_D$, then $s$ 
 freezes out and the late decay of $s$ into $X$ is activated, $X$ yield rises again, eventually producing the second plateau. The horizontal black dotted 
 line represents the central value of the present DM relic abundance. We see that the blue line with $m_s=14$ TeV matches to correct relic, given other 
 parameters fixed as mentioned in the figure heading. Also note here that we choose $x=0.001$ to start the freeze-in production, 
 although ideally the maximum temperature of the bath ($T_{\rm{RH}}$) should be very high $T_{\rm{RH}}\sim T_{U(1)}$. This is simply because, 
 in both decay and scattering dominated freeze-in of $X$ in this model, the yield is independent of $T_{\rm{RH}}$, a typical feature of IR freeze-in.\par
 
 In Fig. \ref{frzin_fig2}, \ref{frzin_fig3} and \ref{frzin_fig4}, we show how the freeze-in of X depends on the parameters $m_X,~g_X$ and $\lambda_{HS}$ 
 when $s$ decay is the main source of $X$ production. In each plot three cases are shown, one for correct relic, one for under abundance and one for 
 over abundance. The values of the parameters kept fixed to achieve them can be read from figure insets and headings. As the resultant yield is 
 proportional to the decay width of $s$, the dependence of these parameters on the decay width solely determine the relic density accumulated. For example, 
 $m_X$ is inversely proportional to $s$ decay width. Therefore, with larger $m_X$, the relic density decreases in Fig. \ref{frzin_fig2}. On the other hand, $s$ decay 
 width is proportional to $g_X$, therefore $X$ relic density increases with larger $g_X$ as is clear from Fig. \ref{frzin_fig3}. In Fig. \ref{frzin_fig4}, we have shown 
 the dependence on $\lambda_{HS}$. The decoupling of $s$ from thermal bath depends on $\lambda_{HS}$. With larger $\lambda_{HS}$, $s$ annihilation 
 cross-section increases, delaying the decoupling of $s$ which in turn reduces the late decay contribution to $X$ yield, as evident from Fig. \ref{frzin_fig4}.

The very fact that the late decay contribution of $s$ is essentially that of freeze-out abundance of $s$ converting into $X$ yield, is clear when we solve the coupled 
BEQ for $s$ freeze-out and $X$ freeze-in together as elaborated in Appendix \ref{sec:bEWSB-details} (see Eq.~\ref{cbeq_sX}) and demonstrated in Fig. \ref{frzin_fig5}. 
Here, the green line represents the variation of $X$ yield ($Y_X$) with $m_s/T$ and the red line represents $Y_s$. $Y_s$ shows the freeze-out of $s$ 
from the equilibrium distribution and then late decay to $X$ (descending part of $s$ yield after freeze out). The freeze-out yield of $Y_s$ matches to $Y_X$ yield completely. 
The vertical green dashed line confirms that the entire phenomenon occurs bEWSB for the chosen parameters of the model.\par 

\begin{figure}[htb!]
	\centering
\subfloat[]{\includegraphics[width=0.5\linewidth]{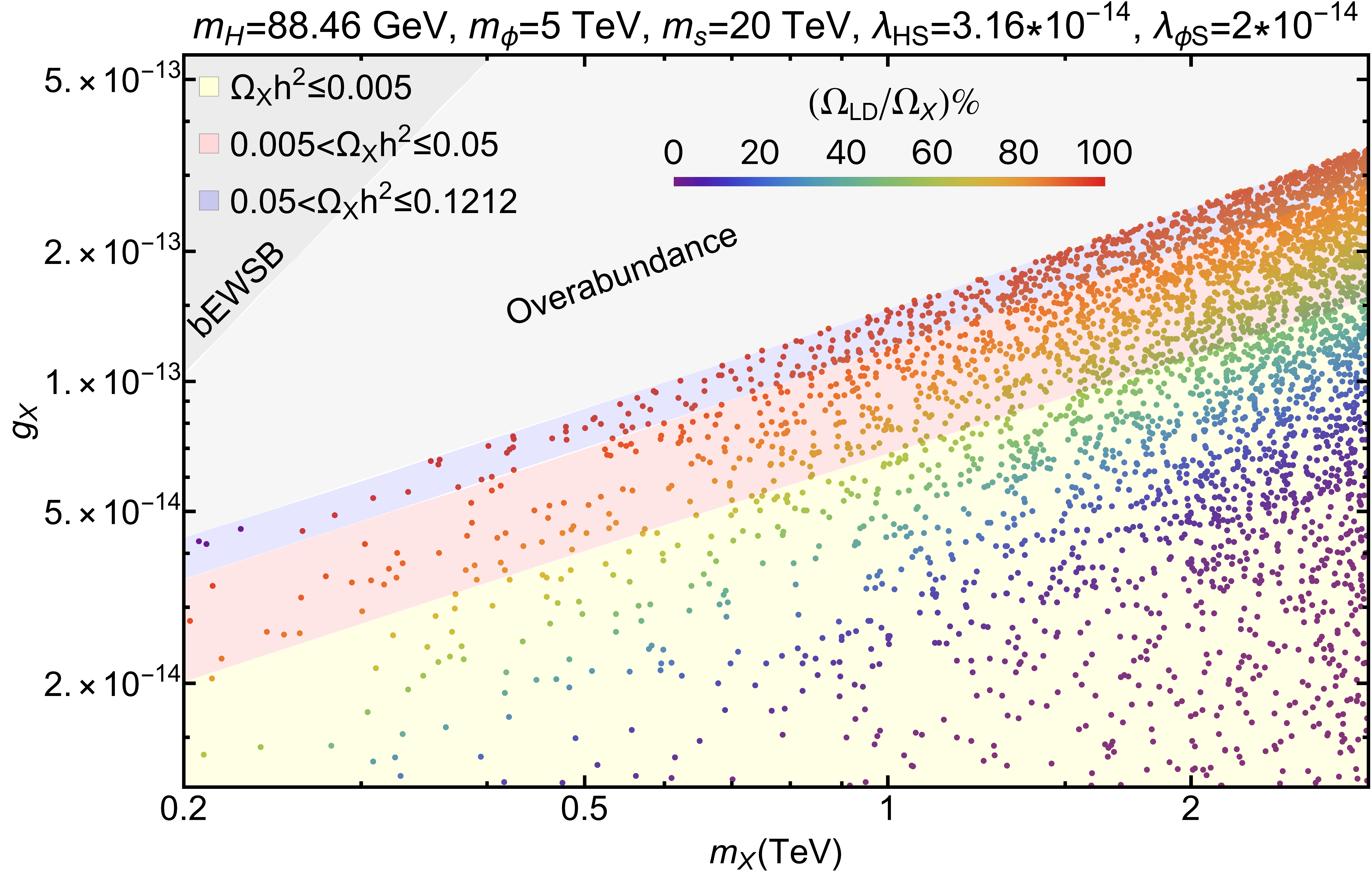}\label{frzin-scan1}}~~
\subfloat[]{\includegraphics[width=0.51\linewidth]{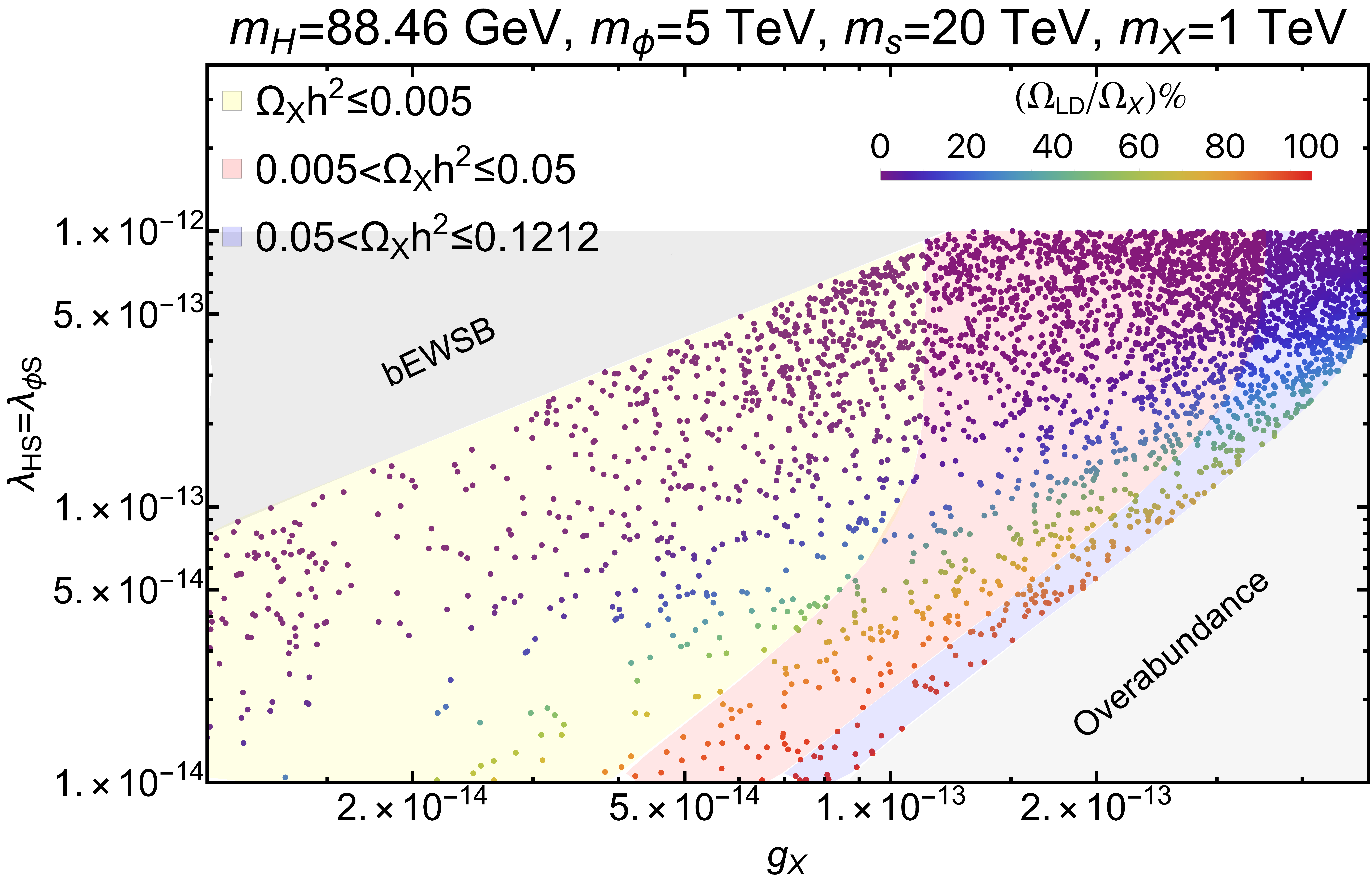}\label{frzin-scan2}}\\
\subfloat[]{\includegraphics[width=0.55\linewidth]{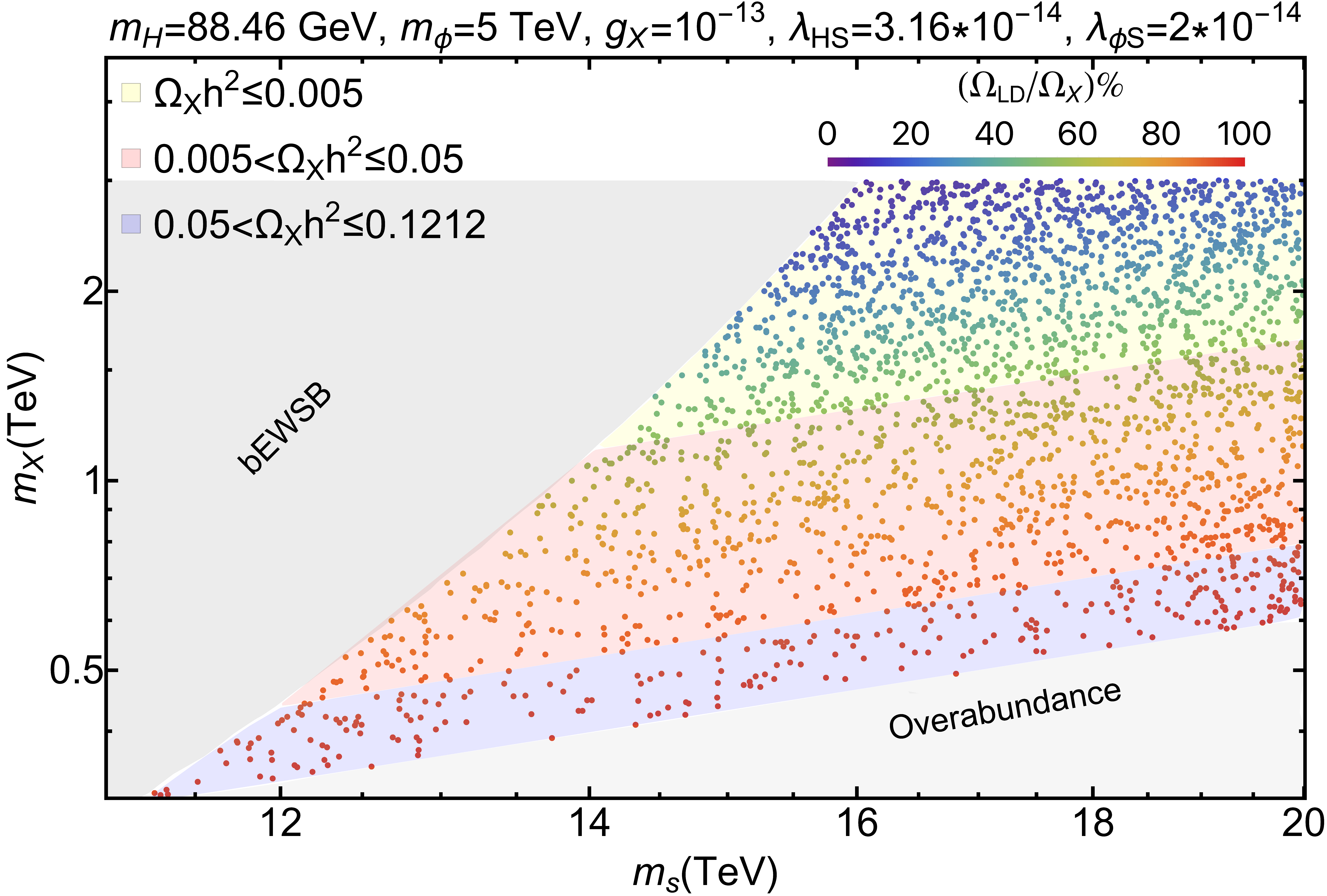}\label{frzin-scan3}}
  \caption{ Numerical scan of the under-abundant region of $X$ relic (Eq.~\ref{omg_X}) for freeze in bEWSB, when $\left(2m_X \leq m_s\right)$. 
  Figs. \ref{frzin-scan1} shows scan in $m_X-g_X$ plane, \ref{frzin-scan2} in $g_X-\lambda_{HS}=\lambda_{\phi S}$ plane and \ref{frzin-scan3} in $m_s-m_X$ plane. 
  Different colour shades indicate different ranges of relic density for fixed values of other parameters within our said bound as mentioned in figure inset 
  and caption. The rainbow colour bar represents the contribution of the late decay of $s$ in $X$ freeze-in by the ratio $\left(\Omega_{\rm{LD}}/\Omega_X\right)$. 
  This kinematic region is free from collider and direct search constraints, see text for details.}
  \label{frzin_plot_2}
\end{figure}

We find out next the relic under abundant parameter space of the model (Eq.~\ref{omg_X}) via numerical scan for the kinematic region 
$2m_X \leq m_s$ in Fig.~\ref{frzin_plot_2}. Fig. \ref{frzin-scan1} shows the parameter space in $g_X$ vs. $m_X$ plane. The color 
shades in light yellow, light red and light blue indicate different ranges of relic density (see Fig. inset). The scattered points with shades 
as in the color bar signify the percentage of `late decay' contribution to the relic density of $X$ ($\Omega_{\rm{LD}}h^2$) with 
respect to the total $X$ relic density ($\Omega_Xh^2$). The variation of relic density with $g_X$ and $m_X$ is consistent with the behaviour 
already noted in Fig. \ref{frzin_fig2} and \ref{frzin_fig3}, as we show that relic density increases with increasing $g_X$ and decreasing $m_X$. 
This is also true for the scattered points, as the functional dependence of the parameters are the same for both in-equilibrium 
decay and the late decay of $s$. In other two correlation plots, i.e., Fig. \ref{frzin-scan2} (scan in $g_X-\lambda_{HS}$ plane) and Fig. \ref{frzin-scan3} 
(scan in $m_s-m_X$ plane), we find that the change in relic density is consistent with Fig. \ref{frzin_fig1} and Fig. \ref{frzin_fig4}. In all these three 
correlation plots, we mark the overabundant region with light grey shaded region and the deep grey area signifies the parameter region where freeze-in 
bEWSB condition is not maintained. We further note that as only decay of $s$ dominates the production of $X$, Higgs mixing does not appear 
aEWSB and so collider bound is mostly absent. DD cross-section (the discussion is postponed to appendix \ref{sec:directsearch} as it is a 
standard exercise) is only affected by $\lambda_{\phi H}$ parameter, which is not very sensitive to the decay dominated production, 
especially when $m_{\phi}$ is in TeV range. This makes the parameter space free from the experimental constraints. We must also note that 
for all plots bEWSB in the kinematic region $2m_X \leq m_s$, the choice of $m_H=m_{h_1}/\sqrt{2}=88.46$ GeV is consistent with a SM 
Higgs with $m_{h_1}=125.1$ GeV.\par

\paragraph{\textbf{\underline{Case-II:}} $\left( m_s\lesssim2m_X\right):$\\}  

Now, we consider a kinematic region where $s$ decay is kinematically forbidden to produce $X$, with $m_s < 2m_X$. Absence of decay (and late decay) 
indicates that scattering, as shown in  Fig. \ref{fig:FD-frzin-bEWSB}, plays a crucial role in the production of $X$. If $X$ is produced through 
scattering or WIMP-FIMP conversion bEWSB, then $s$ remains in the thermal bath to mix with $h$ after EWSB, eventually connecting DM to 
SM. In this case, DD remains a viable option for detection of WIMP (see appendix \ref{sec:directsearch}). Also the mixing angle 
($\sin\theta$) of $s$ and the CP-even neutral component of Higgs doublet is restricted by the upper bound on mixing obtained from collider 
search as $\sin\theta\le \mathcal{O}(0.3)$ \cite{Chalons:2016lyk}. On top of that, following the correlation between $\lambda_{HS}$ and $\sin\theta$ as 
in Eq. \ref{aEWSB_rel}, $\lambda_{HS}$ will get further constrained by the mixing bounds, and constrain the parameter space bEWSB. On the contrary, 
when FIMP production completes bEWSB via $s$ decay (and late decay), $\lambda_{HS}$ remains mostly unconstrained due to absence of $s$ aEWSB.\par

\begin{figure}[htb!]
\centering
\subfloat[]{\includegraphics[width=0.48\linewidth]{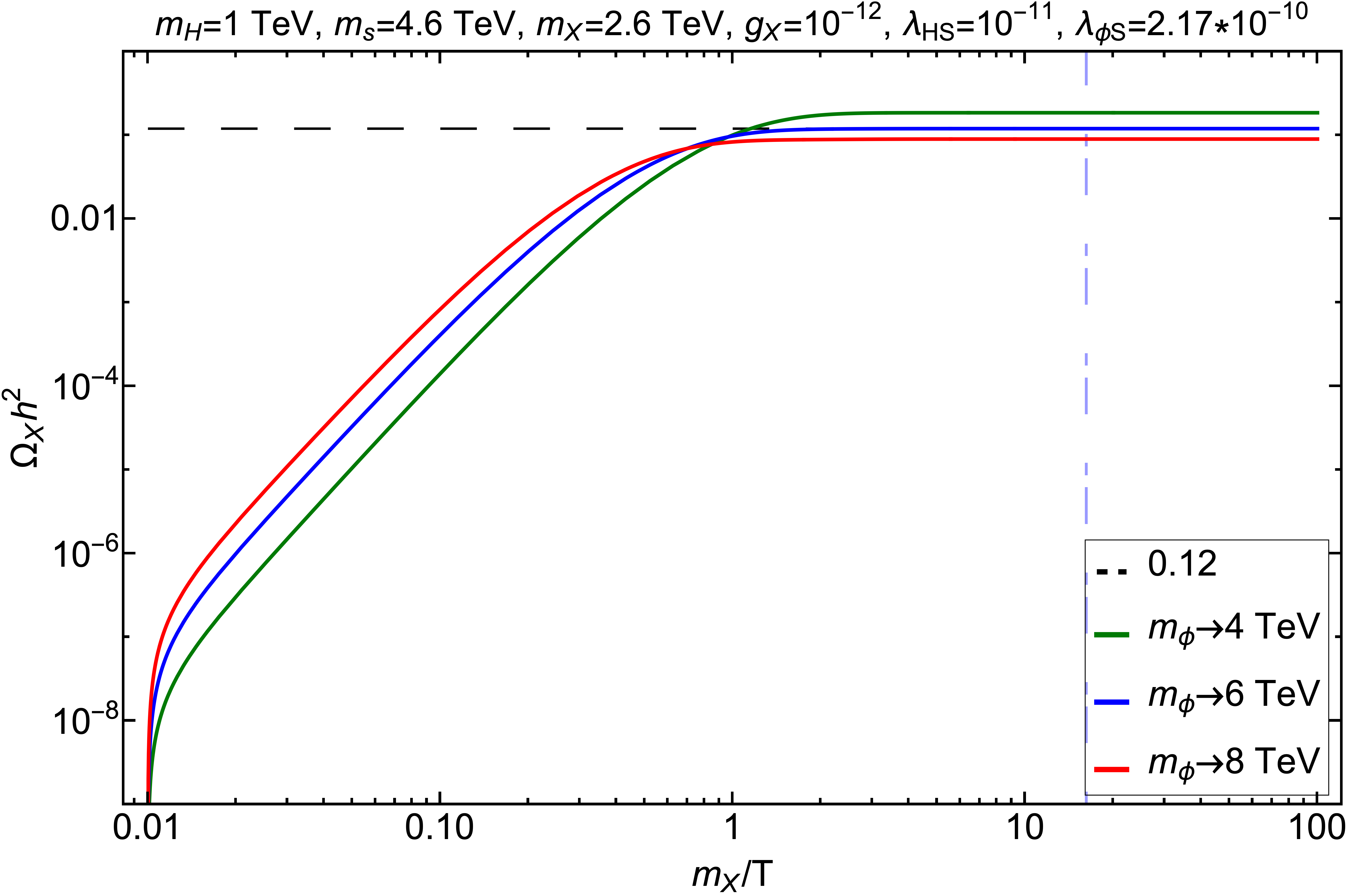}\label{frzin-scatt2}}~~
\subfloat[]{\includegraphics[width=0.48\linewidth]{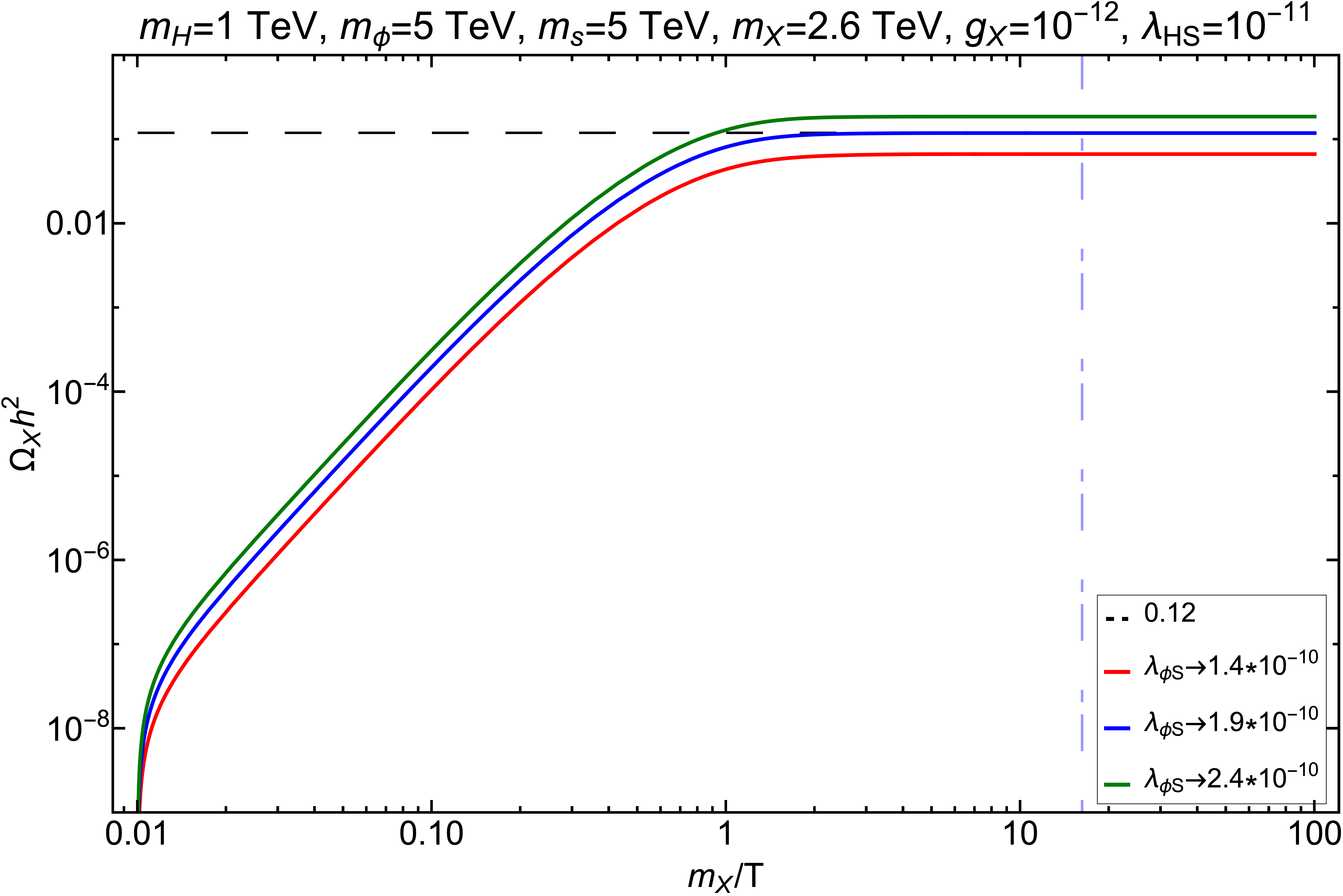}\label{frzin-scatt4}}
 \caption{Freeze-in production for $X$ bEWSB in relic density ($\Omega_Xh^2$) versus $x=\frac{m_X}{T}$ plane for the kinematic region $m_s < 2m_X$, 
 when scattering processes contribute to DM production. Variation with respect to $m_\phi$ (left) and $\lambda_{\phi S}$ (right) for three representative values that 
 provide under, correct and over relic abundance are shown by red, blue, green lines. Horizontal black dashed line shows observed relic density. The vertical dot-dashed 
 lines denote the boundary of EWSB. The parameters kept fixed are written in the figure insets as well as in the figure heading. }
\label{frzin_scattering_plot}
\end{figure}

We first depict the freeze-in patterns in Fig.~\ref{frzin_scattering_plot}, in terms of $\Omega_Xh^2$ as a function of $x=m_X/T$. 
This is similar to Fig.~\ref{frzin_plot_1}, where the vertical dot-dashed lines denote EWSB and in each case we ensure that $X$ 
freezes in bEWSB ($x_{\rm{FI}}<x_{\rm{EW}}$), but for kinematic region $m_s < 2m_X$. In Fig. \ref{frzin-scatt2} and Fig. \ref{frzin-scatt4}
the variation of $\Omega_X h^2$ is shown with respect to $m_{\phi}$ and $\lambda_{\phi S}$ respectively. In each case three choices 
of parameters provide under, correct and over abundance to indicate their role in DM production. For example, with the 
increase of $m_{\phi}$, the $X$ production cross-section decreases, which in turn, decreases FIMP abundance as evident from Fig. \ref{frzin-scatt2}. 
Similarly, larger $\lambda_{\phi S}$ enhances DM production cross-section and FIMP relic, as seen in Fig. \ref{frzin-scatt4}. The parameters kept 
fixed for the plots are mentioned in Figure captions and respect the constraints elaborated in section \ref{sec:constraints}. \par

In both the freeze-in patterns observed in Fig.~\ref{frzin_plot_1} and Fig.~\ref{frzin_scattering_plot}, we see that the abundance builds slowly upto $x\sim 1$ which 
is usually classified as Infra Red (IR) freeze in, where the mass effect turns important. This is contrasted to the Ultra Violet (UV) freeze-in pattern advocated usually 
for DM EFT theories as in \cite{Bhattacharya:2021edh,Fitzpatrick:2012ix,Criado:2021trs,Falkowski:2020fsu}, where the abundance builds up at very high 
temperature or low $x$ and saturates. One may also notice the {\it slight} difference in freeze-in pattern due to decay and annihilation dominated productions; 
for the decay, the yield builds up even slower with late decay contribution adding up as in Fig.~\ref{frzin_plot_1},
compared to the production via scattering as in Fig.~\ref{frzin_scattering_plot}.

\begin{figure}[htb!]
\centering
\subfloat[]{\includegraphics[width=0.48\linewidth]{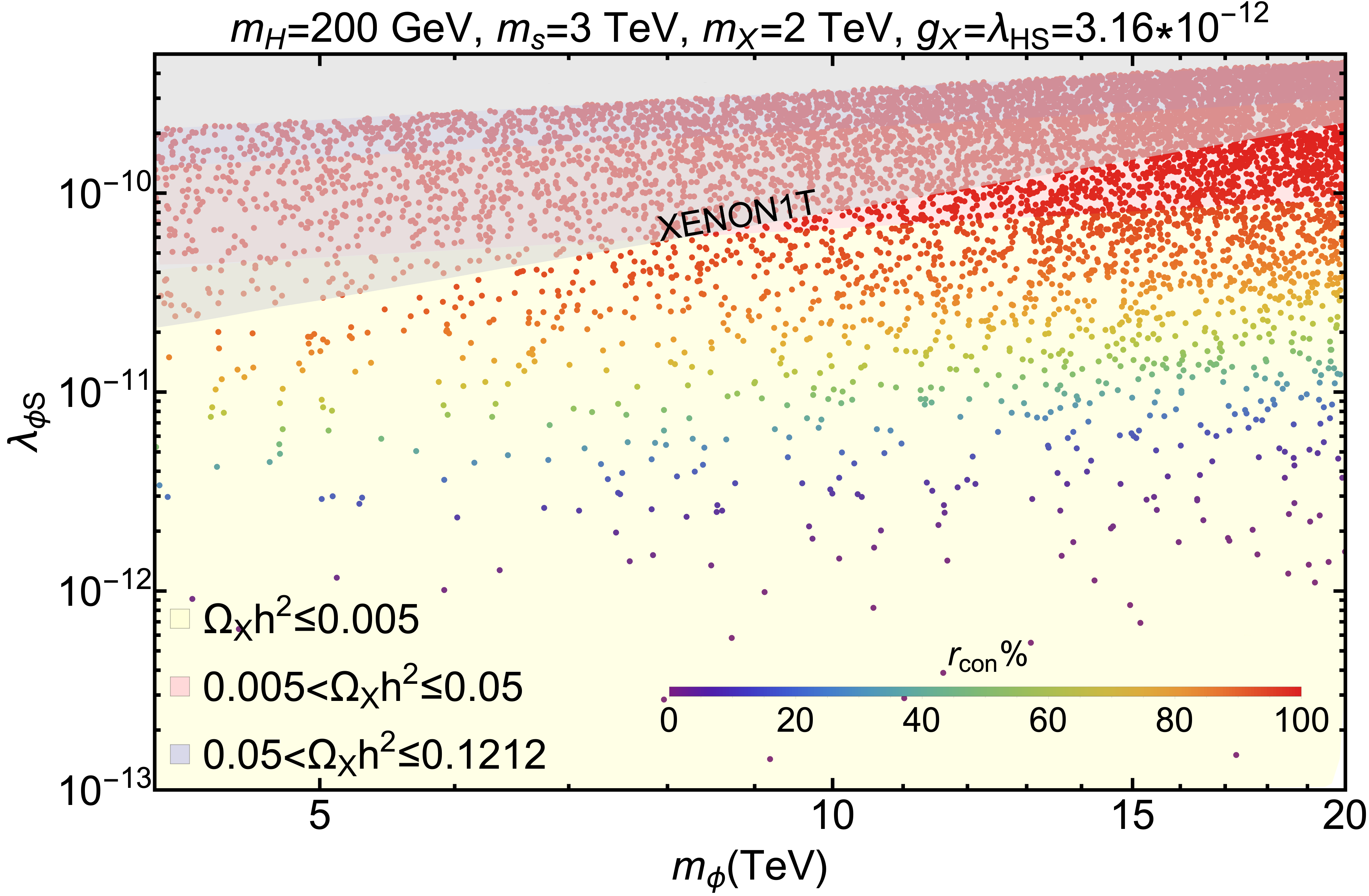}\label{frzin-scatt9}}~~
\subfloat[]{\includegraphics[width=0.48\linewidth]{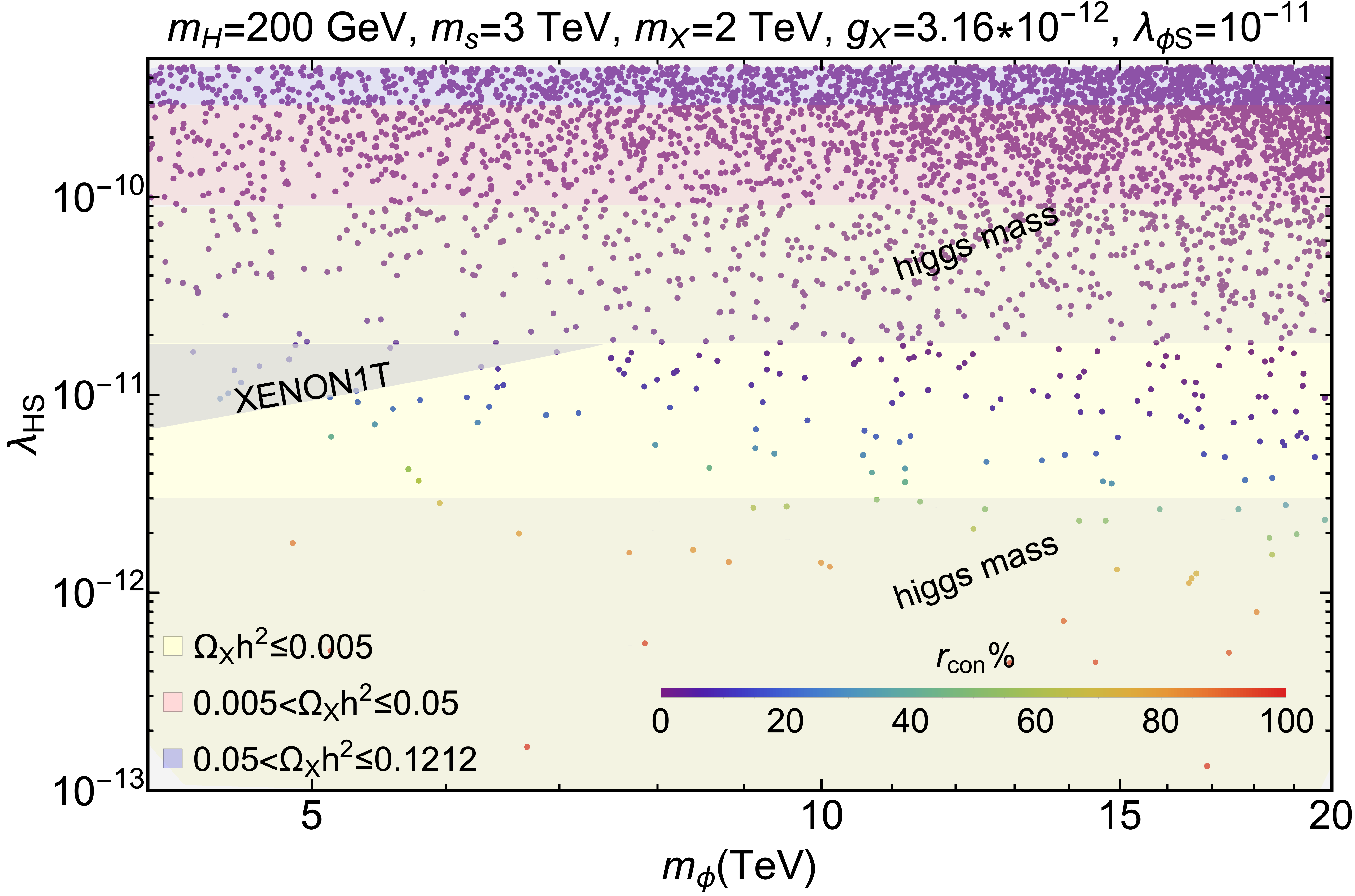}\label{frzin-scatt10}}
\caption{Numerical scan of the under-abundant region of $X$ relic for freeze-in bEWSB, when $m_s < 2m_X$. Fig. \ref{frzin-scatt9} and \ref{frzin-scatt10} show 
correlation in $m_\phi-\lambda_{\phi S}$ and $m_\phi-\lambda_{H S}$ planes respectively. Three different colour shades indicate different ranges of relic density
(see figure inset) for fixed values of other parameters written in figure heading. The rainbow colour bar represents how much the WIMP-FIMP conversion 
($\phi\phi\to XX$) is contributing in production of $X$ by the ratio $r_{\rm{con}}$ (see Eq.~\ref{eq:con}). 
Grey shaded areas signify the regions excluded by direct detection and Higgs mass constraints.}
\label{frzin_scattering_scan_plot}
\end{figure}
\par

We turn next to the parameter space scan for the FIMP under abundance in the kinematic region $ m_s < 2m_X$ as shown in 
Fig. \ref{frzin_scattering_scan_plot}, correlating different parameters relevant for scattering/conversion processes. 
While the light yellow, light red and the light blue shaded regions signify different ranges of relic density (see figure inset), the scattered points 
with different colours as in the colour bar signify the percentage of WIMP-FIMP conversion channels with respect to the total FIMP production via the following ratio:
\bea
r_{\rm{con}}=\frac{\langle\sigma \mathpzc{v}\rangle_{\phi\phi \to XX}}{\langle\sigma \mathpzc{v}\rangle_{\rm{Tot}}} \,.
\label{eq:con}
\eea
As mentioned before, the presence of $s$ after EWSB, when FIMP production is prohibited from the decay of $s$, ensures direct search and collider 
search possibilities of the model, which in turn puts appropriate bounds on the parameter space in absence of a signal. These constraints are superimposed 
on the parameter space by grey shaded exclusion regions. Fig. \ref{frzin-scatt9} shows scan in $m_\phi-\lambda_{\phi S}$ plane and we see that XENON1T 
bound heavily constrains $\lambda_{\phi S} \lesssim 2\times10^{-11}$. WIMP-FIMP conversion is then restricted significantly as $\lambda_{\phi S}$ affects 
conversion contribution directly and FIMP becomes heavily under abundant $\Omega_Xh^2\lesssim 0.005$. In Fig. \ref{frzin-scatt10} we show scan in 
$m_\phi-\lambda_{H S}$ plane. As expected Higgs mass bound plays a crucial role together with DD constraints to limit 
$3\times10^{-12}\lesssim \lambda_{HS} \lesssim 2\times10^{-11}$, again to make the FIMP heavily under abundant ($\Omega_Xh^2\lesssim 0.005$), 
particularly with the choice of $\lambda_{\phi S}=10^{-11}$ as done for the scan. The conversion contribution is large when $m_\phi$ is small, as expected.

We further intend to highlight that in all these scans, the parameters directly affecting Higgs mass and mixing after EWSB, ie, $m_s$, $m_X$ and $g_X$ 
(see Eq.~\ref{eq:mixing}) are all very fined-tuned. Owing to this requirement, there is not enough range to show the variations of these parameters in a scan. 
Therefore, we choose to vary the parameters in the dark sector that does not affect Higgs mass; $\lambda_{HS}$ being the only exception, shows a very narrow 
viable region, as pointed out in Fig. \ref{frzin-scatt10}.

\subsection{Freeze-out of $\phi$}

The scalar singlet dark matter $\phi$ is assumed to be in thermal bath as WIMP, tracking the equilibrium (non-relativistic and Maxwell-Boltzmann) distribution
in early universe. When the bath temperature ($T$) goes below the decoupling temperature of $\phi$, i.e. $T\lesssim T^\phi_D$, the interaction rate of DM 
with the bath particles eventually becomes less than the Hubble expansion rate $H$. This causes the DM to decouple from the thermal bath and freeze out to 
give the saturation abundance. In this section, we assume the freeze-out to occur bEWSB and find the region of parameter space 
where it happens and produces under abundance. As mentioned previously, there are several constraints to ensure freeze-out bEWSB such as $m_\phi\ge$4 TeV. 
Further constraints on model parameters come from DD and collider searches as discussed before. 
We indicate the bounds in resulting parameter space. The annihilation channels of WIMP $\phi$, through which it depletes the number density can be divided 
into two main categories:

\begin{itemize}

\item {\bf Annihilation to visible sector}: 

The channels bEWSB, include $\phi$ pair-annihilation into $s$ and $H$ pairs. The relevant Feynman diagrams are in Fig. \ref{fig:frzout-FD-EWSB}. The couplings 
relevant to the above scatterings are $\lambda_{\phi S}$, $\lambda_{H S}$, $\lambda_{\phi H}$ and $\lambda_S ~(=\frac{m_s^2}{2v_s^2})$. 
As already mentioned, $v_s=m_X/g_X$ must always be very large ( $\sim 10^{14}$ GeV) throughout the analysis in order to have a successful 
FIMP ($X$) production as a CDM. Hence, unless we choose $\lambda_{\phi S} \sim 1/v_s$, couplings like $\phi\phi s\ (\propto v_s\lambda_{\phi S}$), 
will make the annihilation cross-sections very large, resulting in negligible $\phi$ abundance. Hence, in order to get a reasonable annihilation of 
$\phi$, $$\lambda_{\phi S}\lesssim 10^{-12} ~{\rm for}~ g_X \sim 10^{-12}.$$ 
Such a choice is consistent with both the freeze-in of $X$ and direct detection constraints on $\lambda_{\phi S}$. Although such small 
$\lambda_{\phi S}$ makes the four-point scattering cross section, such as the top left channel in Fig. \ref{fig:frzout-FD-EWSB}, 
practically negligible, $\phi$ mainly annihilates via the $s$ and $t$-channel diagrams in Fig. \ref{fig:frzout-FD-EWSB}, where presence of $v_s$ in one of the 
vertices like $\phi\phi s$ make the contribution sizeable. We further note that $\lambda_{\phi H}$ should also be greater than $10^{-3}$ to get a reasonable 
annihilation via $\phi\phi \to HH$.  Although $\phi$ freezes out bEWSB, we recall that $s$ and $h$ mixes due to EWSB. As a result,
$\lambda_{H S}$ is traded off as a parameter dependent on mixing. So, the collider searches of Higgs at the LHC, restricts $\lambda_{H S}$. 
We indicate the effect of such constraints on the allowed parameter space.

\item {\bf Conversion to FIMP DM} : 

$\phi$ annihilates into $X$ pair via $s$ mediation (bottom panel of Fig. \ref{fig:frzout-FD-EWSB}). But since freeze-in requires $g_X$ to be very 
small ($\lesssim 10^{-12}$), one vertex of conversion diagram ($XXs$) proportional to $g_X^2v_s$ is also minuscule; this evidently implies that unless the other vertex 
$\lambda_{\phi S}$ is chosen sufficiently large ($\sim$ 1), the conversion contribution is negligible. However, $\lambda_{\phi S}$ requires to be small from DD, 
makes the conversion very small. Secondly, large conversion cross section to FIMP production automatically implies that $X$ production will be too fast for the 
non-thermal freeze-in and it will drive $X$ towards equilibrium, seizing the FIMP nature of $X$. Hence, WIMP$\leftrightarrow$FIMP conversion is negligible 
in the context of WIMP, but plays an important role in the FIMP production as already demonstrated in previous subsection. 

\end{itemize}

Upon neglecting the WIMP-FIMP conversion, the cBEQ reduces to two individual uncoupled BEQs; the one for $\phi$ is given by 
Eq. \ref{split-coup-BEQ}, which can be easily solved numerically (we use Mathematica 12.3.1.0 \cite{Mathematica}). The parameters 
are chosen in such a way that the freeze-out occurs bEWSB. The relic density for $\phi$ can then be written in terms of freeze-out yield 
\cite{Kolb:1990vq} and we again focus on the under abundant region of the parameter space, given $\phi$ is one of the two DM 
components that we assume to constitute the dark sector:  
$$
\Omega_{\phi} {\rm h^2} \simeq 2.744 \times 10^8\ m_{\phi} Y_{\phi}^{{\rm bEWSB}}; ~~\Omega_{\phi} {\rm h^2} \le  0.1212 \,.
$$
 where WIMP dark matter relic density is written in terms of the reduced Hubble parameter, $h$ in units of 100 km/s/Mpc.
\subsubsection{Phenomenology}

\begin{figure}[htb!]
\centering
  \subfloat[]{\includegraphics[width=0.48\linewidth]{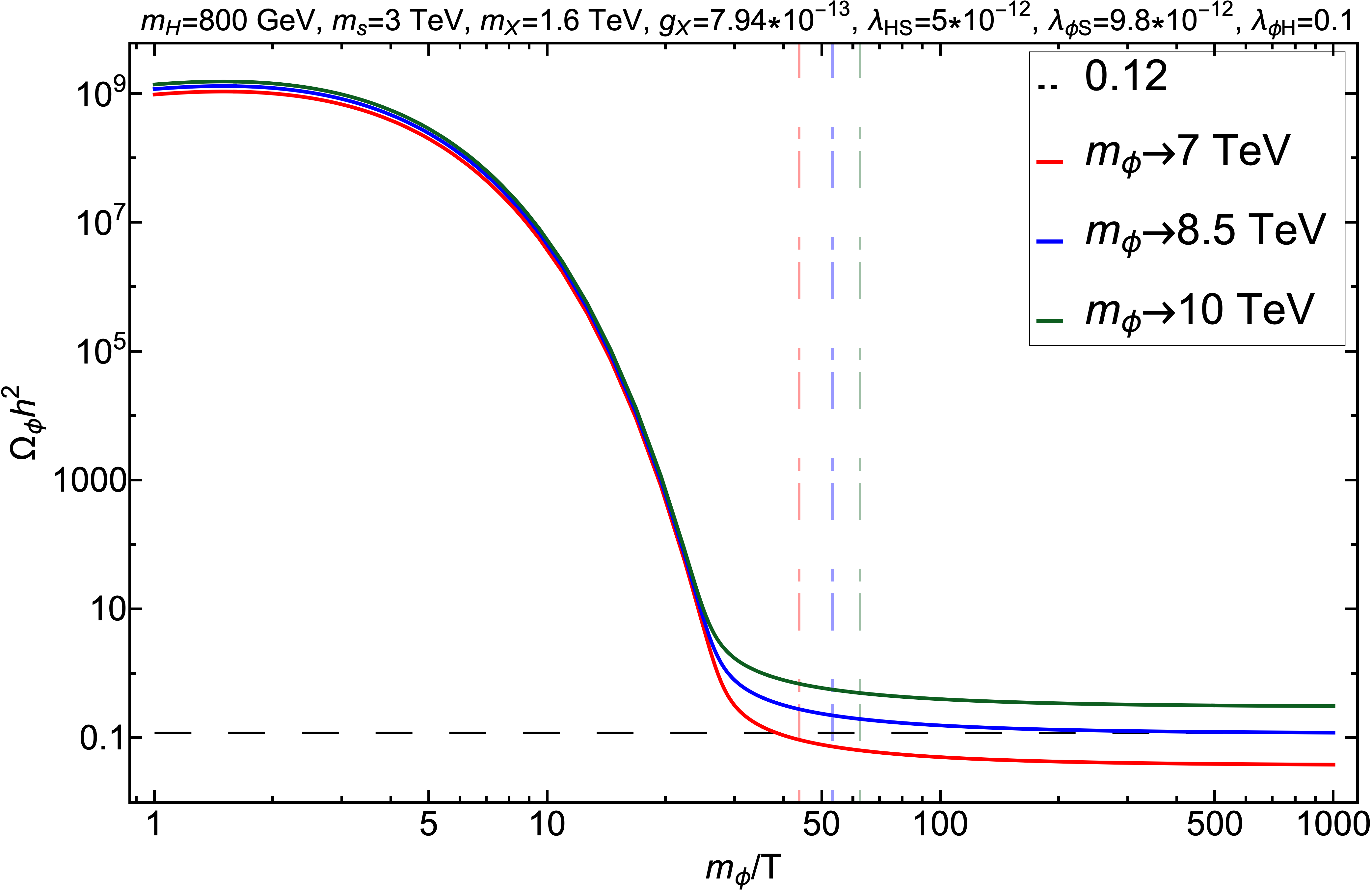}\label{wimp_bEWSB_m2}}~~
  \subfloat[]{\includegraphics[width=0.48\linewidth]{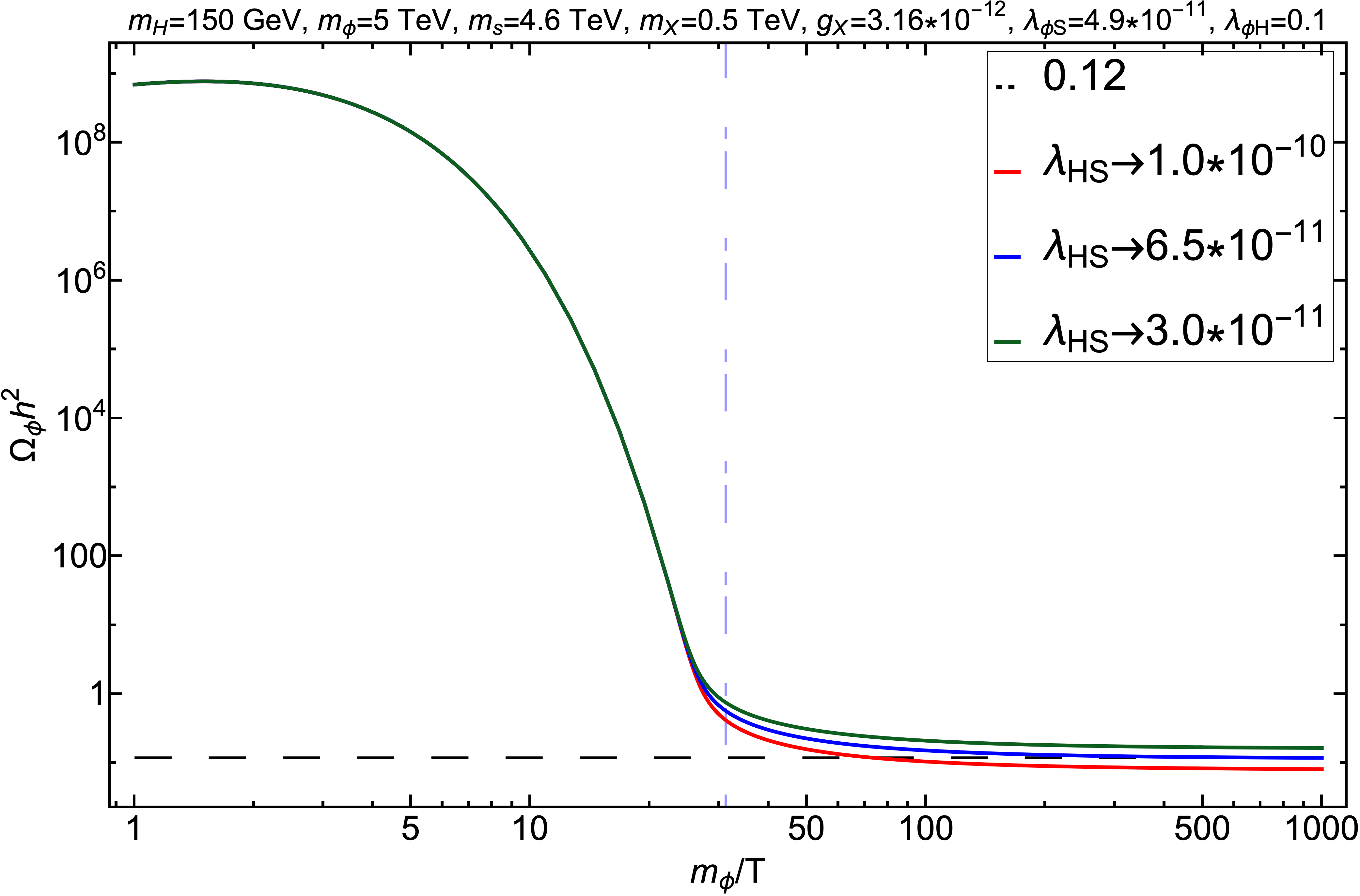}\label{wimp_bEWSB_l1S}}\\
    \subfloat[]{\includegraphics[width=0.48\linewidth]{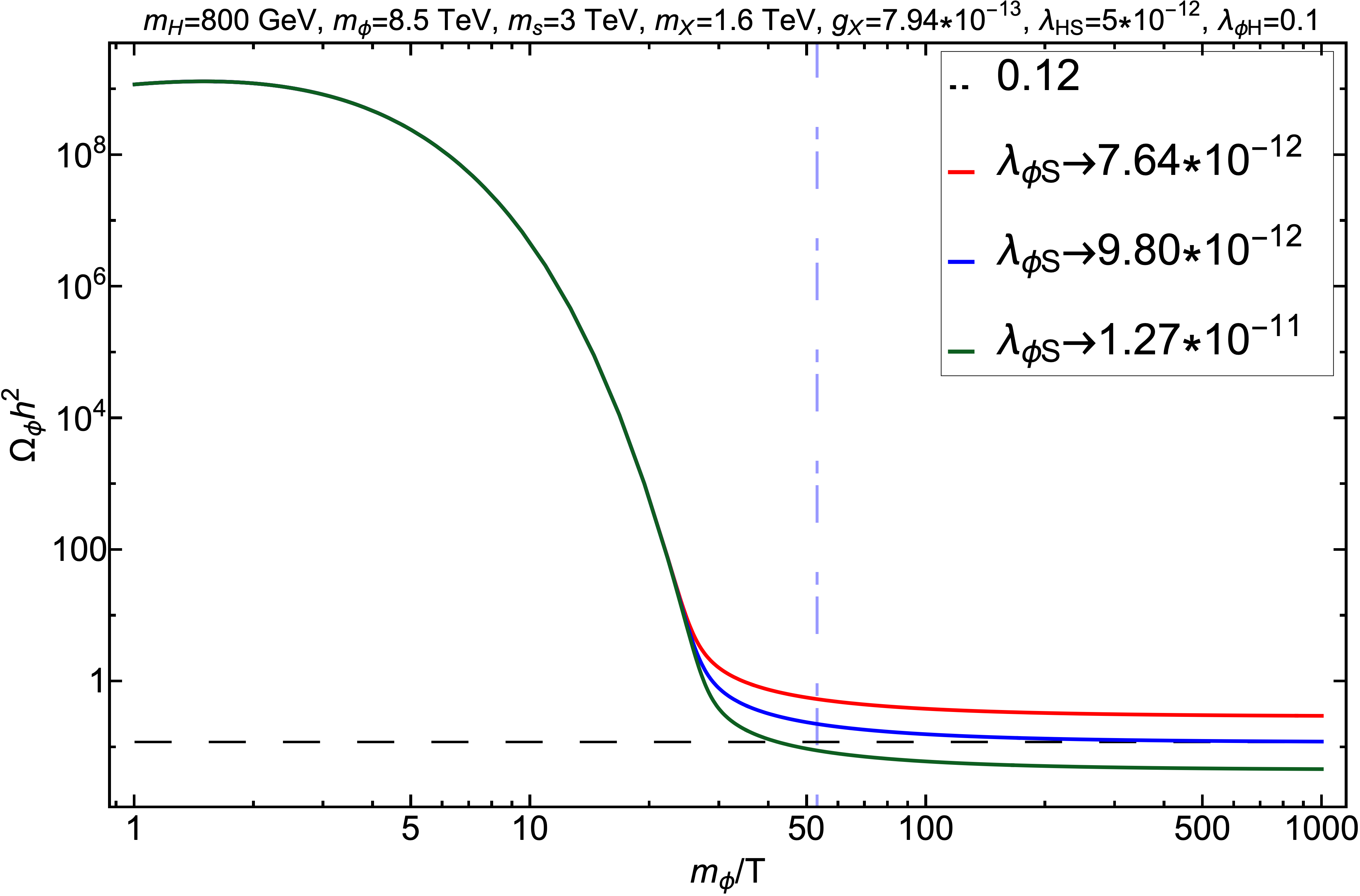}\label{wimp_bEWSB_l2S}}~
  \subfloat[]{\includegraphics[width=0.48\linewidth]{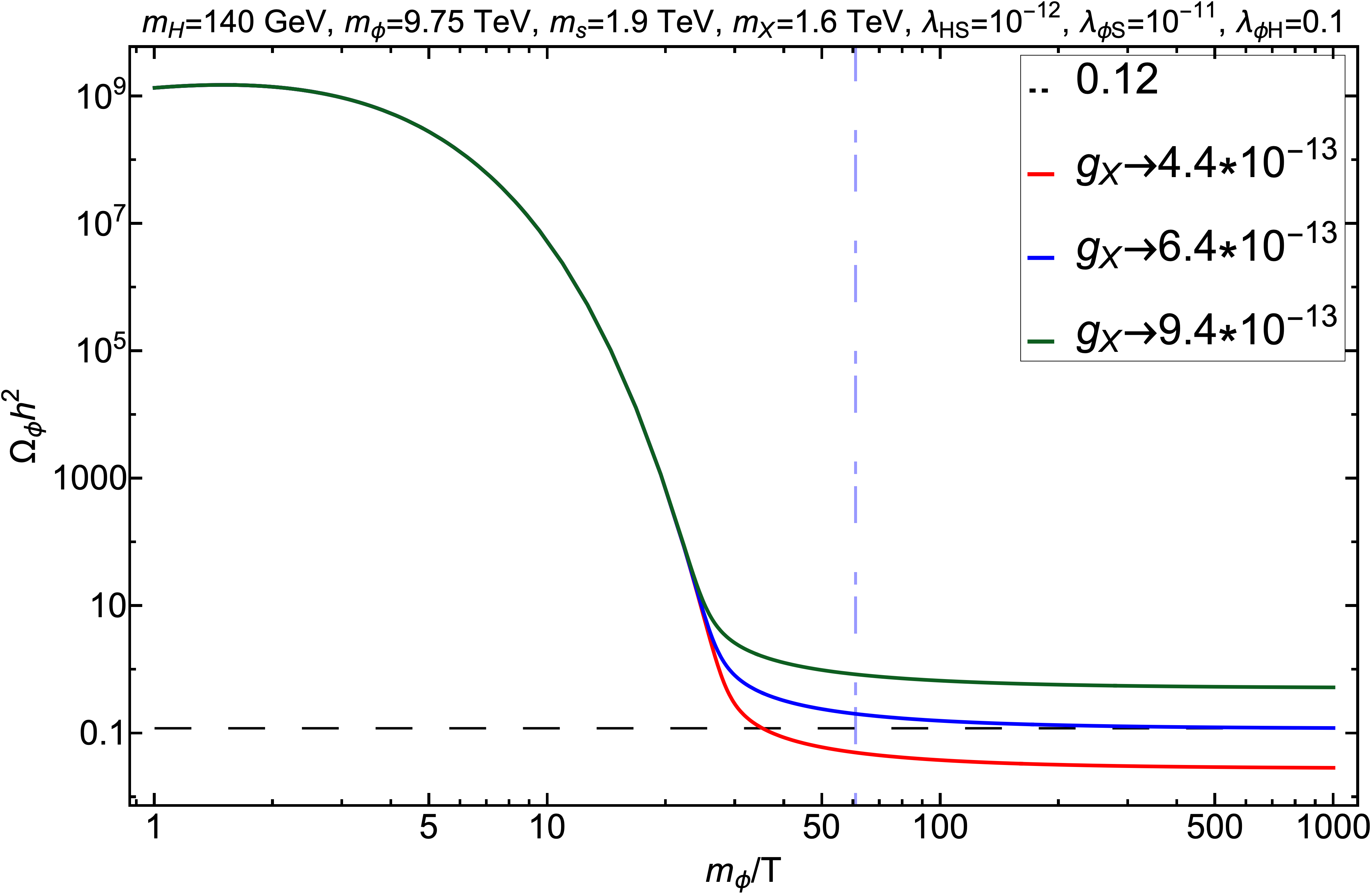}\label{wimp_bEWSB_gx}}
  \caption{WIMP relic density ($\Omega_{\phi}h^2$) in terms of $x=m_{\phi}/T$ for three different values of parameters 
  $\lambda_{HS},\lambda_{\phi S},m_{\phi},g_X$ (from left to right and top to bottom) to provide correct relic, under and over abundance. 
  The parameters kept fixed are mentioned in the insets as well as in the figure headings. The horizontal black dotted line denotes correct 
  relic abundance and the vertical dot-dashed lines depict EWSB, $x_{\rm{EW}}= m_{\phi}/160$. }
  \label{wimp_bEWSB}
\end{figure}

We first study $\phi$ freeze-out bEWSB as a solution of the BEQ \ref{split-coup-BEQ} in Fig.~\ref{wimp_bEWSB}, where we plot 
$\Omega_{\phi}h^2$ with $x=m_{\phi}/T$ for parameters $\lambda_{HS},\lambda_{\phi S},m_{\phi},g_X$. We choose three representative
values of these parameters so that we produce correct relic, under abundance and over abundance. The horizontal black dotted 
line denotes the current central value of DM abundance. Vertical dot-dashed lines indicate EWSB ($x_{\rm{EW}}={m_\phi}/{160~\rm{GeV}}$) 
and each freeze out occurs bEWSB with $x_{\rm{FO}}<x_{\rm{EW}}$. The parameters kept fixed for these plots, as mentioned in the figure 
insets and headings, comply with all the constraints mentioned earlier. 

As already mentioned, to ensure the WIMP freeze-out to take place bEWSB, the allowed mass of $\phi$ is constrained to 
$m_{\phi}\gtrsim 4$ TeV. To comply with this bound, in Fig.~\ref{wimp_bEWSB_m2}, the freeze out of $\phi$ is shown for $m_{\phi}=$ 7 TeV, 
8.5 TeV and 10 TeV, depicted by red, blue and green coloured lines respectively. As annihilation cross-section is inversely proportional to WIMP mass, 
and freeze-out yield is also inversely proportional annihilation cross-section, we see that as $m_\phi$ increases, the WIMP relic density 
also enhances and the case with $m_{\phi}=8.5$ TeV matches with correct relic. In Figs. \ref{wimp_bEWSB_l1S}, \ref{wimp_bEWSB_l2S} and  
\ref{wimp_bEWSB_gx}, we show the effects in WIMP relic due to variation of $\lambda_{HS},\lambda_{\phi S},g_X$ respectively. As the annihilation 
cross-section of WIMP ($\phi$) increases with larger couplings, we see that the relic density reduces expectedly with larger $\lambda_{HS},\lambda_{\phi S}$ 
in Fig. \ref{wimp_bEWSB_l1S} and \ref{wimp_bEWSB_l2S}. The scenario changes in Fig. \ref{wimp_bEWSB_gx}, where variation with respect to $g_X$ is shown. 
In annihilation cross-section, $g_X$ enters inversely through $v_s~(=m_X/g_X)$, as a result, annihilation to $H,s$ reduces with the increase of 
$g_X$, resulting in an enhancement of relic with $g_X$ as shown in Fig. \ref{wimp_bEWSB_gx}. \par

DD of $\phi$ occurs through Higgs mediation (see Appendix \ref{sec:directsearch}). Even if WIMP freezes-out bEWSB, direct search of $\phi$ is possible 
at present epoch, so the constraints apply. However, the constraints depend on kinematical regions: (i) $m_s\geq2m_X$ and (ii) $m_s<2m_X$ in a similar 
vein as discussed before.

 \paragraph{\textbf{\underline{Case-I:}} $\left( m_s\geq2m_X\right):$ \\}
  
  When $m_s\geq 2m_X$, and we ensure $X$ freeze-in to saturate bEWSB, the decay of $s$ totally depletes its number density, 
  so that  any $s-h$ mixing aEWSB is non-existent, resulting only SM Higgs mediating direct search for $\phi$. So the situation 
  is similar to the DD of single component scalar singlet $\phi$.\par

\begin{figure}[htb!]
\centering
\includegraphics[width=0.55\linewidth]{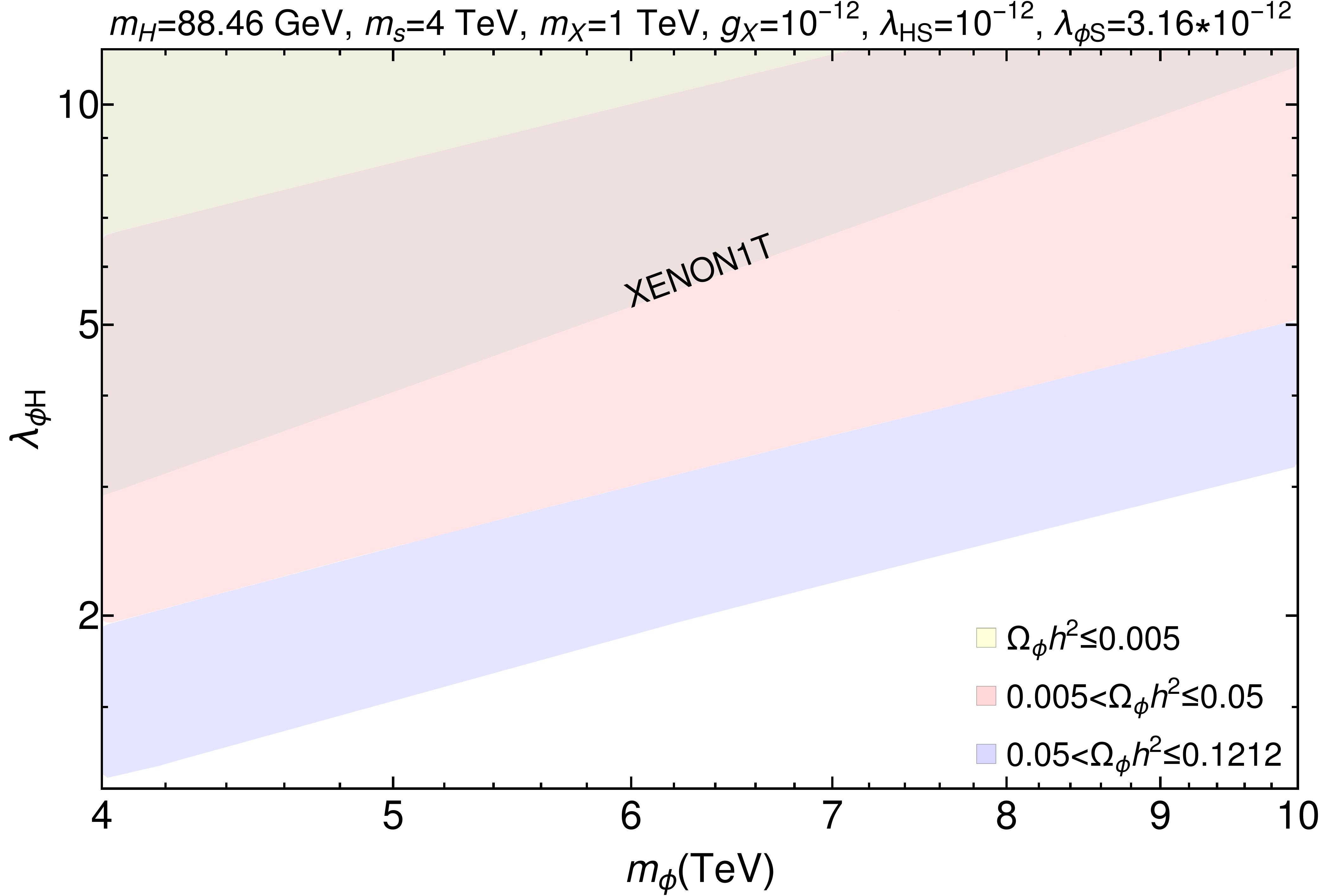}
	\caption{Under abundant region for $\phi$ ($\Omega_{\phi} \rm h^2 \le  0.1212$) with respect to variation of $\lambda_{\phi H}$ vs. $m_\phi$ for kinematic region 
	$m_s\geq2m_X$ when freeze-out occurs bEWSB. Spin independent XENON1T direct search excluded region is showed by the grey shaded area 
	at the top for large $\lambda_{\phi H}$. Light yellow, light red and light blue shades indicate different ranges of $\Omega_{\phi} \rm h^2$ as mentioned in 
	figure inset. Parameters kept fixed for the plot are mentioned in the figure heading and comply with all other constraints.}
	\label{frzout_scan_msg2mx}
\end{figure}

Fig. \ref{frzout_scan_msg2mx} shows the under abundant parameter space in $m_{\phi}-\lambda_{\phi H}$ plane where $\phi$ freezes out bEWSB, 
in the kinematic region $m_s\geq2m_X$. Three colour shades indicate different ranges of $\Omega_\phi h^2$ (mentioned figure inset). The grey shaded 
region is excluded by the present spin-independent XENON1T limit, which restricts only very high values of $\lambda_{\phi H}\gtrsim 3$, given other parameters are
kept constant at values mentioned in the figure heading. As $m_s>4$ TeV (see Eq.~\ref{eq:masslimit}) for late decay to complete bEWSB, 
WIMP annihilation mostly occur through the four point interaction $\phi\phi \to HH^{\dagger}$. The correlation between $m_{\phi}-\lambda_{\phi H}$ 
is consistent with two features already discussed: (a) WIMP annihilation cross-section via four point interaction increases with 
$\lambda_{\phi H}$ which in turn reduces the abundance and (b) WIMP annihilation cross-section decreases with $m_\phi$, which causes 
$\Omega_\phi h^2$ to increase with the WIMP mass. 

\paragraph{\textbf{\underline{Case-II:}} $\left( m_s<2m_X\right):$\\}

\begin{figure}[htb!]  
\centering
	\subfloat[]{\includegraphics[width=0.48\linewidth]{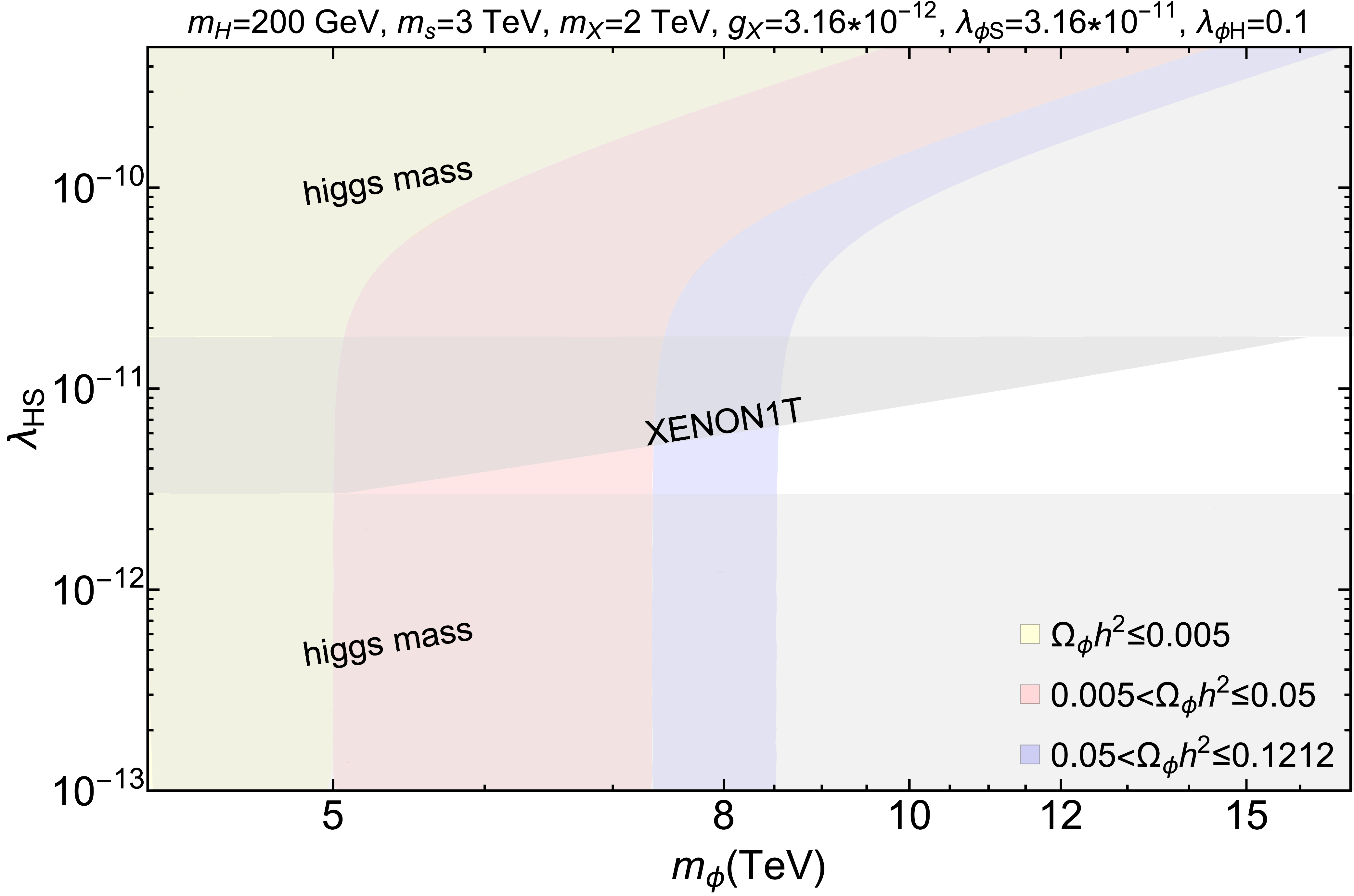}\label{wimp_m2l1s}}~~
	\subfloat[]{\includegraphics[width=0.48\linewidth]{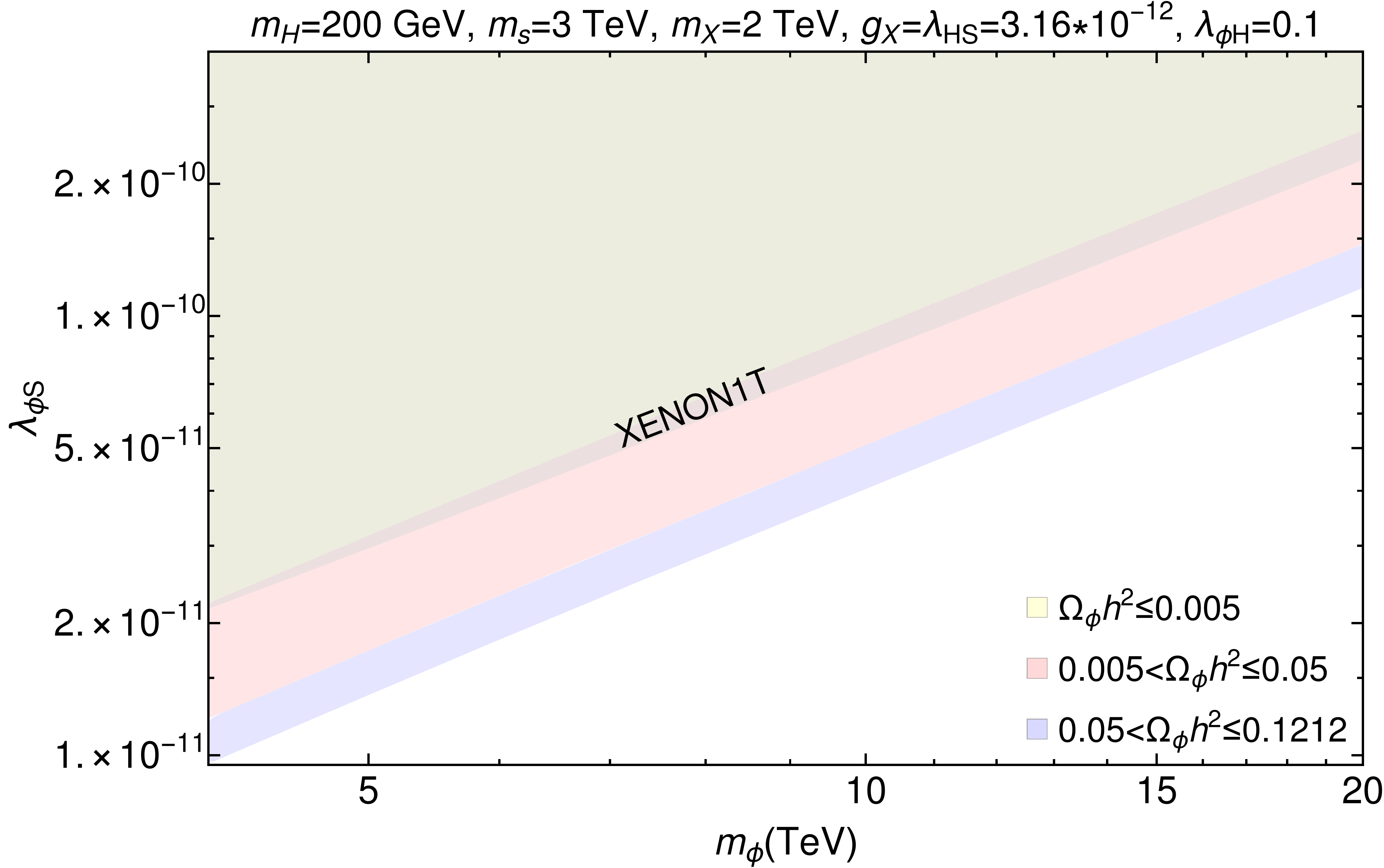}\label{wimp_m2l2s}}
	\caption{Under abundance for $\phi$ ($\Omega_{\phi} \rm h^2 \le  0.1212$) in the kinematic region $m_s<2m_X$. Fig.~\ref{wimp_m2l1s} shows the correlation 
	between $m_{\phi}-\lambda_{HS}$ and Fig.~\ref{wimp_m2l2s} shows the correlation between $m_{\phi}-\lambda_{\phi S}$. Different colour shades in light yellow, 
	light red and light blue indicates under abundance within ranges as mentioned in figure inset. Grey shaded regions are excluded by latest XENON1T bound, 
	Higgs mass and collider bound on scalar mixing (see text for details).}
	\label{frzout_scan_msl2mx}
\end{figure}

When $m_s\leq 2m_X$, $s-h$ mixing occurs aEWSB and direct search occurs via mediation of both physical states $h_1,h_2$. 
Therefore, mixing plays an important role in the direct detection of WIMP. In this case, the parameter space and the constraints are expectedly 
different from the previous case where mixing was absent. Correlations of relevant parameters for under abundance of $\phi$ 
($\Omega_{\phi} \rm h^2 \le  0.1212$) in this kinetic regime is shown in Fig.~\ref{frzout_scan_msl2mx} together with direct search and Higgs mixing constraint.

Fig. \ref{wimp_m2l1s} shows the under abundant parameter space in $m_{\phi}$ vs. $\lambda_{HS}$ plane where grey shaded regions are excluded 
by XENON1T direct search bound and Higgs mass/scalar mixing constraints. The functional dependence of $m_{\phi}$ as in Fig. \ref{wimp_bEWSB_m2} 
and of $\lambda_{HS}$ as in Fig. \ref{wimp_bEWSB_l1S} are retained here. We conclude that $\lambda_{HS}\sim5\times 10^{-12}$ is safe for $m_{\phi}$ varying 
within 7 to 8.5 TeV, given the other model parameters are kept fixed as mentioned in the figure caption. Fig. \ref{wimp_m2l2s} shows the correlation between 
$m_{\phi}-\lambda_{\phi S}$. Recall that $Y_\phi$ increases with larger $\lambda_{\phi S}$ (see Fig. \ref{wimp_bEWSB_l2S}) as well as with larger WIMP 
mass ($m_\phi$), which is also evident in Fig. \ref{wimp_m2l2s}. Once $\lambda_{H S}\sim 3.16\times10^{-12}$ is fixed in Fig. \ref{wimp_m2l2s}, 
it fixes the mixing angle within experimental limit, there is no other constraint on this parameter space excepting for the direct search bounds, depicted in grey shade. 
\begin{figure}[htb!]  
\centering
	\subfloat[]{\includegraphics[width=0.48\linewidth]{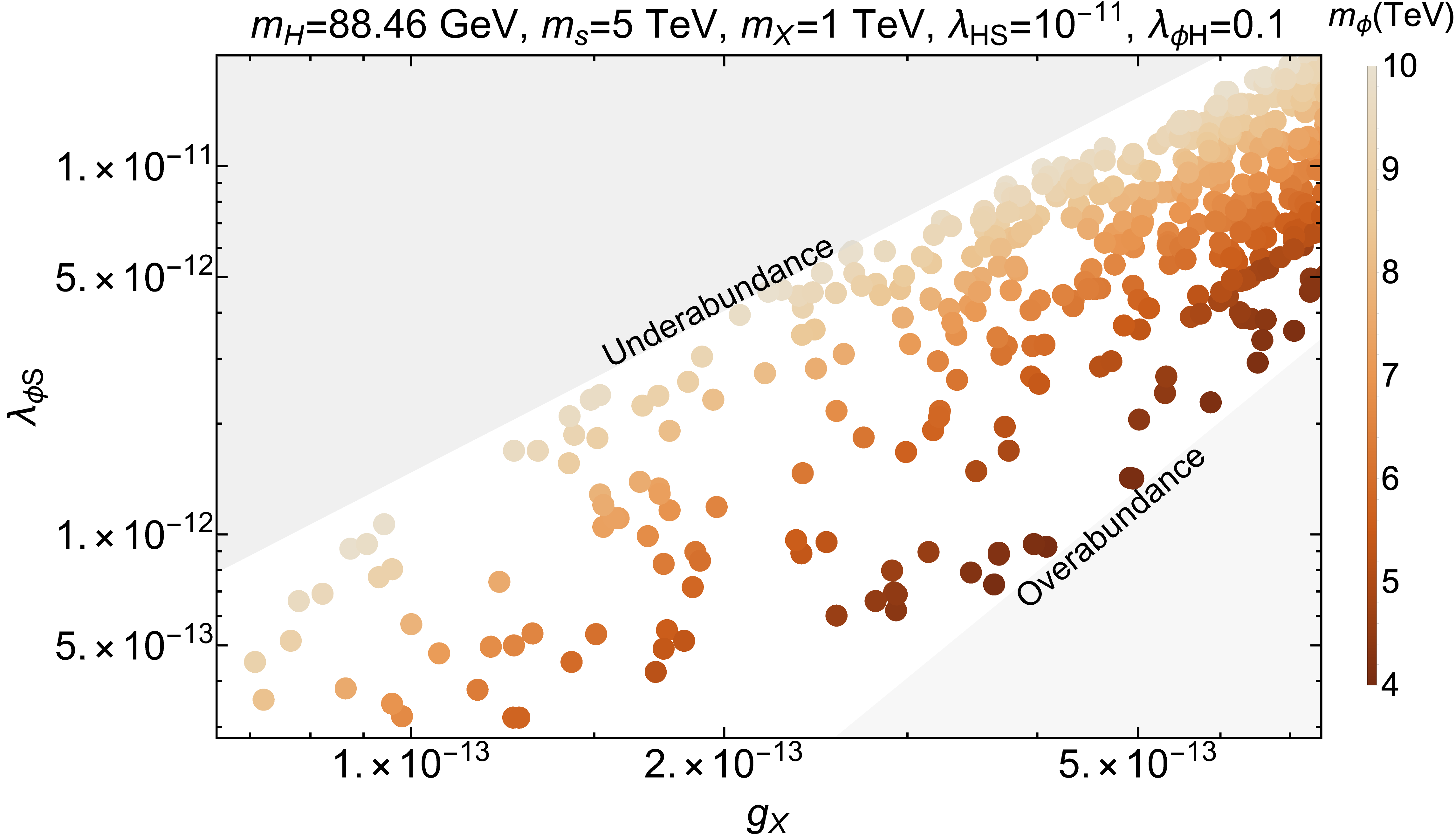}\label{wimp_fimp_bewsb_m2d}}
	\subfloat[]{\includegraphics[width=0.48\linewidth]{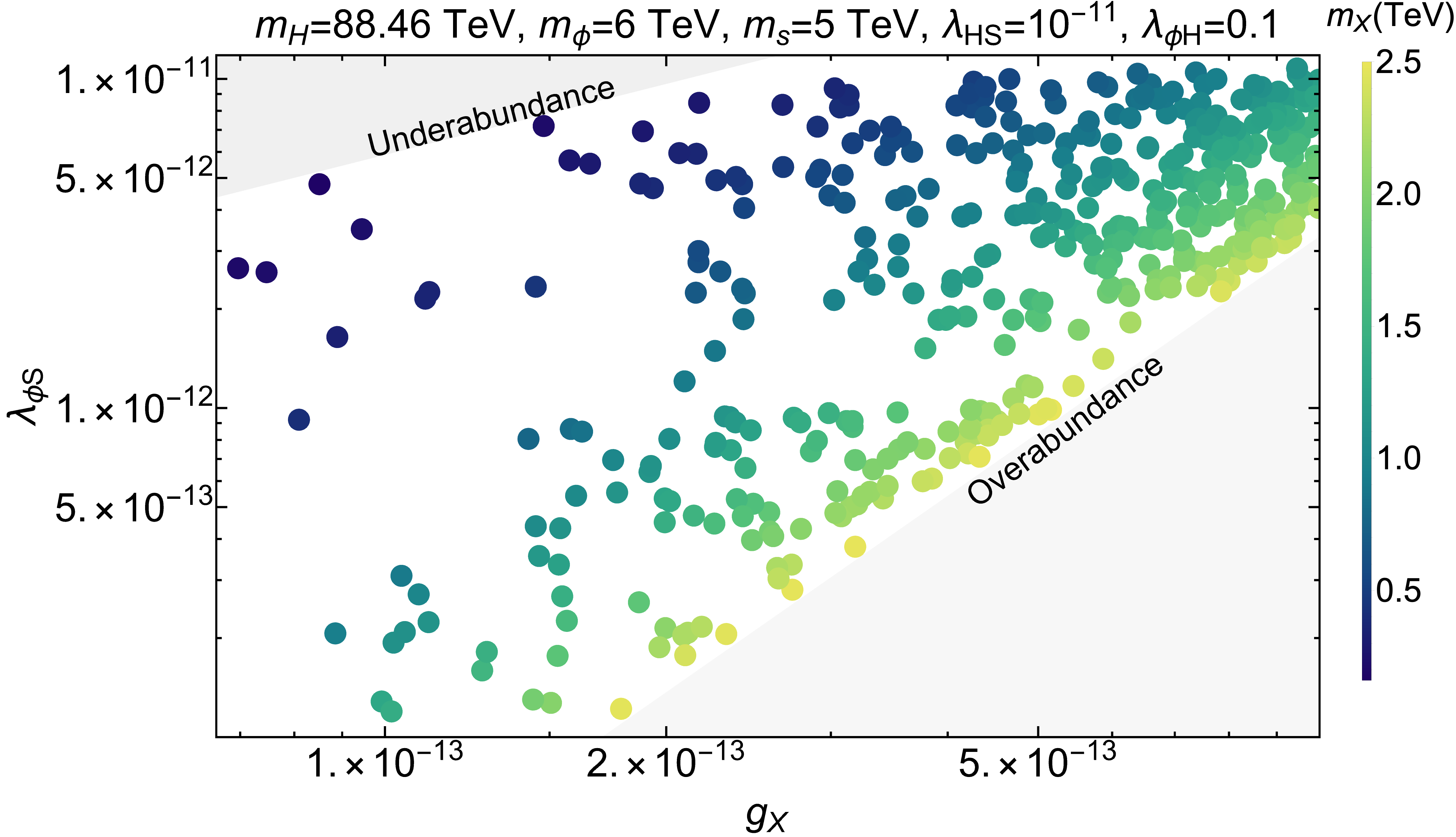}\label{wimp_fimp_bewsb_mxd}}
	\caption{Scan in $g_X-\lambda_{\phi S}$ plane when both WIMP and FIMP  components add to observed relic density, 
	$\Omega_{X}h^2+\Omega_{\phi}h^2=0.1200\pm 0.0012$,
	simultaneously addressing other constraints when both freeze in and freeze out occur bEWSB in kinematic region $m_s\geq2m_X$. In 
	Fig. \ref{wimp_fimp_bewsb_m2d}, we keep FIMP mass fixed at $m_X=1$ TeV and vary WIMP mass $m_\phi$ as shown in the SiennaTones colour bar. 
	In Fig. \ref{wimp_fimp_bewsb_mxd}, we keep $m_\phi=6$ TeV fixed and vary $m_X$ as shown by the BlueGreenYellow colour bar. Other parameters 
	kept fixed are mentioned in the figure heading.Under abundance and over abundance constraints from freeze-in/freeze-out bEWSB are shown by grey shaded 
	regions.}
	\label{wfbewsb}
\end{figure}

\subsection{Putting WIMP and FIMP together}
\label{sec:wimp-fimp-bewsb}

\begin{table}[htb!]
\scalebox{0.68}{
  \renewcommand{\arraystretch}{2.3}
  \begin{tabular}{|p{1.7cm}|c |c |c |c |c |c |c |c |c |c |c |c |c |c| c|}
   \hline
Scenario&\multirow{1}{2cm}{Benchmark\\~~ points}&\multirow{1}{3cm}{$m_{H},m_{\phi},m_s,m_X$\\~~~~~~(TeV)}&$g_X,\lambda_{H S},\lambda_{\phi S},\lambda_{\phi H}$& $\Omega_{\phi}h^2$ & $ \Omega_Xh^2$ & $ \frac{\Omega_{\phi}}{\Omega_T}\%$ & $ \frac{\Omega_{X}}{\Omega_T}\%$&\multirow{1}{1cm}{$\sigma_{\phi_{eff}}^{\text{SI}}$\\$(\text{cm}^2)$} \\ \hline

    \multirow{3}{*}{$m_{s}\geq 2m_X$}&~~~~BP1&0.088, 8.0, 7.0, 1.0 & $ 10^{-13},10^{-11},5.69\times 10^{-13},0.1 $&0.1176&0.0014&98.82&1.18 &$1.44\times 10^{-48}$\\\cline{2-9}
&~~~~BP2&0.088, 7.0, 6.0, 1.5&$10^{-12},10^{-11},1.07\times10^{-11},0.1$&0.0626 &0.0571 &52.30& 47.70 &$ 9.99\times 10^{-49}$\\\cline{2-9}
&~~~~BP3&0.088, 6.0, 5.0, 2.0&$2.33\times 10^{-12},10^{-11},2.6\times10^{-11},0.1$&0.0087&0.1112&7.26&92.74&$1.89\times 10^{-49}$\\\hline

    \multirow{3}{*}{$m_{s}< 2m_X$}&~~~~BP4&0.2, 8.3, 3.0, 2.0  & $10^{-11},10^{-11},10^{-10},0.3$ &0.1062   &0.0138 &88.50  &11.50  &$2.05\times 10^{-45}$\\\cline{2-9} 
&~~~~BP5&0.2, 14.5, 3.0, 2.0,   & $10^{-11},10^{-11},2.74\times 10^{-10},0.3  $ &0.0592  &0.0605 &49.46  &50.54  &$3.09 \times 10^{-45}$\\\cline{2-9}
&~~~~BP6&0.2, 11.0, 3.0, 2.0& $10^{-11},10^{-11},3.32\times 10^{-10},0.2 $ &0.0056  &0.1154 & 4.63 &95.37  &$7.55\times 10^{-46}$\\\hline
\end{tabular}}
\caption{Some sample benchmark points for the WIMP-FIMP model, when both freeze-in of $X$ and freeze-out of $\phi$ occur bEWSB respecting the total relic density, 
direct search, Higgs mass/mixing and other constraints. The benchmark points depict the possibilities when one component dominates over the other as well 
as the case when they have almost equal share for the total DM relic density.}
\label{tab:benchmark-bEWSB}
\end{table}
 So far we discussed the under abundant parameter space for both WIMP ($\phi$) and FIMP ($X$) individually when they freeze-out and freeze-in bEWSB. 
However, the fact that the total DM relic density has to be achieved (Eq.~\ref{eq:totrelic}) from both these components, will correlate these two cases. 
Two such example scans are shown in Fig. \ref{wfbewsb}, where we show the relic density allowed parameter space in $\lambda_{\phi S}-g_X$ 
plane for the kinematic region $m_s\ge 2m_X$, abiding by other relevant constraints. In Fig. \ref{wimp_fimp_bewsb_m2d}, we keep FIMP mass fixed 
at $m_X=1$ TeV and vary WIMP mass $m_\phi$ as shown in the SiennaTones colour bar. In Fig. \ref{wimp_fimp_bewsb_mxd}, we keep $m_\phi=6$ TeV 
fixed and vary $m_X$ as shown by the BlueGreenYellow colour bar. The other parameters kept fixed are mentioned in the figure headings. \par
In Fig. \ref{wimp_fimp_bewsb_m2d}, we see that for a fixed $g_X$, when we make
$\lambda_{\phi S}$ larger, the FIMP ($X$) relic almost remains the same, but WIMP ($\phi$) relic decreases due to larger annihilation cross-section; so 
$m_\phi$ requires to be larger to keep the WIMP relic in the similar ballpark and total relic density constant. This is why we see darker points with smaller 
$m_\phi$ populating smaller $\lambda_{\phi S}$ regions, while for larger $\lambda_{\phi S}$, the WIMP mass( $m_\phi$) requires to be larger with 
brighter points populating such regions. In the same figure, we see that when we enhance $g_X$, 
FIMP relic gets larger, and accordingly WIMP relic needs to be smaller by having larger 
$\lambda_{\phi S}$ as well as small $m_\phi$. Of course, if we keep $\lambda_{\phi S}$ unchanged with larger $g_X$, the total relic density goes beyond the experimental observation 
and provides over abundance, shown by grey shaded region. In a similar way, when $\lambda_{\phi S}$ is larger than a specific value for a given $g_X$, 
then WIMP relic is so tiny that it leads to under abundant total relic, also marked by the grey shaded region.
A complementary behaviour is observed in Fig. \ref{wimp_fimp_bewsb_mxd}. Here, for a fixed $g_X$, with larger $\lambda_{\phi S}$, WIMP relic decreases, 
but with $m_\phi$ kept constant, there is only one way to keep the observed relic density constant, by enhancing FIMP contribution i.e. by decreasing $m_X$. 
This is why we see darker points with small $m_X$ favouring larger $\lambda_{\phi S}$ regions and brighter points with larger $m_X$ populating smaller 
$\lambda_{\phi S}$ regions. Grey shaded over abundance for small $\lambda_{\phi S}$ and under abundance for large $\lambda_{\phi S}$ regions can be 
described in a similar way as in Fig. \ref{wimp_fimp_bewsb_m2d}. A similar correlation can be made when FIMP production occurs dominantly via scattering processes with 
$m_s\le 2m_X$, but the allowed parameter space becomes tinier due to the involvement of $\lambda_{\phi S}$ into both freeze-in and freeze-out processes.
We next furnish some characteristic benchmark points in Table \ref{tab:benchmark-bEWSB}, where the abundance of FIMP ($X$) and WIMP ($\phi$) 
adds to the total observed relic density together with addressing direct search and Higgs mixing constraints ensuring that freeze-in of $X$ and 
freeze-out of $\phi$ both occur bEWSB. The benchmark points BP1 and BP4 depict the possibilities when $\phi$ dominates over $X$, BP3, BP6 show the 
other possible hierarchy when $X$ dominates over $\phi$, while BP2 and BP5 demonstrate the case when they have almost equal share for the relic density.
Before concluding this section, we would like to comment that if both freeze-in/freeze-out has to occur 
bEWSB the masses $m_\phi, m_s$ need to be very heavy, and possibility of any collider production is difficult. The FIMP is 
anyway very feebly coupled to SM. The WIMP can still have a direct search possibility, larger when the FIMP can be produced via 
scattering, smaller when it is produced via $s\to XX$ decay, providing an interesting correlation between the WIMP and FIMP DM components.

\section{Dark Matter phenomenology aEWSB}
\label{sec:aEWSB}

In this section, we address a situation where the freeze-in of $X$ and freeze-out of $\phi$ both occur after EWSB. 
This is equivalent to saying that both the DM components attain saturation at a temperature smaller than $T_{\rm{EW}}$, i.e.:
\bea
T_{U(1)}>T_{\rm{EW}}>T_{\rm{ FI}}; ~~T_{U(1)}>T_{\rm{EW}}>T_{\rm {FO}} \,.
\label{eq:aEWSBtemp}
\eea
The methodology of finding the allowed parameter space for such a situation is similar to the previous case; to solve BEQ for both WIMP and FIMP 
cases individually including all the processes that contribute aEWSB, and choosing model parameters in such a way that we satisfy 
Eq.~\ref{eq:aEWSBtemp}. This is the case usually considered for most of the DM analysis, excepting for checking the validity of 
Eq.~\ref{eq:aEWSBtemp}, which we additionally ensure. However, as the approach remains the same as elaborated in the last section, 
we highlight on the main features that this possibility offers, without going too much of the details. 

\subsection{Physical states and interactions}

The physical particles and interactions aEWSB is obtained when both $S$ and $H$ acquire non-zero VEVs $v_s$ and $v$ respectively. 
In unitary gauge we write, 
\bea
S =\frac{v_s+s}{\sqrt{2}} \to \langle S \rangle =\frac{1}{\sqrt 2} v_s,~H =\begin{pmatrix}0\\\frac{h+v}{\sqrt{2}}\end{pmatrix} \to \langle H \rangle =\frac{1}{\sqrt 2} v, 
~\langle \phi \rangle =0\,.
\eea

Evidently, this induces mixing between the two scalars ($s-h$), the strength of which is dictated by the mixing angle $\theta$. 
Upon diagonalization, two physical scalars $h_1$ and $h_2$ emerge, where $h_1$ is assumed to be the SM Higgs with 
$m_{h_1}\sim$125.1 GeV, whereas $h_2$ may be assumed heavy with mass $m_{h_1}\ll m_{h_2}$. 
The physical and the unphysical fields are related through an orthogonal matrix,

\bea
\begin{pmatrix}
	h_1\\h_2
\end{pmatrix}=\begin{pmatrix}\cos\theta&-\sin\theta\\\sin\theta&\cos\theta\end{pmatrix}\begin{pmatrix}
	h\\s
\end{pmatrix} \,.
\eea

For details, see Appendix \ref{sec:aEWSB-details}, where the minimization of the scalar potential and emergent conditions are specified. 
We may note one point here that $\lambda_{H S}$, which was an external parameter bEWSB, can now be considered as an internal 
parameter and it is dictated by the mixing angle as given below:

\bea
\lambda_{H S}=\frac{\sin2\theta}{2v_sv}\left(m^2_{h_2}-m^2_{h_1}\right).
\label{lambdahs}
\eea

In Table \ref{tab:constraintasb}, we list all the relevant parameters of the model considered for the analysis, classified into 
external (parameters that we choose to vary as input) and internal (or derived) parameters. We further note, that excepting 
for the constraints on dark sector particle masses imposed to make the freeze-in/freeze-out occur bEWSB, we adhere to all the 
other constraints as in section \ref{sec:constraints}.

\begin{table}[htb]
	\centering
	\begin{tabular}{|c |c| } \hline
		External parameters & Internal parameters \\ \hline 
		$m_{h_1}$, $m_{\phi}$, $m_{h_2}$, $m_X$, $g_X$, $\lambda_{\phi}$, $\lambda_{\phi S}$, $\lambda_{\phi H}$, $\sin\theta$
		& $\mu_H$, $\mu_{\phi}$, $\mu_S$, $v_s$,
		$\lambda_H$, $\lambda_S$, $\lambda_{H S}$ \\ \hline
	\end{tabular}
	\caption{The parameters used in the aEWSB analysis.}
	\label{tab:constraintasb}
\end{table}

We further note here that WIMP mass for $\phi$ aEWSB is changed due to the additional contribution proportional to DM-Higgs portal interaction 
$\lambda_{\phi H}$. See Eq. \ref{awimp_mass} in Appendix \ref{sec:aEWSB-details}, where $\mathfrak{m}_{\phi}$ refers to WIMP mass aEWSB, 
although we have used the same notation $m_{\phi}$ in the text to avoid clutter. This essentially does not affect the phenomenology to a great extent. 
\subsection{BEQ in aEWSB scenario}
The BEQ does not change aEWSB, the change is only in the processes of DM production and annihilation, and in the limit of $x$ which goes beyond 
$x_{\rm{EW}}$. First point to note that even aEWSB, the WIMP-FIMP conversion is still small to keep $X$ out-of-equilibrium, so that it is only 
relevant for FIMP production, and the cBEQs reduce to two independent BEQs as before, 
\bea\begin{split}
\frac{dY_X}{dx}&=\Biggl\{\frac{45}{3.32\pi^4}\frac{g_s\, {\rm M_{Pl}}\,m_s^2\Gamma_{s\to XX}}{m_X^4}\frac{x^3 K_1\left[\frac{m_s}{m_X}x\right]}{g^s_*(x)\sqrt{g_*^{\rho}(x)}}\\&+\frac{4\pi^2 \rm{M_{ Pl}}}{45\times 1.66}\frac{g^s_*(x)}{\sqrt{g^{\rho}_*(x)}}\frac{m_X}{x^2}\left(\sum_{i=s,H}\,\langle\sigma \mathpzc{v} \rangle_{ii\to XX}(Y_i^{eq})^2 +\langle \sigma \mathpzc{v} \rangle_{\phi\phi \to XX}Y_{\phi}^2\right)\Biggr\}\Theta[x_{\rm{EW}}-x]\\&+
\Biggl\{\frac{45}{3.32\pi^4}\sum_{A=h_1,h_2}\frac{g_A\, {\rm M_{ Pl}}\, m_A^2\Gamma_{A\to XX}}{m_{ X}^4}\left(\frac{x^3 K_1\left[\frac{m_A}{m_{ X}}x\right]}{g^s_*(x)\sqrt{g_*^{\rho}(x)}} \Theta[x_D^A-x]
\right.  \\
&\left. +\ e^{-\frac{0.602\, {\rm M_{Pl}}\, \Gamma_{A \to XX}}{m_{ X}^2\sqrt{g^{\rho}_*(x)}}(x^2-x_D^{A^2})}\frac{ x^2x_D^A}{\eta(x,x_D^A)}K_1\left[\alpha(x,x_D^A)\frac{m_A}{m_{ X}}\frac{x^2}{x_D^A}\right]e^{\frac{m_A}{m_{ X}}\left(\alpha(x,x_D^A)\frac{x^2}{x_D^A}-x_D^A\right)} \Theta[x-x_D^A]\right)  \\
& + \frac{4\pi^2\, {\rm M_{Pl}}}{45\times 1.66}\frac{g^s_*(x)}{\sqrt{g^{\rho}_*(x)}}\frac{m_{ X}}{x^2}\Biggl(\sum_{i=h_2,\rm{SM}}\langle\sigma \mathpzc{v} \rangle_{ii\to XX}(Y_i^{eq})^2 +\langle \sigma \mathpzc{v} \rangle_{\phi\phi \to XX}Y_{\phi}^2\Biggr)\Biggr\}\Theta[x-x_{\rm{EW}}],
\label{xaewsb}
\end{split}\eea

\bea\begin{split}
	\frac{dY_{\phi}}{dx}
	=&
	-\frac{2\pi^2\, {\rm M_{Pl}}}{45\times 1.66}\frac{g_*^s(x)}{\sqrt{g_*^{\rho}(x)}}\frac{m_{\phi }}{x^2}\Biggl[ \Biggl\{\sum_{i=H,s} \langle\sigma   v \rangle_{\phi\phi\to ii}\Theta[x_{\rm{EW}}-x]\hspace{4.8cm}\\&+\sum_{j=h_2,\rm{SM}} \langle\sigma   v \rangle_{\phi\phi\to jj}\Theta[x-x_{\rm{EW}}]\Biggr\}(Y_{\phi}^2-Y_{\phi}^{eq^2})
	\Biggr]\,.
	\end{split}
	\label{phiaewsb}
\eea
It is worthy mentioning that $x=\frac{m_{ X}}{T},~\frac{m_{\phi}}{T}$ in BEQ of FIMP and WIMP respectively and $\rm{SM}=h_1,W^{\pm},Z,\ell,q$ includes 
all possible massive particles. Note that the $\Theta[x-x_{\rm{EW}}]$ functions present in both Eqs.~\ref{xaewsb} and \ref{phiaewsb} denote processes that take part in DM 
production/annihilation before and after EWSB. While the $\Theta$ function separates the FIMP production into two distinct regions, before 
and after EWSB, this does not include the third possibility of DM production during $\rm EWSB$ at $x\sim x_{\rm EW}$. Such contribution may arise in certain models as 
explored in \cite{Heeba:2018wtf,Redondo:2008ec,Baker:2017zwx}, where it is shown that a significant amount of FIMP production via oscillations from Higgs is possible 
during phase transition when $m_{h}(T)\sim m_{DM}$, if the DM remains out of equilibrium, has a Higgs portal coupling and having mass less than the Higgs mass. 
However, in our case, such contributions do not arise. This is because the scalar ($s$) having a Higgs portal, is not a DM, rather a particle present in the thermal bath 
producing $X$ via in-equilibrium or late decays or scattering. If $s$ remains in thermal bath during EWSB, the oscillations cannot help to enhance the number density of 
$s$ and therefore of $X$; on the other hand, if it needs to be out-of-equilibrium during EWSB, the mass turns out to be pretty heavy $\sim$ 4 TeV, as pointed out in Eq.~\ref{eq:masslimit}, 
way beyond the Higgs mass ($\gg m_h$) for the oscillations to produce additional $s$.

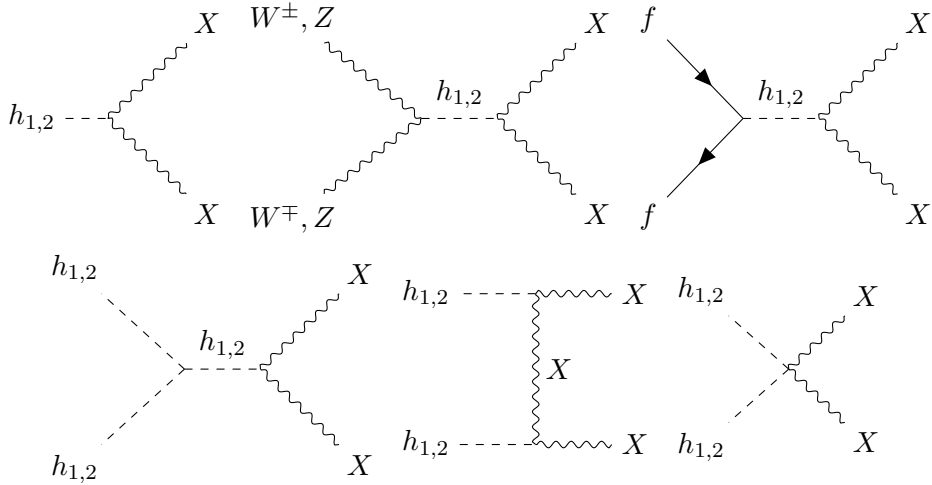
\begin{figure}[htb!]
	\centering
	\begin{tikzpicture}[baseline={(current bounding box.center)}]
	\begin{feynman}
	\vertex (a){\(h_{1,2}\)};
	\vertex [right=1cm of a] (b);
	\vertex[above right=1cm and 1cm of b] (a1){\(X\)}; 
	\vertex[below right=1cm and 1cm of b] (a2){\(X\)}; 
	\diagram* {	(a) -- [scalar] (b) -- [boson] (a1),	(a2) -- [boson]	(b)};
	\end{feynman}
	\end{tikzpicture}
	\begin{tikzpicture}[baseline={(current bounding box.center)}]
	\begin{feynman}
	\vertex (a);
	\vertex[right=1cm of a] (b);
	\vertex[above left=1cm and 1cm of a] (a1){\(W^{\pm},Z\)}; 
	\vertex[below left=1cm and 1cm of a] (a2){\(W^{\mp},Z\)}; 
	\vertex[above right=1cm and 1cm of b] (b1){\(X\)}; 
	\vertex[below right=1cm and 1cm of b] (b2){\(X\)}; 
	\diagram* {
		(a) -- [boson] (a1),(a) -- [boson] (a2),
		(b1) -- [boson]	(b) -- [boson] (b2),(a) --[scalar, edge label={\(h_{1,2}\)}] (b)
	};\end{feynman}
	\end{tikzpicture}
	\begin{tikzpicture}[baseline={(current bounding box.center)}]
	\begin{feynman}
	\vertex (a);
	\vertex[right=1cm of a] (b);
	\vertex[above left=1cm and 1cm of a] (a1){\(f\)}; 
	\vertex[below left=1cm and 1cm of a] (a2){\(f\)}; 
	\vertex[above right=1cm and 1cm of b] (b1){\(X\)}; 
	\vertex[below right=1cm and 1cm of b] (b2){\(X\)}; 
	\diagram* {
		(a1) -- [fermion] (a) -- [fermion] (a2),
		(b1) -- [boson]	(b) -- [boson] (b2),(a) --[scalar, edge label={\(h_{1,2}\)}] (b)
	};\end{feynman}
	\end{tikzpicture}
	\begin{tikzpicture}[baseline={(current bounding box.center)}]
	\begin{feynman}
	\vertex (a);
	\vertex[right=1cm of a] (b);
	\vertex[above left=1cm and 1cm of a] (a1){\(h_{1,2}\)}; 
	\vertex[below left=1cm and 1cm of a] (a2){\(h_{1,2}\)}; 
	\vertex[above right=1cm and 1cm of b] (b1){\(X\)}; 
	\vertex[below right=1cm and 1cm of b] (b2){\(X\)}; 
	\diagram* {
		(a) -- [scalar] (a1),(a) -- [scalar] (a2),
		(b1) -- [boson]	(b) -- [boson] (b2),(a) --[scalar, edge label={\(h_{1,2}\)}] (b)
	};\end{feynman}
	\end{tikzpicture}
	\begin{tikzpicture}[baseline={(current bounding box.center)}]
	\begin{feynman}
	\vertex (a);
	\vertex[below=2cm of a] (b);
	\vertex[left=1cm and 1cm of a] (a1){\(h_{1,2}\)}; 
	\vertex[right=1cm and 1cm of a] (a2){\(X\)}; 
	\vertex[left=1cm and 1cm of b] (b1){\(h_{1,2}\)}; 
	\vertex[right=1cm and 1cm of b] (b2){\(X\)}; 
	\diagram* {
		(a) -- [scalar] (a1),(a2) -- [boson] (a),
		(b1) -- [scalar] (b) -- [boson] (b2), (a) --[boson, edge label={\(X\)}] (b)
	};\end{feynman}
	\end{tikzpicture}
	\begin{tikzpicture}[baseline={(current bounding box.center)}]
	\begin{feynman}
	\vertex (a);
	\vertex[above left=1cm of a] (a1){\(h_{1,2}\)}; 
	\vertex[below left=1cm of a] (a2){\(h_{1,2}\)}; 
	\vertex[above right=1cm of a] (b1){\(X\)}; 
	\vertex[below right=1cm of a] (b2){\(X\)}; 
	\diagram* {
		(a) -- [scalar] (a1),(a) -- [scalar] (a2),
		(b1) -- [boson]	(a) -- [boson] (b2)
	};\end{feynman}
	\end{tikzpicture}
	\caption{Feynman diagrams showing non-thermal production channels of X aEWSB}
	\label{feyn_fimp_aewsb}
\end{figure}

Let us discuss a few salient features of freeze-in/freeze out aEWSB here. For example, in scattering dominated FIMP production regime, with $m_s< 2 m_X$, all scattering processes $(ss\to XX; \phi\phi\to XX {\rm~ and ~} HH\to XX)$ play important role in FIMP ($X$) production bEWSB ($x<x_{\rm{EW}})$; but aEWSB ($x>x_{\rm{EW}}$) new scattering channels open up, as shown in 
Fig. \ref{feyn_fimp_aewsb}. Again, one needs to remember, that the Goldstone degrees of freedom for $H$ is now converted to massive gauge 
bosons $W^\pm,Z$, but contributions from massive fermions add to the production. Now, consider $m_s\geq 2m_X$ but $m_s\lesssim 4$ TeV, 
then FIMP is dominantly produced from $s$ decay bEWSB, but the decoupling of $s$ occurs aEWSB and $s-h$ mixing occurs to produce 
$h_1,h_2$. Then dominating FIMP production aEWSB comes from the decay (and late decay) of $h_{1,2}$ as shown in Fig. \ref{feyn_fimp_aewsb}. 
Both the processes bEWSB and aEWSB contribute to the freeze-in yield aEWSB as indicated in Eq. \ref{xaewsb}. For WIMP ($\phi$) however, 
when it freezes out aEWSB, annihilation channels bEWSB do not matter much as they only maintain the WIMP in thermal bath, the freeze-out 
(or decoupling) of WIMP as well as the consequent relic density ($\Omega_\phi h^2$) are mainly governed by the processes aEWSB. The 
corresponding Feynman graphs for $\phi$ freeze-out aEWSB is shown in Fig.~\ref{feyn_wimp_aewsb}. WIMP-FIMP conversion aEWSB is shown in Fig.~\ref{feyn_cons_aewsb}.

\begin{figure}[htb!]
\centering
	\begin{tikzpicture}[baseline={(current bounding box.center)}]
	\begin{feynman}
	\vertex (a);
	\vertex[right=1cm of a] (b);
	\vertex[above left=1cm and 1cm of a] (a1){\(\phi\)}; 
	\vertex[below left=1cm and 1cm of a] (a2){\(\phi\)}; 
	\vertex[above right=1cm and 1cm of b] (b1){\(h_{1,2}\)}; 
	\vertex[below right=1cm and 1cm of b] (b2){\(h_{1,2}\)}; 
	\diagram* {
		(a) -- [scalar] (a1),(a) -- [scalar] (a2),
		(b1) -- [scalar]	(b) -- [scalar] (b2),(a) --[scalar, edge label={\(h_{1,2}\)}] (b)
	};\end{feynman}
	\end{tikzpicture}
	\begin{tikzpicture}[baseline={(current bounding box.center)}]
	\begin{feynman}
	\vertex (a);
	\vertex[below=2cm of a] (b);
	\vertex[left=1cm and 1cm of a] (a1){\(\phi\)}; 
	\vertex[right=1cm and 1cm of a] (a2){\(h_{1,2}\)}; 
	\vertex[left=1cm and 1cm of b] (b1){\(\phi\)}; 
	\vertex[right=1cm and 1cm of b] (b2){\(h_{1,2}\)}; 
	\diagram* {
		(a) -- [scalar] (a1),(a2) -- [scalar] (a),
		(b1) -- [scalar] (b) -- [scalar] (b2), (a) --[scalar, edge label={\(\phi\)}] (b)
	};\end{feynman}
	\end{tikzpicture}
	\begin{tikzpicture}[baseline={(current bounding box.center)}]
	\begin{feynman}
	\vertex (a);
	\vertex[above left=1cm of a] (a1){\(\phi\)}; 
	\vertex[below left=1cm of a] (a2){\(\phi\)}; 
	\vertex[above right=1cm of a] (b1){\(h_{1,2}\)}; 
	\vertex[below right=1cm of a] (b2){\(h_{1,2}\)}; 
	\diagram* {
		(a) -- [scalar] (a1),(a) -- [scalar] (a2),
		(b1) -- [scalar]	(a) -- [scalar] (b2)
	};\end{feynman}
	\end{tikzpicture}
	\begin{center}
		\begin{tikzpicture}[baseline={(current bounding box.center)}]
		\begin{feynman}
		\vertex (a);
		\vertex[right=1cm of a] (b);
		\vertex[above left=1cm and 1cm of a] (a1){\(\phi\)}; 
		\vertex[below left=1cm and 1cm of a] (a2){\(\phi\)}; 
		\vertex[above right=1cm and 1cm of b] (b1){\(f\)}; 
		\vertex[below right=1cm and 1cm of b] (b2){\(f\)}; 
		\diagram* {
			(a) -- [scalar] (a1),(a) -- [scalar] (a2),
			(b2) -- [fermion] (b) -- [fermion] (b1),(a) --[scalar, edge label={\(h_{1,2}\)}] (b)
		};\end{feynman}
		\end{tikzpicture}
		\begin{tikzpicture}[baseline={(current bounding box.center)}]
		\begin{feynman}
		\vertex (a);
		\vertex[right=1cm of a] (b);
		\vertex[above left=1cm and 1cm of a] (a1){\(\phi\)}; 
		\vertex[below left=1cm and 1cm of a] (a2){\(\phi\)}; 
		\vertex[above right=1cm and 1cm of b] (b1){\(W^{\pm},Z\)}; 
		\vertex[below right=1cm and 1cm of b] (b2){\(W^{\mp},Z\)}; 
		\diagram* {
			(a1) -- [scalar] (a) -- [scalar] (a2),
			(b1) -- [boson]	(b) -- [boson] (b2),(a) --[scalar, edge label={\(h_{1,2}\)}] (b)
		};\end{feynman}
		\end{tikzpicture}
	\end{center}
	\caption{Feynman diagrams showing annihilation channels of $\phi$ aEWSB}\label{feyn_wimp_aewsb}
	\end{figure}
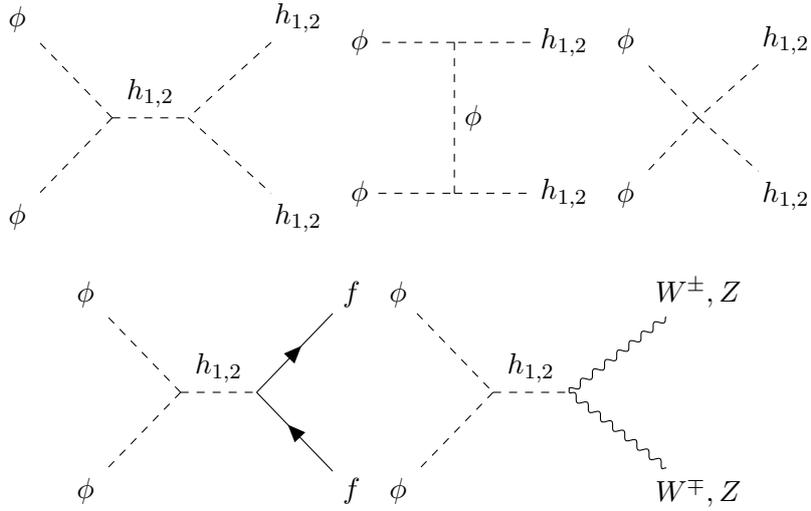
	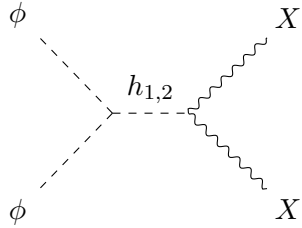
\begin{figure}[htb!]
	\begin{center}
		\begin{tikzpicture}[baseline={(current bounding box.center)}]
		\begin{feynman}
		\vertex (a);
		\vertex[right=1cm of a] (b);
		\vertex[above left=1cm and 1cm of a] (a1){\(\phi\)}; 
		\vertex[below left=1cm and 1cm of a] (a2){\(\phi\)}; 
		\vertex[above right=1cm and 1cm of b] (b1){\(X\)}; 
		\vertex[below right=1cm and 1cm of b] (b2){\(X\)}; 
		\diagram* {
			(a) -- [scalar] (a1),(a) -- [scalar] (a2),
			(b1) -- [boson]	(b) -- [boson] (b2),(a) --[scalar, edge label={\(h_{1,2}\)}] (b)
		};\end{feynman}
		\end{tikzpicture}\caption{WIMP-FIMP conversion channel aEWSB.}
		\label{feyn_cons_aewsb}
	\end{center}
\end{figure}
\subsection{Freeze-in of $X$}
\label{sec:finaEWSB}
\begin{figure}[htb!]  
	\centering
	\subfloat[]{\includegraphics[width=0.48\linewidth]{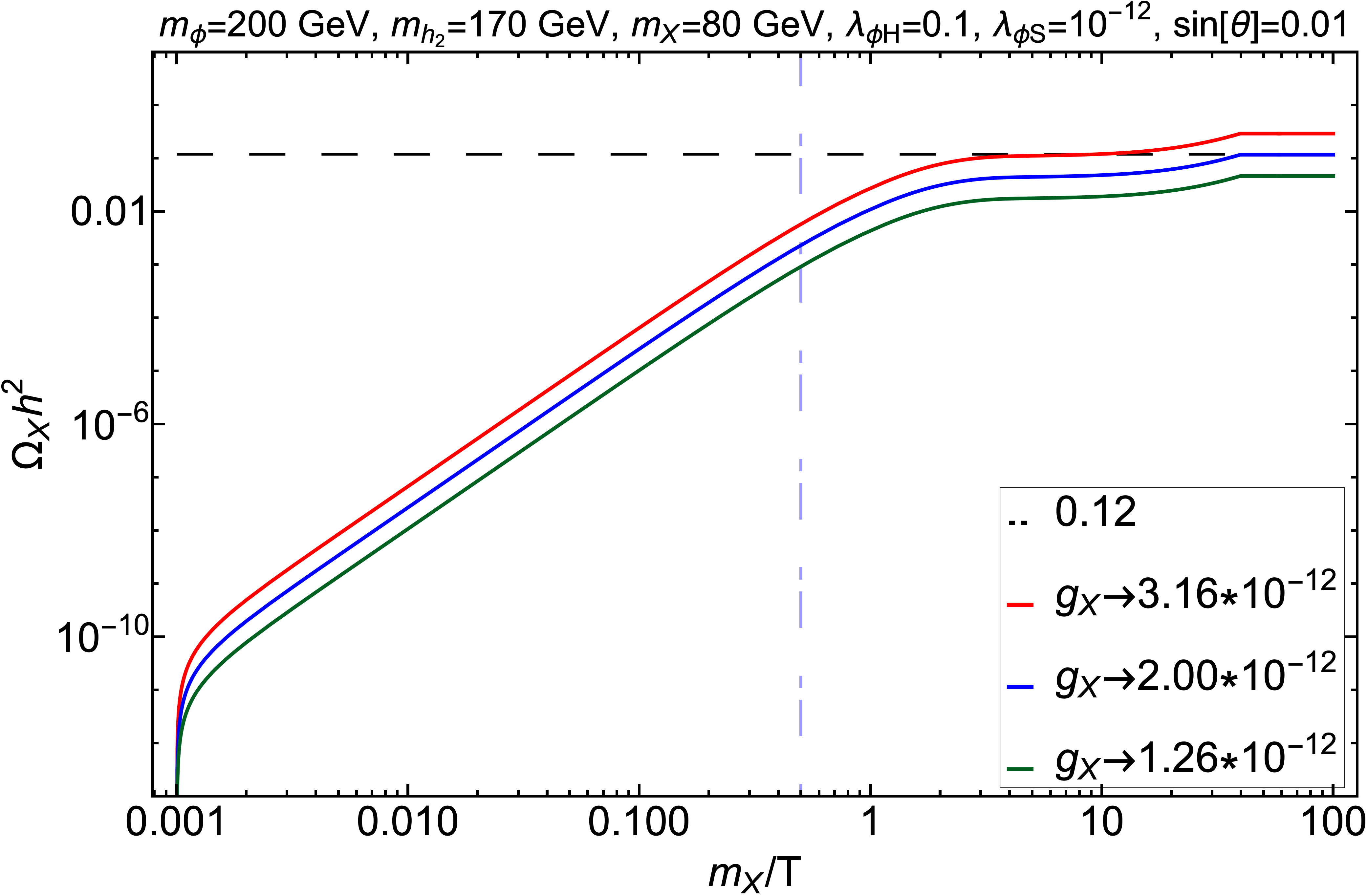}\label{fimp_aewsb_gx}}~~
	\subfloat[]{\includegraphics[width=0.48\linewidth]{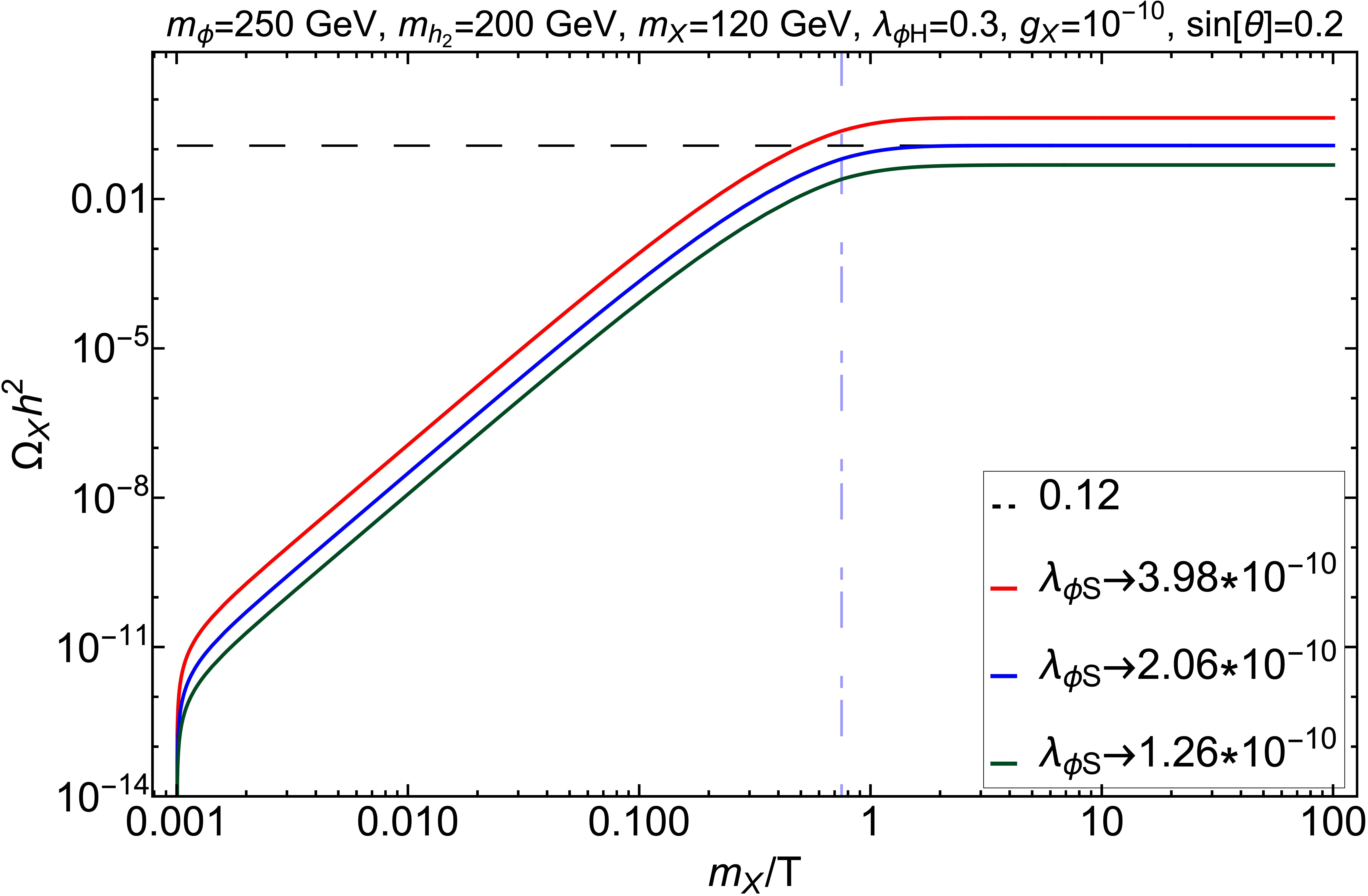}\label{relic_aewsb_l2s}}
		\caption{Variation of $X$ relic density as a function of $x=m_X/T$ to demonstrate freeze-in aEWSB; In left panel (Fig. \ref{fimp_aewsb_decay_mx_gx}) 
		we show decay dominant case where freeze-in production aEWSB occurs via $h_2 \to XX$ for three discrete $g_X$ values. In the right panel 
		Fig. \ref{fimp_aewsb_scattering_m2_l2s}, decay is kinematically forbidden and scattering processes dominate freeze-in production. Vertical dot-dashed line 
		indicates EWSB and black dashed line shows the central value of observed relic density. Parameters kept fixed for the plot are mentioned in the figure heading.}
	\label{fimp_relic_aewsb}
\end{figure}

As evident from Eq.~\ref{xaewsb}, $X$ freeze-in has an important contribution accumulated from processes bEWSB, while aEWSB 
($x>x_{\rm{EWSB}}$), $X$ production occurs mainly via $h_2$ decay ($h_1$ decay to $XX$ is assumed kinematically forbidden by considering 
$m_{h_1}<2m_X$), and scattering processes as shown in Fig. \ref{feyn_fimp_aewsb}, in absence of decay. Freeze-in aEWSB is ensured by checking 
$Y_{x_{\rm EW}}<Y_{x>x_{\rm EW}}$.
Since the essential phenomenology of aEWSB freeze-in is not entirely different from bEWSB, we show a few representative plots to demonstrate the viable parameter 
space in this region. We show first freeze-in production of $X$ in Fig.~\ref{fimp_relic_aewsb} in terms of $\Omega_X h^2$ as a function of $m_X/T$. In 
Fig. \ref{fimp_aewsb_gx}, we show the case where $h_2\to X X$ is the dominant DM production channel, as the decay is kinematically allowed. 
Here we show the freeze-in pattern for three different $g_X$ values (mentioned in the figure inset) by red, blue and green 
coloured lines respectively. FIMP relic density increases with $g_X$, which is already discussed and correct relic density is obtained for $g_X=2\times10^{-12}$. 
The  horizontal dashed line depicts the central value of the observed DM relic. The vertical dot-dashed line refers to EWSB and we ensure the freeze-in to happen 
aEWSB. We again see that in decay dominated production, late decay adds significantly to the FIMP yield $Y_X$. In Fig. \ref{relic_aewsb_l2s}, we show the same 
$\Omega_X h^2$ vs. $m_X/T$ variation, but for scattering dominated production, absent the kinematically forbidden $h_2 \to XX$ decay mode for different 
$\lambda_{\phi S}$ represented by the red, blue and green lines. Expectedly, FIMP relic is enhanced with larger $\lambda_{\phi S}$, similar to the bEWSB case. 
Again, freeze-in abundance to settle aEWSB is explicitly seen when compared to vertical dot-dashed lines depicting EWSB ($x_{\rm{EW}}$) .

\begin{figure}[htb!]  
	\centering
	\subfloat[]{\includegraphics[width=0.48\linewidth]{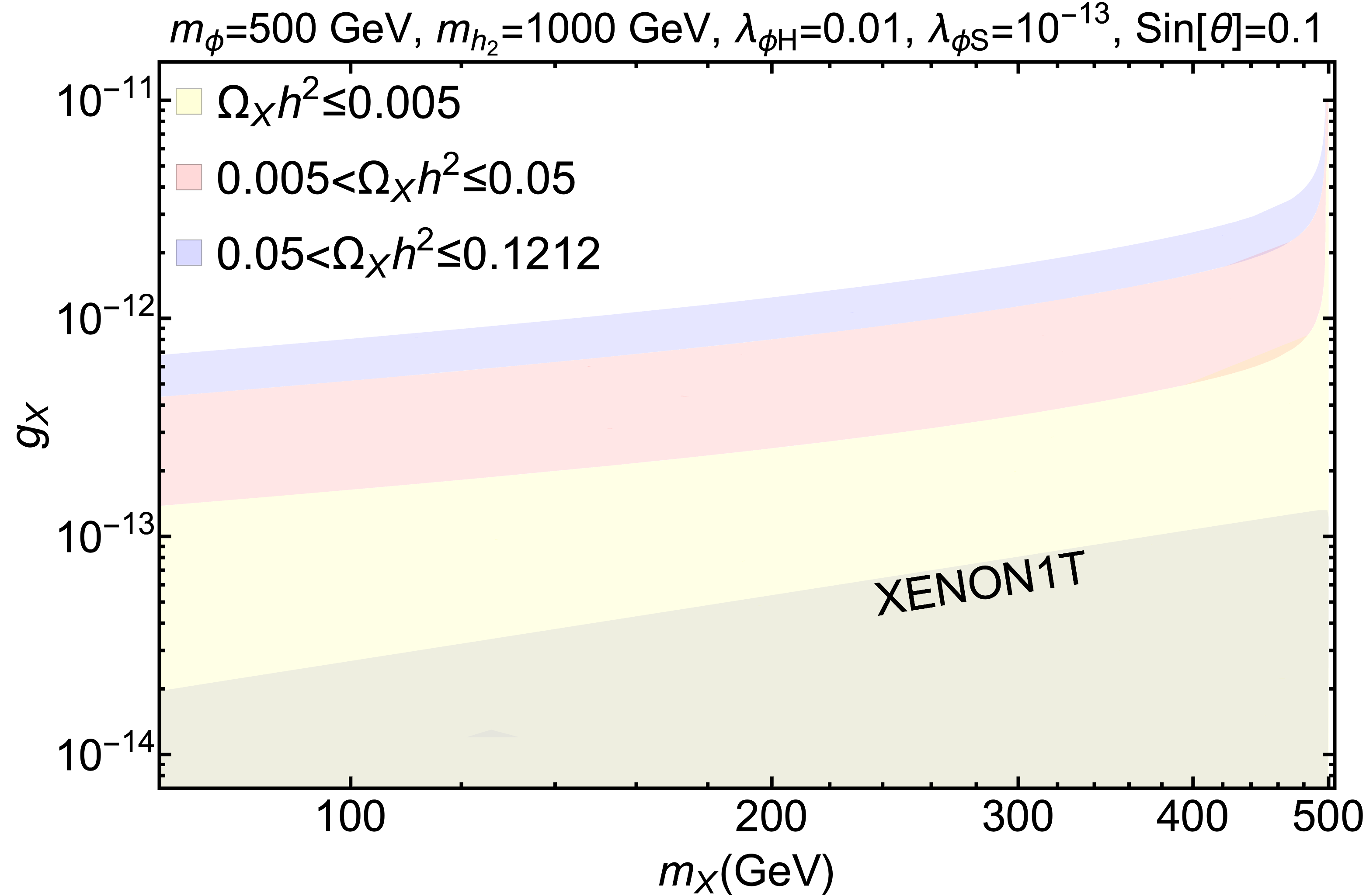}\label{fimp_aewsb_decay_mx_gx}}~~
    \subfloat[]{\includegraphics[width=0.48\linewidth]{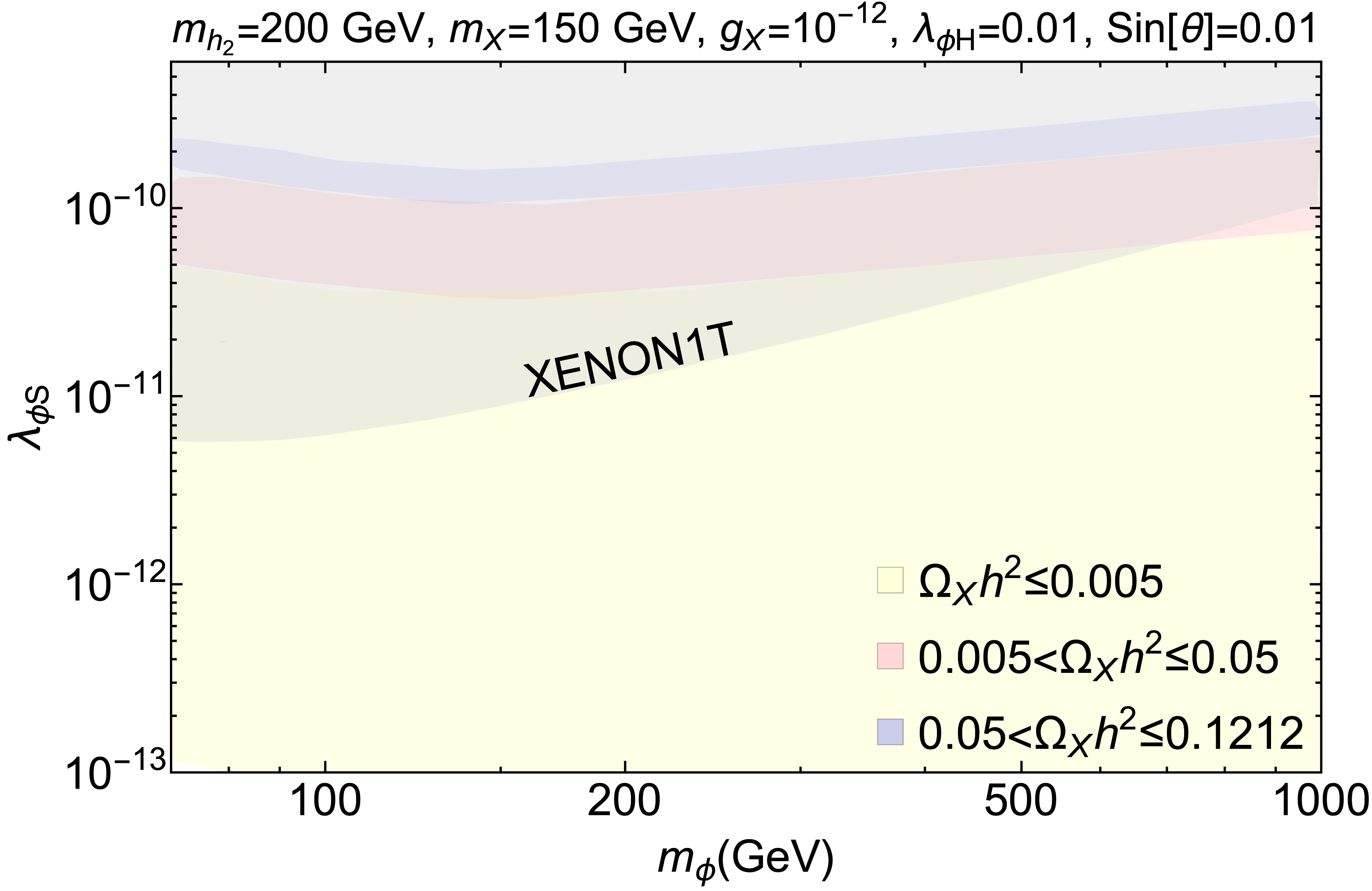}\label{fimp_aewsb_scattering_m2_l2s}}\\
    \subfloat[]{\includegraphics[width=0.48\linewidth]{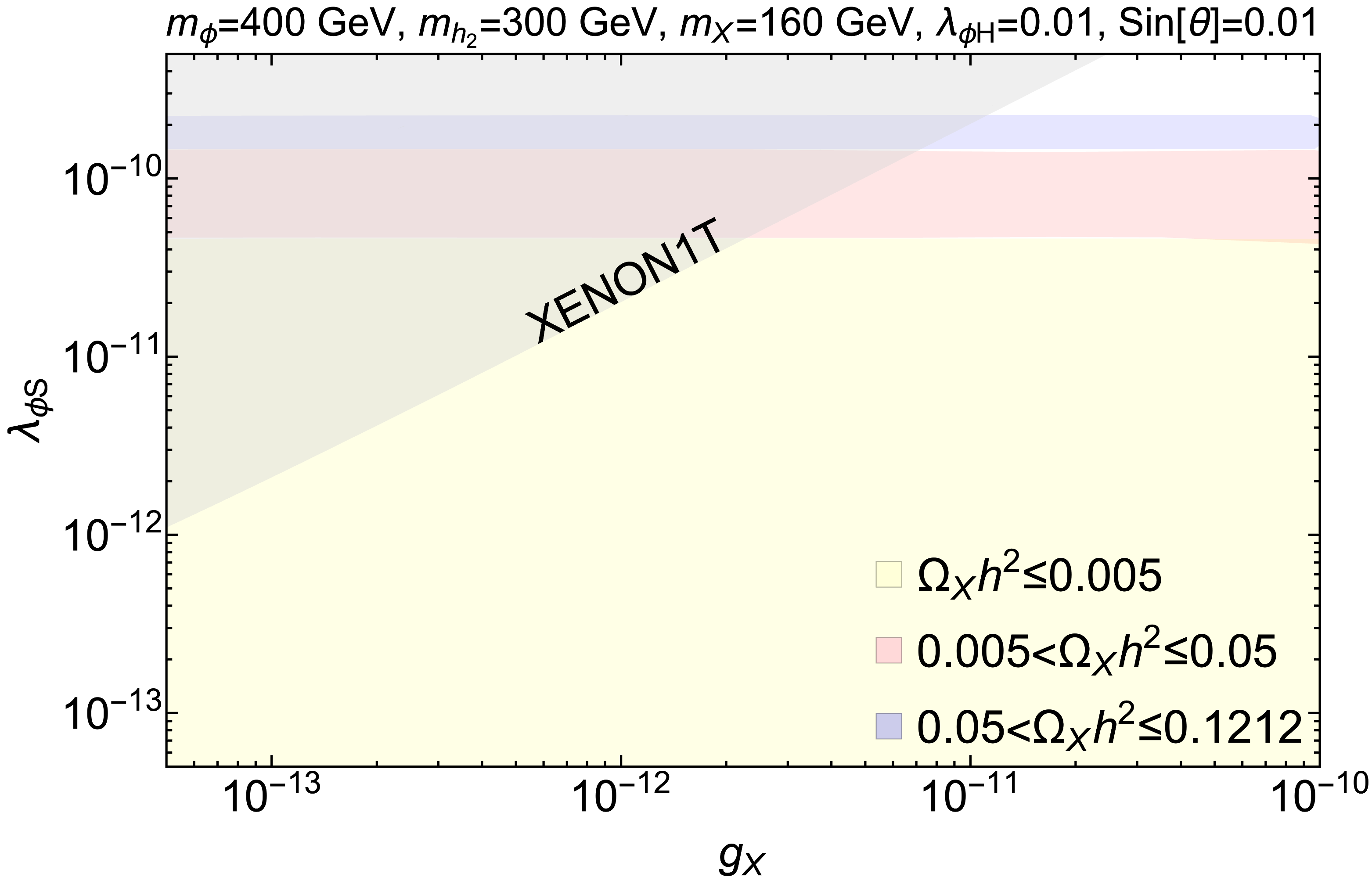}\label{relic_aewsb_gx_l2s}}
	\caption{Under abundant ($\Omega_X h^2\le \Omega_T h^2$) parameter space for DM freeze-in aEWSB. Fig. \ref{fimp_aewsb_decay_mx_gx} shows a 
	correlation in $g_X$ vs. $m_X$ plane with direct search constraint restricting the smaller $g_X$ values shown by the grey shaded region. 
	In Fig.\ref{fimp_aewsb_scattering_m2_l2s}, the under abundant region in $\lambda_{\phi S}$ vs. $m_\phi$ plane is shown, with grey shaded 
	region ruled out by the XENON1T direct search bound. The relic density of FIMP DM is indicated by color code in the figure inset.
	In Fig.\ref{relic_aewsb_gx_l2s}, we show the viable parameter space in $\lambda_{\phi S}$ vs. $g_X$ plane for the scattering dominated 
	production of FIMP. Direct search limits weaken for smaller $\lambda_{\phi S}$ and larger $g_X$. Parameters kept fixed are mentioned in figure headings.}
	\label{fimp_scan_aewsb}
\end{figure}

As the dependence of freeze-in relic density on the parameters remain almost the same aEWSB, it is needless to repeat all the features 
here once again. Nevertheless, in order to demonstrate the viable parameter space complying with aEWSB freeze-in, we show three plots in 
Fig. \ref{fimp_scan_aewsb}. The top left plot, i.e., Fig. \ref{fimp_aewsb_decay_mx_gx} shows a correlation in $m_X$ vs. $g_X$ plane and corresponds 
to the decay dominant FIMP production. We find that excepting for very small $g_X$ regions constrained by direct search of $\phi$ (with 
the direct search cross-section being proportional to $v_s\sim 1/g_X$), the rest of the parameter space shows under relic abundance indicated by color codes 
as mentioned in the figure inset. This is an important contrast to the bEWSB case, where the parameter space for the decay dominant FIMP production is completely 
unconstrained (see Figs.\ref{frzin_plot_2}). Also, one can conclude from this plot that after EWSB, large part of parameter space can be saved from DD 
limits if the FIMP production is decay dominated. The right panel plot, i.e., Fig. \ref{fimp_aewsb_scattering_m2_l2s} shows a correlation in 
$m_\phi-\lambda_{\phi S}$ plane, which corresponds to scattering dominant FIMP production absent $h_2 \to XX$ decay. Here we see that 
a large region of the parameter space is ruled out by DD data particularly for larger $\lambda_{\phi S}$. Since the DD cross-section has very 
strong dependence on both $\lambda_{\phi S}$ and $g_X$, in Fig.\ref{relic_aewsb_gx_l2s} at the bottom panel, we show a correlation in the 
$\lambda_{\phi S}$ vs. $g_X$ plane. Here, FIMP relic density, although increases with $\lambda_{\phi S}$, remains almost constant with the 
variation of $g_X$, as the scattering dominant production cross-section has no explicit dependence on $g_X$. On the other hand, the direct 
detection cross-section, having explicit dependence on $\lambda_{\phi S}$ and $v_s\sim 1/g_X$, shows weaker bounds for small $\lambda_{\phi S}$ 
and large $g_X$. Parameters kept fixed for the scans, are mentioned in the respective figure headings and ensure all the other constraints.

\subsection{Freeze-out of $\phi$ }
\begin{figure}[htb!]  
	\centering
    \subfloat[]{\includegraphics[width=0.48\linewidth]{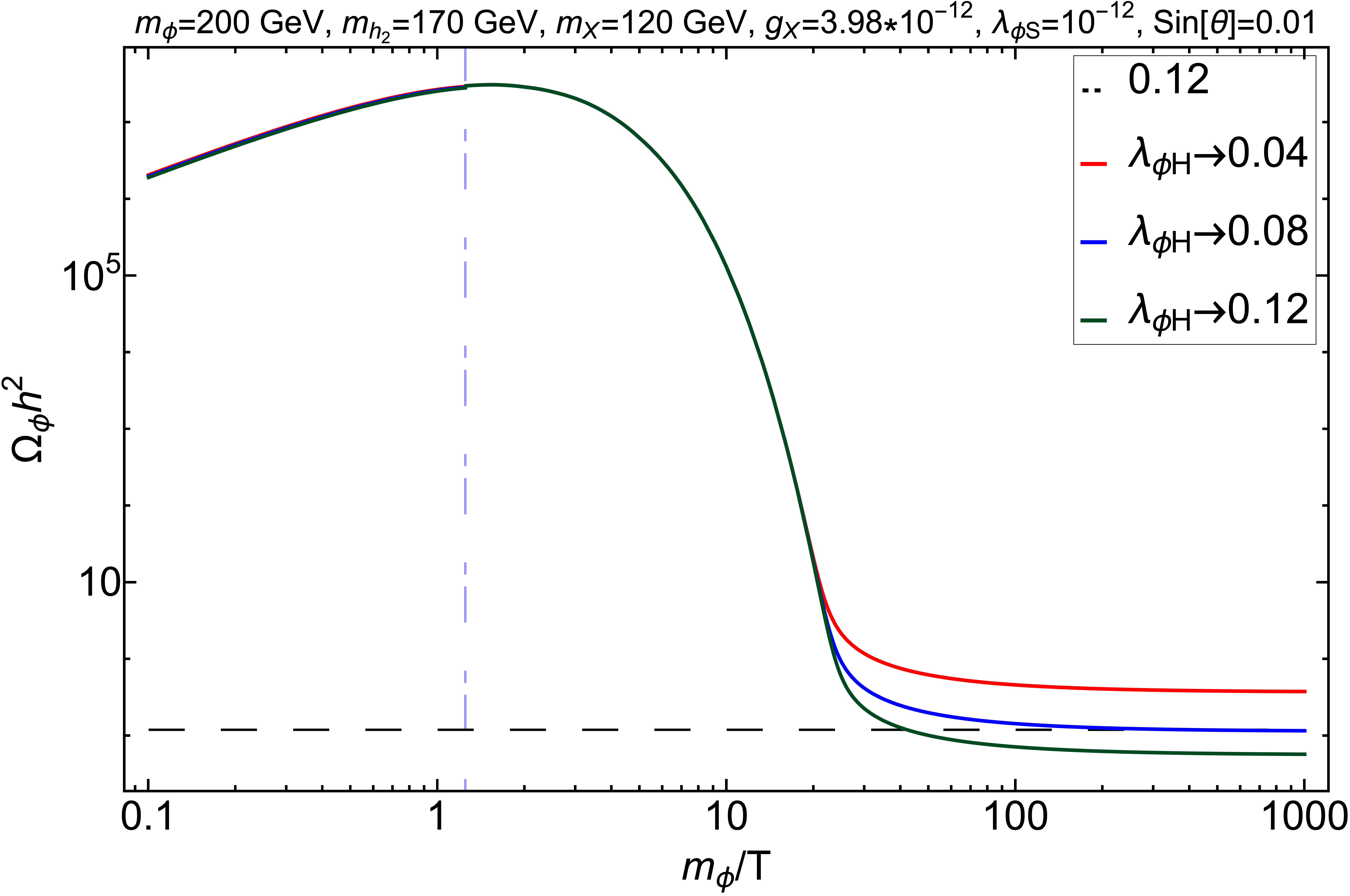}\label{wimp_aewsb_l12}}~~	
	\subfloat[]{\includegraphics[width=0.5\linewidth]{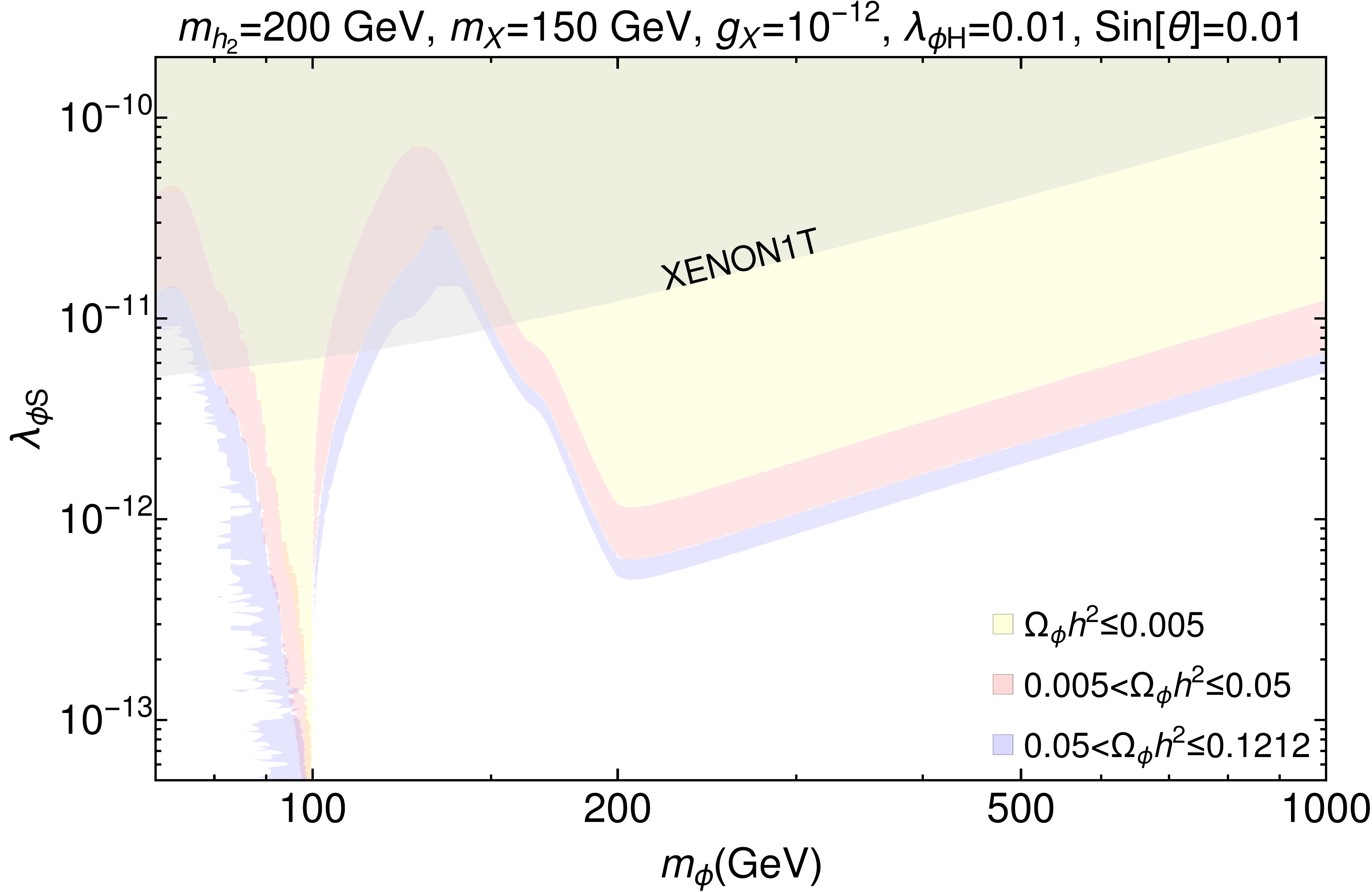}\label{wimp_aewsb_m2_l2s}}\\
	\subfloat[]{\includegraphics[width=0.5\linewidth]{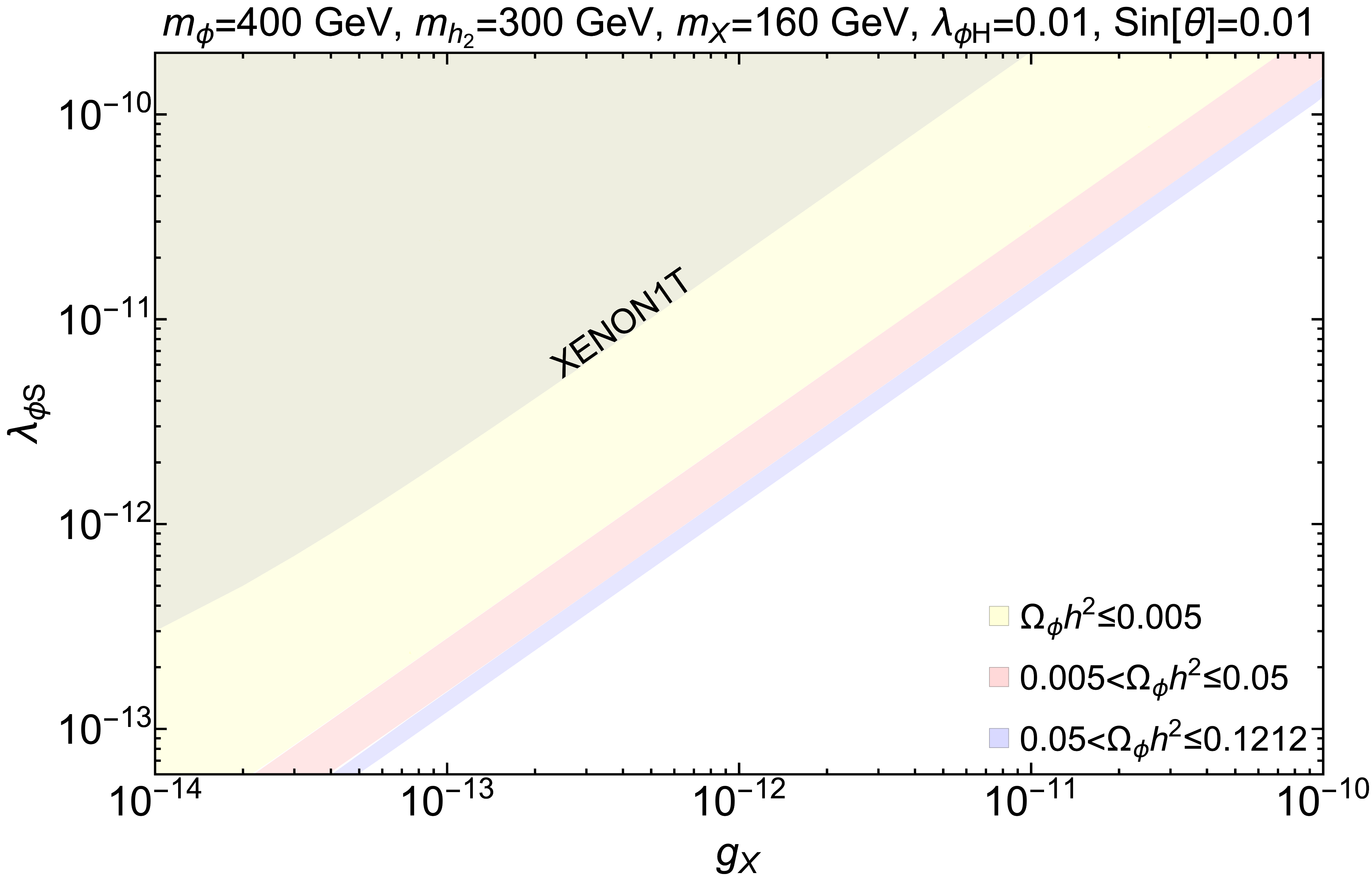}\label{wimp_aewsb_gx_l2s}}
	\caption{Figure \ref{wimp_aewsb_l12} shows freeze-out pattern for WIMP ($\phi$) relic density as a function of $m_{\phi}/T$ 
	for some fixed values of $\lambda_{\phi H}$, which also ensures freeze-out aEWSB. The vertical dot-dashed line corresponding to EWSB ($x_{\rm{EW}}$). 
	Fig. \ref{wimp_aewsb_m2_l2s} denotes a correlation plot for the under-abundant $\phi$ in $m_{\phi}-\lambda_{\phi S}$ plane. The constraints from spin independent 
	DD cross-section obtained from XENON1T data are shown in grey shades. The parameters kept fixed for are mentioned in figure heading. 
	The colour shades in light blue, light red and light yellow show the ranges of under abundance as mentioned in figure inset.
	In Fig. \ref{wimp_aewsb_gx_l2s}, we show the under abundant $\phi$ in $\lambda_{\phi S}$ vs. $g_X$ plane.}
\label{wimp_scan_aewsb}
\end{figure}

$\phi$ freezes out aEWSB through the annihilation channels as shown in Fig. \ref{feyn_wimp_aewsb}. New annihilation channels open up through for {\it e.g :} 
$h\phi\phi$ vertex aEWSB. The trilinear couplings of $\phi$ with Higgs become relevant in the DM phenomenology, in contrast to only quartic 
DM-Higgs interaction bEWSB for $\phi$ freeze-out. We demonstrate aEWSB freeze-out with three representative plots in Fig.~\ref{wimp_scan_aewsb}.
Fig. \ref{wimp_aewsb_l12} shows the evolution of WIMP abundance ($\Omega_\phi h^2$) with $m_{\phi}/T$ for three discrete values (mentioned in the figure inset) represented by red, blue and green coloured lines respectively. If we increase $\lambda_{\phi H}$, this enhances 
the annihilation cross-section and in turn decrease the relic abundance, which we show in the figure. The blue one with $\lambda_{\phi H}=0.08$ satisfies the 
correct relic. The vertical dot-dashed line ensures that the $\phi$ freeze out occurs aEWSB and the horizontal dashed line represents the central value of the observed DM relic.
Also note in Fig. \ref{wimp_aewsb_l12}, a small bump appearing in the equilibrium distribution due to the change of WIMP mass at EWSB boundary as given by 
Eq. \ref{awimp_mass} in Appendix \ref{sec:aEWSB-details}.\par

In Fig. \ref{wimp_aewsb_m2_l2s}, we show the (under-) relic and direct search allowed parameter space in $m_{\phi}-\lambda_{\phi S}$ plane. 
The three shades light yellow, light red and light blue represent different ranges for under-abundance, as mentioned in the legend. 
The grey shaded region is excluded by present spin independent direct search (XENON1T) bound. With smaller $\lambda_{\phi S}$, relic density expectedly 
 increases. Also, we see the maximum annihilation around the $m_{h_2}$ resonance at 100 GeV, as we fixed $m_{h_2}$ at 200 GeV for this scan. 
 Owing to the fact that $m_{h_2}$ is unknown and loosely constrained, a large amount of relic density allowed parameter space can be brought under the 
 direct search bound if one focuses on the $m_{h_2}$ resonance. We also see that a large parameter space opens up whenever the annihilation channel to 
 $h_2$ pair opens up with $m_\phi>m_{h_2}$. To demonstrate the effect of the two relevant couplings $g_X$ and $\lambda_{\phi S}$ on WIMP relic density 
 and DD, we show a correlation plot in the bottom panel Fig. \ref{wimp_aewsb_gx_l2s}. DD limits show the same trend as Fig. \ref{relic_aewsb_gx_l2s}, 
 whereas the WIMP annihilation cross-section, also being proportional to $\lambda_{\phi S}\,v_s$, shows under abundance for large $\lambda_{\phi S}$ and 
 small $g_X$. Importantly we see that for WIMP, under abundant regions face more exclusion from direct search limit while it is the other way round for 
 FIMP, which obviously stems from the reverse dependence on the cross-section to the DM yield for these two cases.

\begin{figure}[htb!]  
\centering
\subfloat[]{\includegraphics[width=0.5\linewidth]{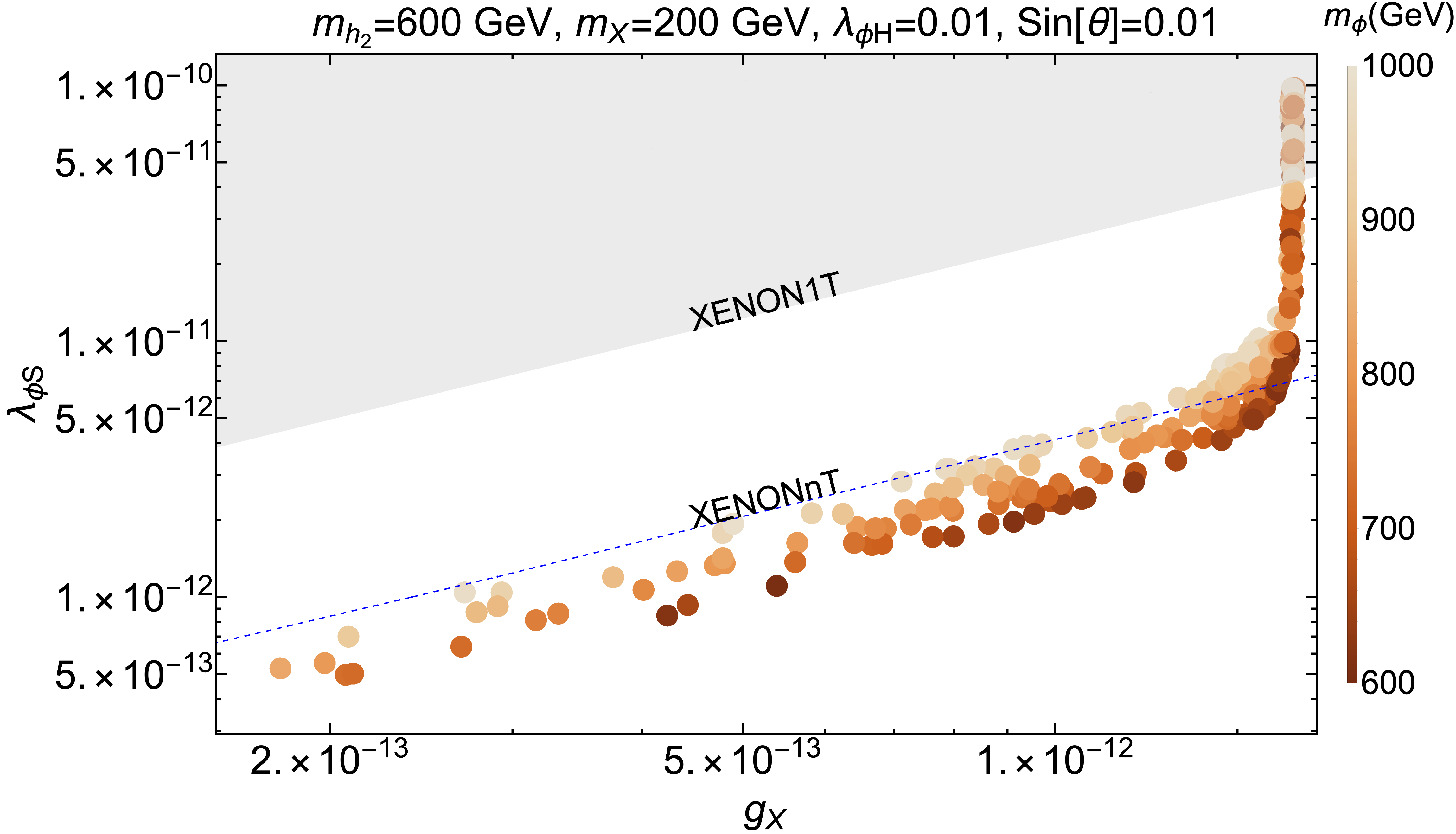}\label{wfaewsb_m2}}	
\subfloat[]{\includegraphics[width=0.5\linewidth]{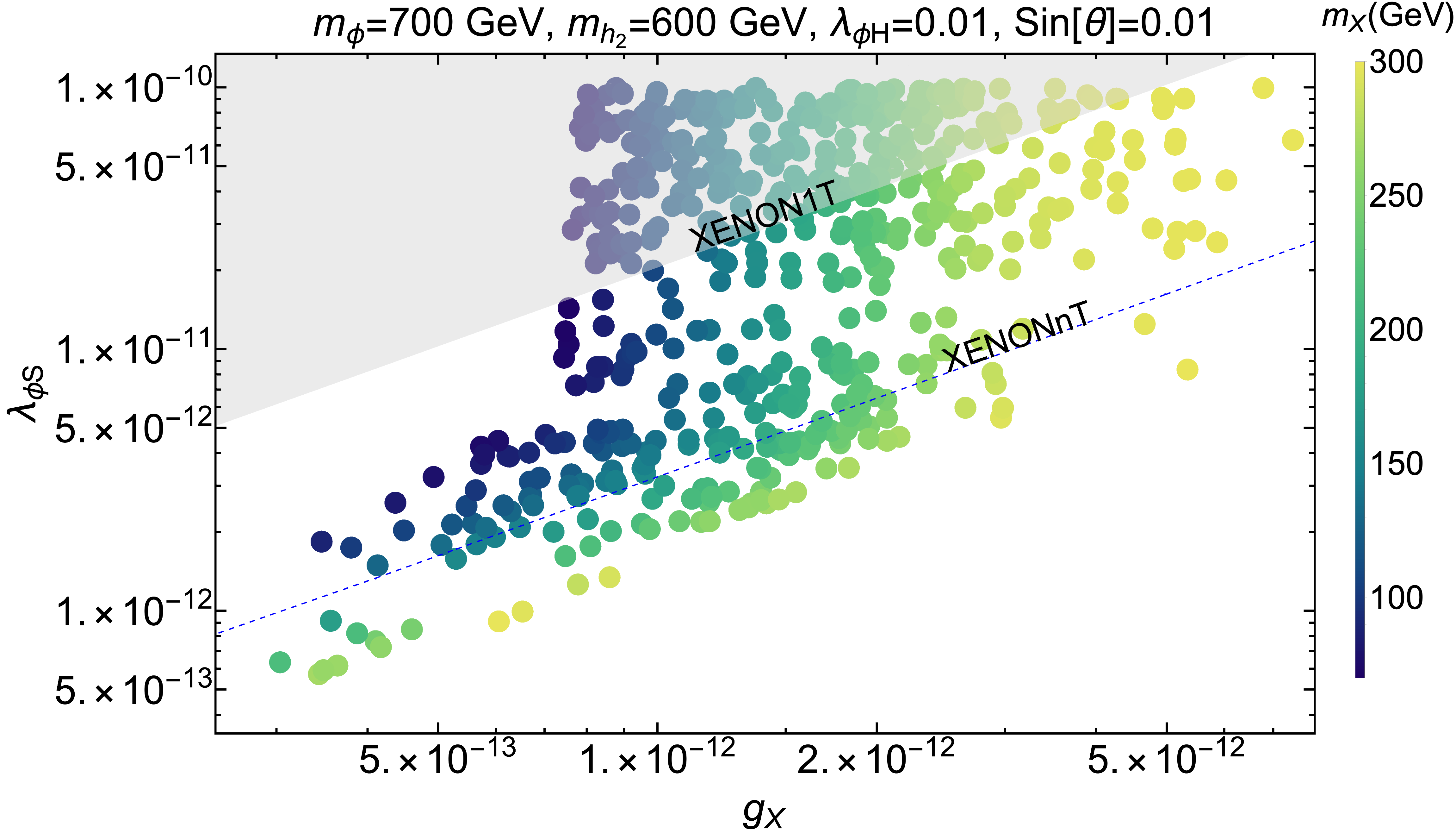}\label{wfaewsb_mx}}	
\caption{Scan in $g_X-\lambda_{\phi S}$ plane when both WIMP and FIMP components add to observed relic density, 
$\Omega_{X}h^2+\Omega_{\phi}h^2=0.1200\pm 0.0012$, simultaneously addressing other constraints when both freeze 
in and freeze out occur aEWSB in kinematic region $m_{h_2}\geq2m_X$. In 
	Fig. \ref{wfaewsb_m2}, we keep FIMP mass fixed at $m_X=200$ GeV and vary WIMP mass $m_\phi$ as shown in the SiennaTones colour bar. 
	In Fig. \ref{wfaewsb_mx}, we keep WIMP mass $m_\phi=700$ GeV fixed and vary FIMP mass ($m_X$) as shown by the BlueGreenYellow colour bar. 
	Other parameters kept fixed are mentioned in the figure heading. Direct search constraint from XENON1T is shown by grey shaded 
	regions, while the future sensitivity of XENONnT is shown by blue dotted line.}
\label{wfaewsbm2mx}
\end{figure}

\subsection{Putting WIMP and FIMP together}

 We again discuss a couple of example plots where the WIMP ($\phi$) and FIMP ($X$) add to the total observed DM relic density. 
In Fig. \ref{wfaewsbm2mx}, we show the scan in $g_X-\lambda_{\phi S}$ plane where both freeze-in of $X$ and freeze-out of $\phi$
occur aEWSB. In Fig. \ref{wfaewsb_m2}, we keep FIMP mass ($m_X$) fixed, while vary WIMP mass ($m_\phi$) as shown by the 
SiennaTones color bar. In Fig. \ref{wfaewsb_mx}, we instead keep WIMP mass ($m_\phi$) fixed, while 
vary FIMP mass ($m_X$) as shown by the BlueGreenYellow color bar. In both cases, we adhere to a parameter space where FIMP production is decay 
dominated. Observations are pretty similar to what we got bEWSB. 

In Fig.~\ref{wfaewsb_m2}, we again see that with larger $g_X$, FIMP relic enhances, which in turn requires $\lambda_{\phi S}$ to enhance as well, 
so that $\Omega_\phi$ decreases (the inclined region). Now, $g_X$ can maximally enhance to $\sim 2\times 10^{-12}$, when FIMP relic completely 
dominates over WIMP, with a sharp rise in $\lambda_{\phi S}$ to bring WIMP relic to very small value. We also see that keeping the $g_X$ fixed (so that 
FIMP relic remains almost unchanged), if we enhance $\lambda_{\phi S}$, WIMP relic decreases unless we adjust $m_\phi$ to larger values to keep the  
total relic within experimental observed value. In Fig. \ref{wfaewsb_mx}, larger $\lambda_{\phi S}$ diminishes $\Omega_\phi$, which is adjusted by larger 
FIMP contribution, by having smaller $m_X$ for a fixed $g_X$. Further, when $g_X$ is enhanced, FIMP contribution becomes larger, then WIMP contribution is 
adjusted to smaller values by larger $\lambda_{\phi S}$. Importantly, present direct search bound from XENON1T (on spin independent cross-section) 
plays an important role here, shown by grey shaded region, which discards part of large $\lambda_{\phi S}$ region. Future projected direct search sensitivity 
of XENONnT experiment is also shown by the blue dashed line, which will probe a large part of the allowed parameter space. \par

 As can be easily seen, that the phenomenology aEWSB is richer and as the masses ($m_\phi,m_s, m_X$) turn out to be much smaller. Such regions are 
 prone to both direct search and collider experiments for the WIMP, which interestingly correlates to the FIMP under abundance as some of the parameters 
 are common. We leave the exercise, where collider signal and direct search sensitivity of WIMP-FIMP model will be discussed, for a separate work. 
 We finally tabulate some characteristic benchmark points for this scenario in table \ref{tab:benchmark_aewsb}, where the total relic obtained from 
 $X$ and $\phi$ adds to the observed one abiding by all the other constraints. Here also, AP1 and  AP4 point out to the cases when $X$ dominates over $\phi$, 
 AP3, AP6 show when $\phi$ dominates over $X$ and AP2, AP5 depict the case when both DM contribute equally.

 
\begin{center}
\begin{table}[htb!]
\scalebox{0.78}{
\scriptsize
  \renewcommand{\arraystretch}{2.45}
  \begin{tabular}{|p{1.5cm}|c|c|c|c |c |c |c |c |c |c |c |c |c |c |c|} \hline
Scenario&\multirow{1}{1.5cm}{Benchmark\\~~ points}&\multirow{1}{2cm}{$m_{\phi},m_{h_2},m_X$\\ ~~~~$\rm{(GeV)}$}  & $g_X,\lambda_{\phi S},\lambda_{\phi H},\sin\theta$ & $\Omega_{\phi}h^2$ & $ \Omega_Xh^2$ & $ \frac{\Omega_{\phi}}{\Omega_T}\left(\%\right)$ & $ \frac{\Omega_{X}}{\Omega_T}\left(\%\right)$&\multirow{1}{0.6cm}{$\sigma_{\phi_{eff}}^{\text{SI}}$\\$(\text{cm}^2)$}\\\hline

    \multirow{3}{*}{$m_{h_2}\geq 2m_X$}&~~~~AP1&200, 150, 70&  $1.82\times10^{-12},3.30\times10^{-12} ,10^{-2},10^{-2}$&0.0117  & 0.1088 & 9.71  & 90.29 &$1.60\times 10^{-48}$\\\cline{2-9}
    
 &~~~~AP2&400, 300, 130&$1.51\times10^{-12},2.80\times10^{-12},10^{-2},10^{-2} $&0.0545& 0.0663 &45.08  & 54.92 &$9.52\times 10^{-50}$ \\\cline{2-9}
 
&~~~~AP3&350, 300, 100&$3.00\times10^{-13},4.85\times10^{-13} ,10^{-2},10^{-2}$& 0.1141  & 0.0053  & 95.54  & 04.46 &$1.52\times 10^{-48}$ \\\hline
    \multirow{3}{*}{$m_{h_2}< 2m_X$} &~~~~AP4&700, 400, 250&$5.22\times 10^{-11},1.58\times 10^{-10},10^{-2},10^{-2}$&0.0186 &0.1002 &15.68  &84.32 &$9.41\times 10^{-49} $ \\\cline{2-9}
 &~~~~AP5&1000, 300, 200&  $3.33\times 10^{-11},1.63\times 10^{-10},10^{-2},10^{-2} $&0.0599& 0.0604 &49.77   &50.23  &$2.29\times 10^{-48} $\\\cline{2-9}
 &~~~~AP6&600, 250, 150&$9.97\times 10^{-12},2.50\times 10^{-11},10^{-2},10^{-2}$&0.1187 &0.0018 &98.54&01.46 &$5.45\times 10^{-50} $ \\\hline
\end{tabular}}
\caption{Some sample benchmark points for the WIMP-FIMP DM model, when both freeze-in and freeze-out occur aEWSB respecting the total relic density, 
direct search, Higgs mass/mixing and other constraints. The benchmark points depict the possibilities when one component dominates over the other as well 
as the cases when they have almost equal share for the relic density.}
\label{tab:benchmark_aewsb}
\end{table}
\end{center}

\section{Summary and Conclusions }
\label{sec:conclusions}

 In this analysis, we show that EWSB plays an important boundary condition for DM freeze-out and freeze-in. To be specific, we ask, if it is possible to 
identify the region of parameter space where saturation of DM yield occurred bEWSB or aEWSB. We see that there are mainly two effects to this end; the main 
point is the mass of the DM or the decaying particle, which plays an important role to saturate the freeze-in or freeze-out abundance before or after EWSB, 
and second is the change in the depletion or production channels for DM across EWSB, particularly for those DM particles that couple to visible sector 
via Higgs portal. Here we have demonstrated the changes in relic density allowed parameter space of a two component WIMP-FIMP model when both
freeze-in and freeze-out occurs before EWSB to that when both occur aEWSB. Some broad characteristics emerge from the study. 
For example, when FIMP freezes in bEWSB, we see that the requirement that even `late decay' of the bath particle to occur bEWSB 
puts constraints on the mass on the decaying particle to be larger than some threshold. This is equivalent to freeze-out of a particle to occur 
bEWSB, where the WIMP mass requires to be sufficiently heavy ($\sim$ 4 TeV); the exact limit depends on the nature and interactions of the DM 
considered. On the other hand for freeze-in or freeze-out to occur aEWSB, the mass of the bath particle for FIMP, or the mass of DM for WIMP can 
be in the range of $\sim \mathcal{O} (100)$ GeV. We have demonstrated the above features by solving appropriate Boltzmann equations, taking care 
of all constrains on the model parameters . This in turn provides with a nice distinguishability of the parameter space of the model; the case bEWSB 
is difficult to probe at collider having heavier masses (be it WIMP mass or the particle in thermal bath that decays to FIMP), direct search (for WIMP) 
serves as the only viable option, while the case aEWSB is more accessible to both collider and direct search prospects with masses of the order of TeV.

Regarding the model we choose for illustration, we have a vector DM ($X$) transforming under additional $U(1)_X$ symmetry, 
which remains out-of-equilibrium and freezes in, while a scalar singlet DM $\phi$ remains in thermal bath and freezes-out to 
acquire correct relic. The effect of EWSB is pronounced in such a case with a larger $U(1)_X$ breaking scale necessitated by the freeze-in of $X$. 
On the contrary, if we imagine a situation, where $\phi$ is WIMP and $X$ is FIMP, then unnatural fine tuning will only be applicable to the portal 
couplings $\lambda_{\phi H}, \lambda_{\phi S}$, bringing down the $U(1)_X$ breaking scale close to EWSB, leaving a very small region of parameter 
space for a massive $X$ freeze-out bEWSB. Further, in such a two component set up, WIMP and FIMP parameters get correlated to produce the 
observed relic. For example, given a $U(1)_X$ coupling $g_X$, it is possible to adjudge a portal $\lambda_{\phi S}$ and vice versa. 
Direct search constraints on the WIMP $\phi$ also limits the FIMP abundance due to the presence of some common coupling 
parameters like $\lambda_{\phi S}$.

While both the models as single component DM have been studied in literature, we find out that the presence of both DM components together 
provides additional features for their respective abundances. For example, $\phi$ has channels to deplete its number density due to the scalar 
$s$ required to break $U(1)_X$ symmetry, which allows a larger allowed parameter space for $\phi$ in the resonance region 
$m_\phi\sim m_s/2\sim m_{h_2}/2$. Further, regions where annihilation to this scalar $m_\phi>m_s$ (or the physical one $h_2$ after mixing) opens up, 
it is easier to satisfy relic under abundance after adhering to direct search bounds. For FIMP ($X$), conversion from WIMP ($\phi$) plays a major role, 
which is only possible in a two component set up like this. On the contrary, in order to keep $X$ out-of-equilibrium, the WIMP-FIMP conversion is never 
sizeable enough to alter the effective annihilation cross-section and relic density of $\phi$. This feature is generic beyond the specific model taken up here. 
We also note that, the coupled BEQ required to address a two component WIMP-FIMP case, reduces to two uncoupled BEQs in the limit of tiny 
annihilation from WIMP to FIMP. However, one may think of a situation where the FIMP has very tiny coupling with SM, but a sizeable one with the WIMP, 
which may bring it to thermal bath and then freezes it out. Consequences for such a situation is interesting and will be addressed elsewhere.

\acknowledgments
SB  would like to acknowledge DST-SERB grant CRG/2019/004078 from Govt. of India. SC acknowledges support from the Indo-French Centre for the Promotion of Advanced Research (CEFIPRA Grant no: 6304-2). DP thanks University Grants Commission for research fellowship and Heptagon, IITG for useful discussions. 

\appendix 

\section{Decoupling of the bath particle decaying to FIMP}
\label{sec:bEWSB-details}
 In our analysis, we have seen that $s \to XX$ (bEWSB) and $h_2 \to XX$ (aEWSB) plays a major role for $X$ yield ($\Omega_X$) when kinematically 
accessible with $m_s\gtrsim2m_X$ or $m_{h_2} \gtrsim2m_X$. However, it is important to note that the decay can occur when $s(h_2)$ is in thermal bath and also 
after it decouples from the bath. The contribution to $X$ yield after the decoupling is referred to as `late decay'. The late decay is taken care of in the 
FIMP BEQ via an additional interaction as shown in Eq. \ref{frzin_int}. The same is shown for freeze-in production of $X$ aEWSB in Eq.~\ref{xaewsb}.
This additional term stems from a cBEQ involving $s$ and $X$, where $s$ freezes-out and $X$ freezes-in. The relevant cBEQs are given by:
\begin{align}\label{cbeq_sX}
\begin{aligned}
 \frac{dY_s}{dx}
 =&
 - \frac{0.264\, {\rm  M_{ Pl}}\,m_s}{x^2}
 \frac{g_*^s}{\sqrt{g_*^{\rho}}}
 \left[ Y_s^2 - (Y_s^{eq})^2 \right]
 \left[\sum_{\substack{i = H,\\ \phi, X}} \ev{\sigma_{ss\to i i}\,v}
 + \frac{\ev{\Gamma_{s\to XX}}}{Y_s^{eq}\dfrac{2\pi^2}{45} g_*^s \left(\dfrac{m_s}{x}\right)^3}\right] ;
 \\
 \frac{dY_X}{dx}
 =&
 \frac{45\, {\rm M_{ Pl}}\, \Gamma_{s\to XX}}{1.67\times2\pi^4\, m_s^2} \frac{K_1[x]\, x^3}{g_*^s\sqrt{g_*^{\rho}}}
 + \frac{0.264\, {\rm M_{ Pl}}\, m_s}{x^2}
 \frac{g_*^s}{\sqrt{g_*^{\rho}}}
 \left[ Y_s^2 - (Y_s^{eq})^2 \right]
 \frac{\ev{\Gamma_{s\to XX}}}{Y_s^{eq}\dfrac{2\pi^2}{45} g_*^s \left(\dfrac{m_s}{x}\right)^3} ;
\end{aligned}
\end{align}
where $x=m_s/T$. As shown in Fig. \ref{frzin_fig5}, the late decay contribution essentially comes from the freeze-out yield of $s (h_2)$. The freeze-out yield 
of $s$, in turn depends on the annihilation cross-sections of $s$ via the channels as shown in Feynman graphs in Fig.~\ref{S-annihilation}. The cross-sections 
at the threshold (denoted by $\sigma^0$) where  center of mass energy $(\mathtt{s})$ is just enough for the production process to occur are given below :

\begin{align}
 \sigma^0_{ss\to H H^\dagger}\vert_{\mathtt{s}=4m_s^2}
 =&
 \frac{\lambda_{ H S}^2 }{8 \pi  g_X^4 m_s^7}\sqrt{m_s^2-m_{H}^2} \left(g_X^4 m_s^4-2 g_X^2 m_s^2 m_X^2 \lambda_{ H S}+m_X^4 \lambda_{HS}^2\right)\,,
 \nonumber \\ 
 \sigma^0_{ss\to \phi\phi}\vert_{\mathtt{s}=4m_{s}^2}
 =&
 \frac{ \lambda_{\phi S}^2}{16 \pi  g_X^4 m_s^7}\sqrt{m_s^2-m_{\phi}^2} \left(-2 g_X^2 m_s^2 m_X^2 \lambda_{ \phi S}+g_X^4 m_s^4+m_X^4 \lambda_{ \phi S}^2\right)\,,
 \\
 \sigma^0_{ss\to XX}\vert_{\mathtt{s}=4m_s^2}
 =&
 \frac{g_X^4}{4 \pi  m_s^7} \sqrt{m_s^2-m_X^2} \left(-20 m_s^2 m_X^2+11 m_s^4+12 m_X^4\right)\,.\nonumber
 \label{sigma0sstosm}
\end{align}
Note that the annihilation cross-section at threshold has the most dominant contribution for freeze out. The expressions for cross-sections of $h_2$ decoupling 
is pretty similar, involves additionally the mixing angle ($\sin\theta$) between the SM isodoublet and singlet.

\section{Higgs mass and constraints}
\label{sec:aEWSB-details}

The scalar potential bEWSB (in terms of $m_H, m_s, m_\phi$) is given by,

\bea\begin{split}
\rm{V(H,s,\phi)\bigg|_{\rm bEWSB} }&=\underbrace{(\mu_{H}^2+ \frac{1}{2}\lambda_{H S}v_s^2)}_{m_H^2}( H^{\dagger}H)+\lambda_{H}( H^{\dagger}H)^2 +\frac{1}{2}\underbrace{(\mu_{\phi}^2+ \frac{1}{2}\lambda_{\phi S}v_s^2)}_{m_{\phi}^2}\phi^2+
\frac{1}{4!}\lambda_{\phi}\phi^4 \\&+\frac{1}{2}\underbrace{(\mu_{S}^2+3\lambda_{ S}v_s^2)}_{m_{s}^2}s^2 +\frac{1}{4}\lambda_{S}s^4+\lambda_{S}v_ss^3   + \frac{1}{2}\lambda_{\phi H} (\phi^2H^{\dagger}H)+\lambda_{H S}v_s(sH^{\dagger}H)\\& + \frac{1}{2}\lambda_{\phi S}v_s(s\phi^2)+ \frac{1}{2}\lambda_{H S}(s^2H^{\dagger}H) + \frac{1}{4}\lambda_{\phi S}(s^2\phi^2)+(\lambda_Sv_s^3-\mu_S^2v_s)s\\&+(\frac{1}{4}\lambda_Sv_s^4-\frac{1}{2}\mu_S^2v_s^2).
\label{}
\end{split}\eea

The potential aEWSB can be written (in terms of $m_H, m_s, m_\phi$) as:

\bea\begin{split}
\rm{V(h,s,\phi)\bigg|_{\rm aEWSB} }&=\frac{1}{2}m_H^2(v+h)^2+\frac{1}{4}\lambda_{H}(v+h)^4 +\frac{1}{2}m_{\phi}^2\phi^2+
\frac{1}{4!}\lambda_{\phi}\phi^4 \\&+\frac{1}{2}m_{s}^2s^2 +\frac{1}{4}\lambda_{S}s^4+\lambda_{S}v_ss^3   + \frac{1}{4}\lambda_{\phi H} \phi^2(v+h)^2+\frac{1}{2}\lambda_{H S}v_s(v+h)^2s\\& + \frac{1}{2}\lambda_{\phi S}v_s\phi^2s+ \frac{1}{4}\lambda_{H S}(v+h)^2s^2 + \frac{1}{4}\lambda_{\phi S}\phi^2s^2+(\lambda_Sv_s^3-\mu_S^2v_s)s\\&+(\frac{1}{4}\lambda_Sv_s^4-\frac{1}{2}\mu_S^2v_s^2).
\label{}
\end{split}\eea

Following above, there is a mixing between $h,s$ fields aEWSB. The mass matrix $\mathcal{M}$ in the basis $\{h,s,\phi\}$ can be written as,
\bea
\mathcal{M}^2 = \left(
\begin{array}{ccc}
m_H^2+3\lambda_H v^2 & \lambda_{HS} v v_s& 0 \\
\lambda_{HS} v v_s& m_s^2+ \frac{1}{2}\lambda_{HS} v^2 & 0 \\
 0 & 0 &  m_{\phi}^2+ \frac{1}{2}\lambda_{\phi H} v^2 \\
\end{array}
\right)\;.
\eea

\begin{figure}[htb!]  
\centering
\subfloat[]{\includegraphics[width=0.35\linewidth]{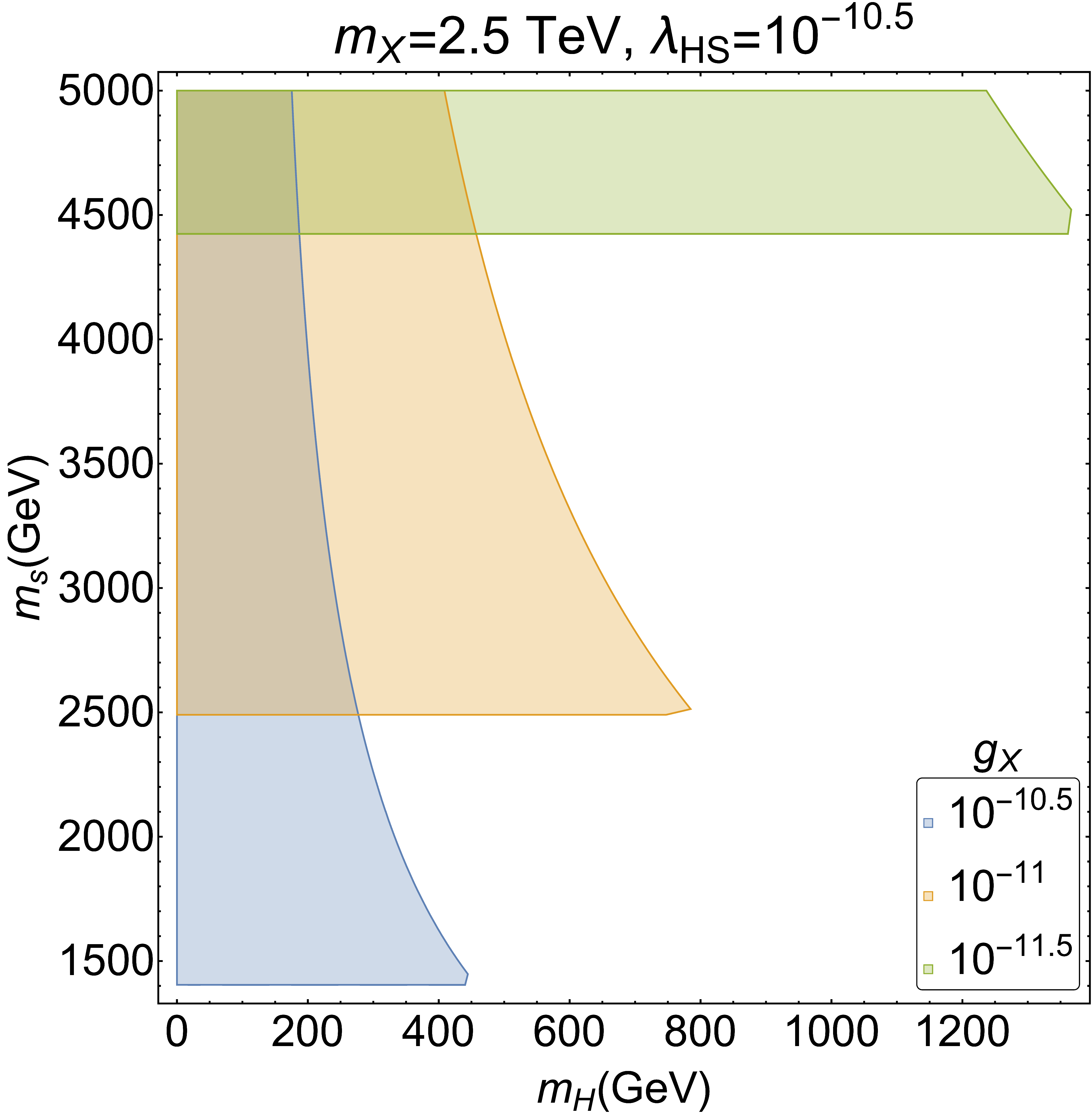}\label{bagxl1s1}}
\subfloat[]{\includegraphics[width=0.36\linewidth]{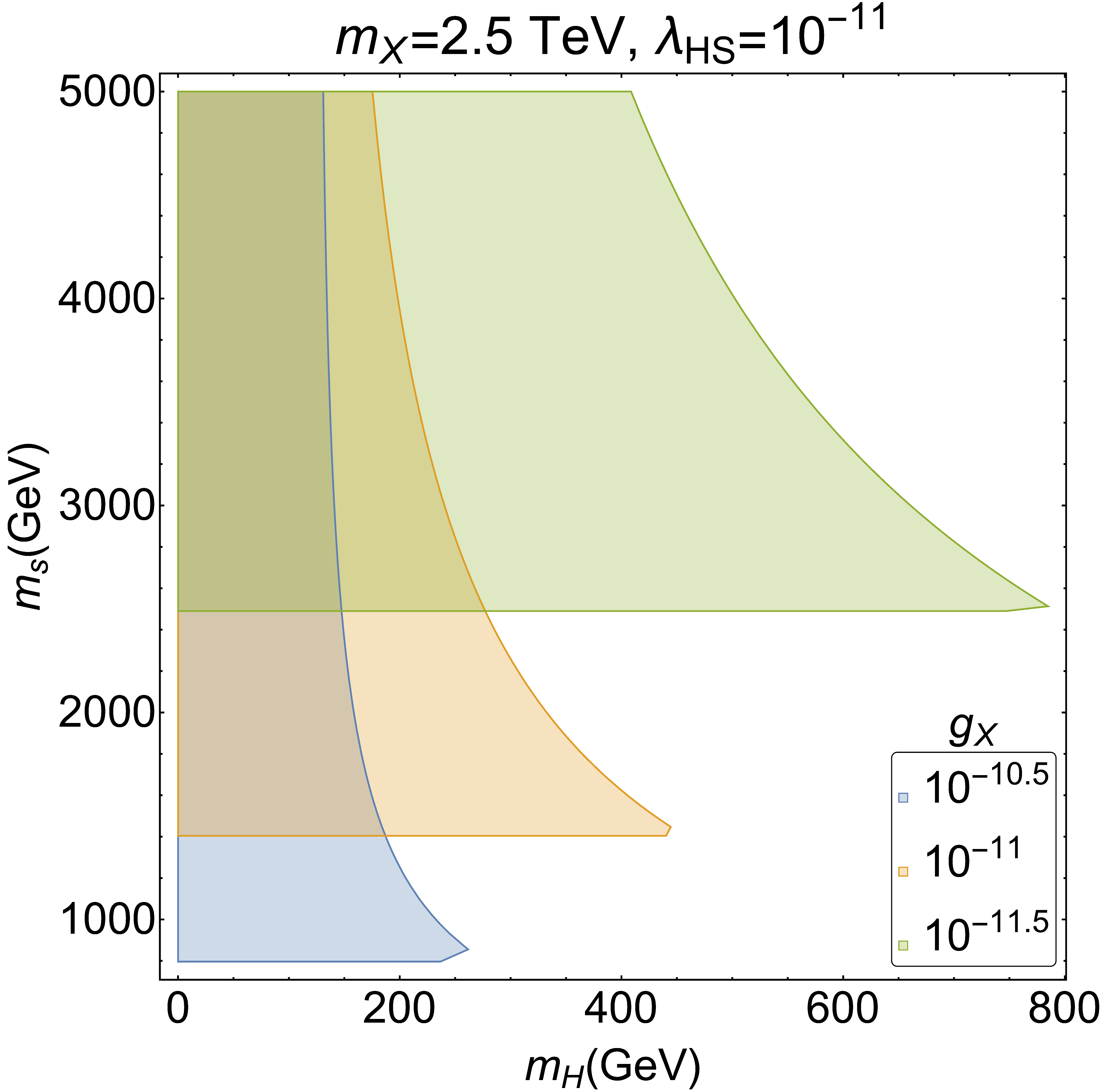}\label{bagxl1s2}}
\subfloat[]{\includegraphics[width=0.35\linewidth]{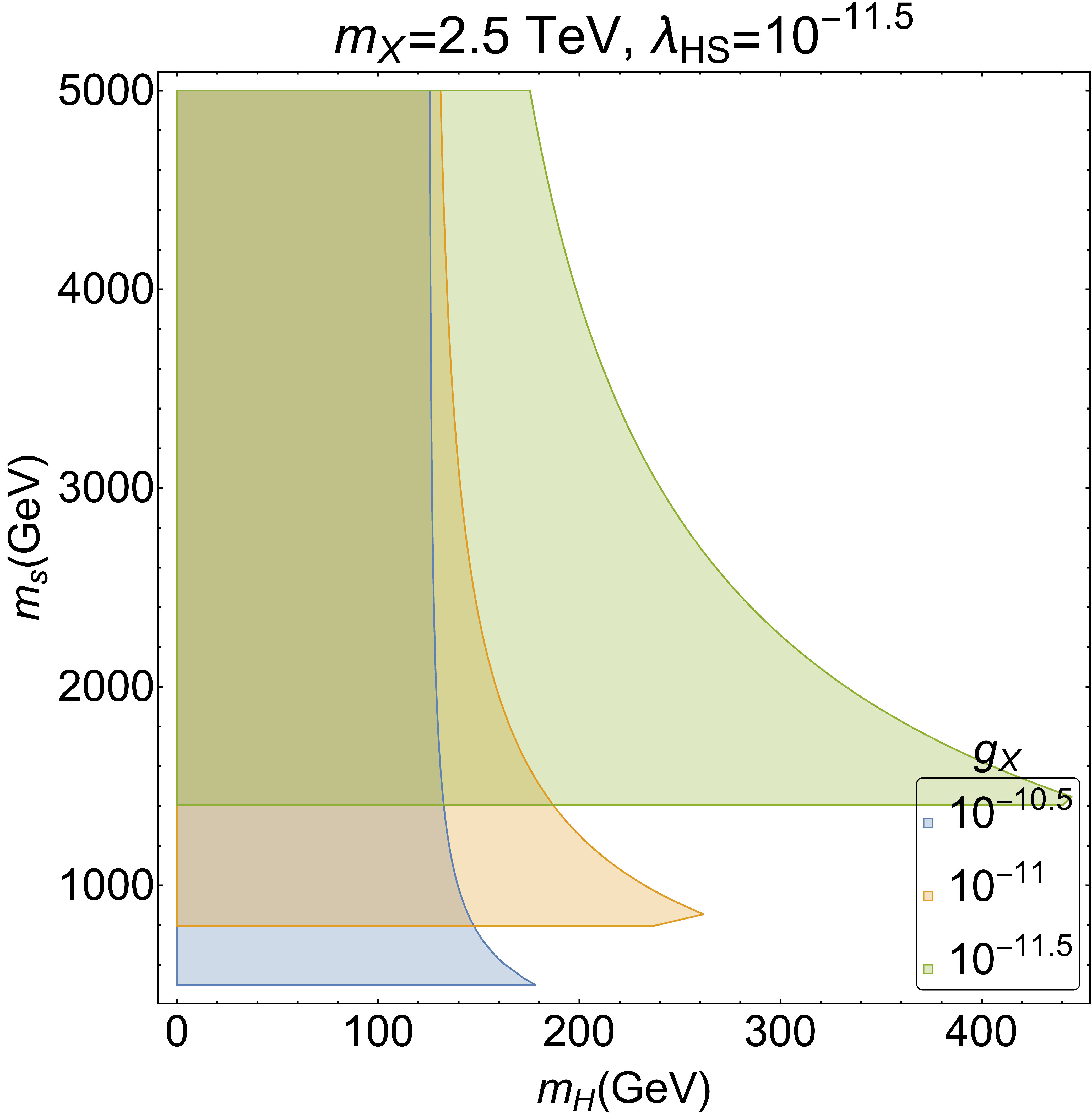}\label{bagxl1s3}}\\
\subfloat[]{\includegraphics[width=0.34\linewidth]{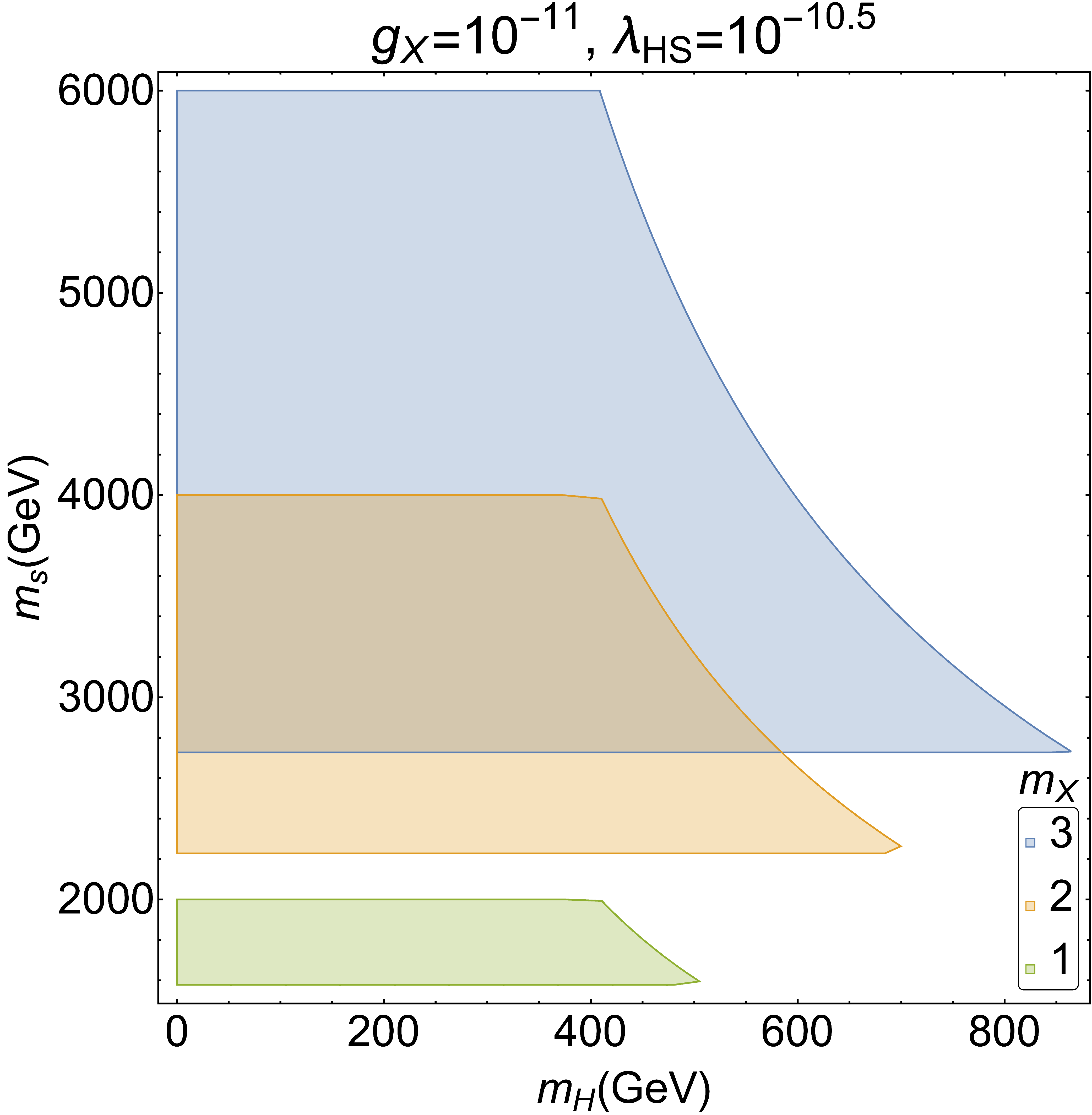}\label{bamxl1s1}}
\subfloat[]{\includegraphics[width=0.35\linewidth]{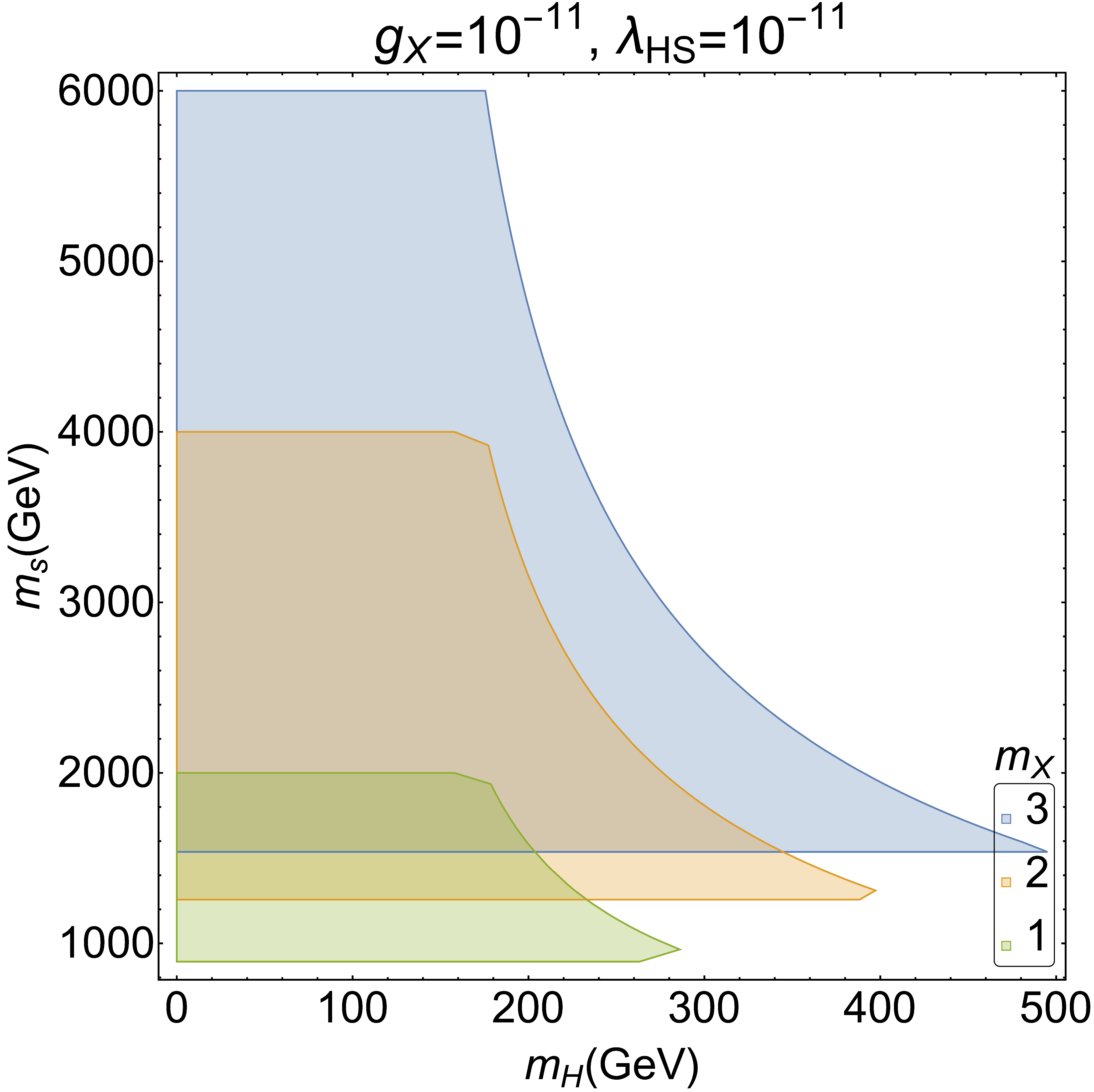}\label{bamxl1s2}}
\subfloat[]{\includegraphics[width=0.35\linewidth]{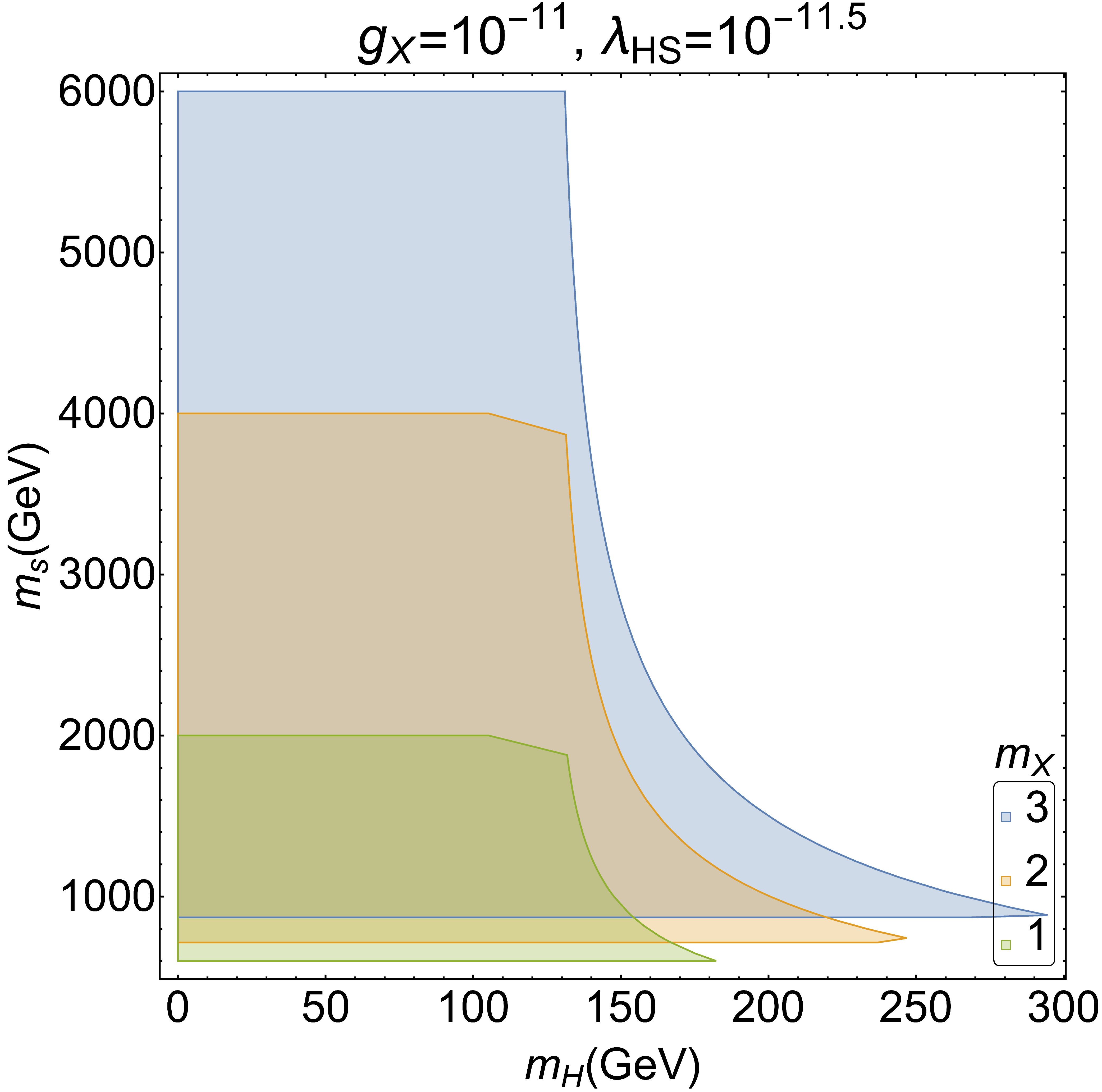}\label{bamxl1s3}}\
\caption{
Allowed parameter space in $m_H-m_s$ plane constrained by the requirement of obtaining 125.1 GeV SM Higgs after EWSB. $|\sin[\theta]|\lesssim 0.3$ and $0<\lambda_H\leq 4\pi$. FIMP mass ($m_X$) is varied in TeV scale in Fig. \ref{bamxl1s1}, \ref{bamxl1s2} and \ref{bamxl1s3}.
}
\label{mHmslhs}
\end{figure}

Upon diagonalisation the mass eigenvalues of the matrix are given by,
\bea\begin{split}
2m_{h_{1,2}}^2&=m_H^2+m_s^2+3\lambda_Hv^2+\frac{1}{2}\lambda_{HS}v^2\mp \sqrt{(m_s^2+\frac{1}{2}\lambda_{HS}v^2-m_H^2-3\lambda_Hv^2)^2+4\lambda_{HS}^2v^2v_s^2};\\
\mathfrak{m}_{\phi}^2&=m_{\phi}^2+\frac{1}{2}\lambda_{\phi H}v^2.
\end{split}\eea
The physical eigenstates are given by
\bea
h_1 &=& \cos\theta~ h -\sin\theta ~s \,, \nonumber \\
h_2&=& \sin\theta~ h +\cos\theta~ s \,;
\eea 
where the mixing angle can be written as, 
\bea
\tan 2\theta=\frac{2\lambda_{HS}v v_s}{m_s^2+ \frac{1}{2}\lambda_{HS} v^2-m_H^2-3\lambda_H v^2} \,.
\label{eq:mixing}
\eea
The mixing is restricted by LHC data as $|\sin[\theta]|\lesssim 0.3$ (see text). Now, $h_1$ is identified with SM Higgs so that $m_{h_1}=125.1$ GeV 
and $h_2$ is assumed to be another neutral scalar, which is dominantly a singlet and can be heavy or light. When $m_s<2m_X$, 
immaterial to whether the freeze-in or freeze-out occurs bEWSB, $h, s$ mixing as stated above occurs and results in a SM Higgs as observed currently. 
So, even in bEWSB epoch, $m_H,m_s$ needs to be chosen in such a way that we obtain correct Higgs mass and respect the mixing angle limit. This 
is what we have done for the scans done in the DM analysis. The correct choices of parameters $\{m_H,m_s,\lambda_{HS},v_s\}$ 
is indicated in Fig. \ref{mHmslhs}, which shows the correlation between $m_H-m_s$ bEWSB allowed by these mass constraints for fixed values of other 
couplings.

Here, we would also like to point out to a caveat that when $m_s\gtrsim 2m_X$ and FIMP freezes-in before EWSB by in-equilibrium decay and late decay of 
$s$, due to complete depletion of $s$ before EWSB, mixing does not arise after EWSB $(\sin \theta \sim 0)$ and therefore we only obtain one physical 
scalar $h_1$ with mass $m_{h_1}=125.1$ GeV. Then the relation between $m_H$ bEWSB and $m_{h_1}$ aEWSB is simply given by
\bea
\begin{split}
m_{h_1}^2=m_H^2+3\lambda_Hv^2,~~m_H^2=\frac{1}{2}m_{h_1}^2.
\end{split}
\eea 


The scalar potential can be written only in terms of $h_1,h_2,\theta$ is given by
\begin{align}
 V(h_1,h_2,\phi)\bigg|_{\rm aEWSB}
 =&
  \frac{\mu_{H}^2}{2}(v+h_1 \cos\theta + h_2\sin\theta)^2
 + \frac{\lambda_H}{4}(v+h_1 \cos\theta +h_2\sin\theta)^4
 + \frac{\mu_{\phi}^2}{2}\phi^2  \notag\\&
 + \frac{\lambda_{\phi}}{4!}\phi^4
 +\frac{\mu_{S}^2}{2}(v_s-h_1\sin\theta+h_2\cos\theta)^2
 + \frac{\lambda_{S}}{4}(v_s-h_1\sin\theta+h_2\cos\theta)^4
 \notag\\
 &
 + \frac{\lambda_{\phi H}}{4}\phi^2(v+h_1 \cos\theta +h_2\sin\theta)^2
 +\frac{\lambda_{\phi S}}{4}\phi^2(v_s-h_1\sin\theta+h_2\cos\theta)^2
 \notag\\
 &
 +\frac{\lambda_{H S}}{4}(v+h_1 \cos\theta +h_2\sin\theta)^2(v_s-h_1\sin\theta+h_2\cos\theta)^2 
\end{align}
Using the extremization condition of the potential
\begin{align}
 \left(\frac{\partial V(h_1,h_2,\phi)}{\partial h_1}\right),\left(\frac{\partial V(h_1,h_2,\phi)}{\partial h_2}\right)\bigg|_{ h_{1,2},\phi=0}
 =
 0
 \label{eq:extrema}
\end{align}
we obtain the following conditions,
\begin{align}
\begin{split}
 v \cos \theta  \left(\lambda_{HS} v_s^2+2 \mu_H^2+2 \lambda_H v^2\right)- v_s\sin \theta  \left(v^2 \lambda_{H S}+2 \lambda_S v_s^2+2 \mu_S^2\right)=0
 \\
 v \sin \theta  \left(\lambda_{HS} v_s^2+2 \mu_H^2+2 \lambda_H v^2\right)+v_s\cos \theta   ( v^2\lambda_{HS} +2\lambda_S v_s^2+2\mu_S^2)=0
\end{split}
\end{align}

\begin{align}
\begin{split}
 \frac{\partial^2 V_{\rm scalar}}{\partial h_1^2}\bigg|_{h_{1,2}=\phi=0}
 =&
 m^2_{h_1}
 =
 (3v^2\lambda_H+\frac{1}{2}v_s^2\lambda_{HS}+\mu_H^2)\cos^2\theta-v v_s\lambda_{HS}\sin2\theta
 \\
 &\qquad
 +(3v_s^2\lambda_S+\frac{1}{2}v^2\lambda_{HS}+\mu_S^2)\sin^2\theta
 \\
 \frac{\partial^2 V_{\rm scalar}}{\partial {\phi}^2}\bigg|_{h_{1,2}=\phi=0}
 =&
\mathfrak{m}^2_{\phi}
 =
 \frac{1}{2} \left(2 \mu_{\phi}^2+\lambda_{ \phi S} v_s^2+v^2 \lambda_{ \phi H}\right)
 \\
 \frac{\partial^2 V_{\rm scalar}}{\partial h_2^2}\bigg|_{h_{1,2}=\phi=0}
 =&
 m_{h_2}^2
 =
 (3v^2\lambda_H+\frac{1}{2}v_s^2\lambda_{HS}+\mu_H^2)\sin^2\theta+v v_s\lambda_{HS}\sin2\theta
 \\
 &\qquad
 + (3v_s^2\lambda_S+\frac{1}{2}v^2\lambda_{HS}+\mu_S^2)\cos^2\theta
\end{split}\label{awimp_mass}
\end{align}
After mixing, off diagonal mass terms of physical fields $h_1$ and $h_2$ are absent; then
$\partial^2V_{\rm scalar}/\partial h_1\partial h_2 = 0$ gives us,
\begin{align}
 \cos 2 \theta ~v v_s \lambda_{H S }+\frac{1}{2}\sin 2 \theta  \left[ 3(v^2\lambda_H-v_s^2\lambda_S)+\frac{1}{2}\lambda_{HS}(v_s^2-v^2)-(\mu_S^2-\mu_H^2)\right]
 =&
 0
\end{align}
Finally, the expressions of mixing angle $(\theta)$ and internal parameters $(\mu_H,\mu_{\phi},\mu_S,\lambda_H,\lambda_S,\lambda_{HS})$ 
in terms of the external parameters are given by:
\begin{align}
\begin{split}
\mu_H^2
 &= -
 (\lambda_H v^2+\frac{1}{2}\lambda_{HS}v^2_s),
 \\
 \mu_{\phi}^2
 &=
\mathfrak{m}^2_{\phi}-\frac{1}{2}\lambda_{ \phi S} v_s^2-\frac{1}{2} \lambda_{\phi H}v^2,
 \\
 \mu_S^2
 &=
- (\lambda_S v^2_s+\frac{1}{2}\lambda_{HS}v^2),
 \\
 v_s
 &=
 \frac{m_X}{g_X},
 \\
 \lambda_H
 &=
 \frac{1}{2v^2}\left(m^2_{h_1}\cos^2\theta+m^2_{h_2}\sin^2\theta\right),
 \\
 \lambda_S
 &=
 \frac{1}{2v_s^2}\left(m^2_{h_2}\cos^2\theta+m^2_{h_1}\sin^2\theta\right),
 \\
 \lambda_{H S}
 &=
 \frac{\sin2\theta}{2v_sv}\left(m^2_{h_2}-m^2_{h_1}\right).
\end{split}
\label{aEWSB_rel}
\end{align}

\section{Invisible decay width of Higgs }
\label{sec:invisible}

In Higgs portal scenarios, where DM couples to SM Higgs, Higgs boson can always decay to a pair of DM particles when kinematically accessible, 
contributing to invisible Higgs decay width. In our model, the possible invisible decay channels of Higgs include $h_1 \to \phi\phi, h_1\to XX, h_1\to h_2h_2$ with
decay widths given by:
\bea\begin{split}
&\Gamma_{h_1\to\phi\phi}=\frac{(\lambda_{\phi S} m_X \sin\theta-\lambda_{\phi H} g_X v \cos\theta)^2}{32 \pi g_X^2 m_{h_1}^2}(m_{h_1}^2-4\mathfrak{m}_{\phi}^2)^{1/2}\Theta(m_{h_1}-2\mathfrak{m}_{\phi})\\
&\Gamma_{h_1\to XX}=\frac{g_X^2\sin^2\theta}{32 \pi m_{h_1}^2 m_X^2}(m_{h_1}^2-4m_{X}^2)^{1/2}(m_{h_1}^4-4m_{h_1}^2m_X^2+12m_X^4)\Theta(m_{h_1}-2m_{X})\\
&\Gamma_{h_1\to h_2h_2}=\frac{(m_X \sin\theta -v g_X  \cos\theta )^2}{32 \pi v^2 m_{h_1}^2 m_X^2}\sin^2\theta\cos^2\theta(m_{h_1}^2+2m_{h_2}^2)^2(m_{h_1}^2-4m_{h_2}^2)^{1/2}\Theta(m_{h_1}-2m_{h_2})
\end{split}
\label{higgs_invisible}
\eea
The expression for the Higgs invisible decay branching ratio is,
\bea
\Gamma_{h_1\to \rm{inv}}=\frac{\Gamma_{h_1\to{\phi\phi}}+\Gamma_{h_1\to{XX}}+\Gamma_{h_1\to{h_2h_2}}}{\Gamma_{h_1}^{\rm{SM}}+\Gamma_{h_1\to{\phi\phi}}+\Gamma_{h_1\to{XX}}+\Gamma_{h_1\to{h_2h_2}}}.
\eea
Invisible Higgs decay widths and branching ratio is heavily restricted by the observed Higgs data at LHC as mentioned in Eq.~\ref{higgs_invisible_decay} 
and therefore, we do not scan the parameter space that comes within. 

 \section{Direct Search possibilities}
 \label{sec:directsearch}

\begin{figure}[htb!]
	\centering	
	\begin{tikzpicture}[baseline={(current bounding box.center)}]
\begin{feynman}
\vertex (a1){\(\phi\)};
\vertex [below right=1.5cm of a1] (a);
\vertex [above right=1.2cm of a] (a2){\(\phi\)};
\vertex [below=1.5cm of a] (b);
\vertex [blob, below left =1.5cm of b] (c) {N};
\vertex [blob, below right =1.5cm of b] (d) {N};
\path (c.30) ++ (00:2) node[vertex] (hf2);
\path (c.-30-|hf2.center) node[vertex] (hf3);

\diagram* {
    (a1) -- [scalar] (a) -- [scalar] (a2),
  (c) -- [fermion] (b),
    (a) -- [scalar, edge label={\(h_{1,2}\)}] (b) -- [fermion] (d),
    (c.30) -- [fermion] (hf2),(c.{-30}) -- [fermion] (hf3)
};
\end{feynman}
\end{tikzpicture}
	\caption{Feynman diagrams for the direct detection of wimp DM $\phi$.}
    \label{fig_direct-search}
\end{figure}
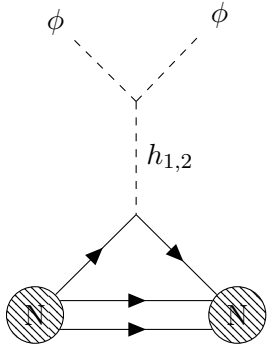

In this two component WIMP-FIMP DM model, FIMP $X$ coupling to SM Higgs ($H$) (via $s$) $\lambda_{HS}$ is very small in order to facilitate 
non-thermal production. Therefore, FIMP-nucleon cross-section is negligible. In case of WIMP $\phi$, it can talk to SM through the portals 
$\lambda_{\phi H}$ and $\lambda_{\phi S}$, given the mixing between $s-h$ present after EWSB, where the physical states become $h_1$ and $h_2$, 
out of which $h_1$ is assumed as SM Higgs, and $h_2$ is dominantly a singlet as explained in Appendix \ref{sec:aEWSB-details}. 
The Feynman graph for direct search cross-section is shown in Fig.~\ref{fig_direct-search}. The relative dominance of the mediators $h_1,h_2$ 
in the DM-nucleon scattering cross-section depends on the mass of new scalar $h_2$, which can be either heavy or light.

The spin-independent scattering cross section of $\phi$-Nucleon, mediated by both the physical scalars after mixing, is given by,

\begin{align}
\sigma^{\rm{SI}}_{n_\phi}=\frac{\Omega_{\phi}}{\Omega_{\phi}+\Omega_{X}}\frac{f_N^2 \mu_n^2 m_n^2}{4\pi v^2 \mathfrak{m}^2_\phi}\left(\cos\theta \frac{\lambda_{h_1\phi\phi}}{m_{h_1}^2}+\sin\theta \frac{\lambda_{h_2\phi\phi}}{m_{h_2}^2}\right)^2\,,
\label{eq:DD}
\end{align}
where $f_N=0.308\pm0.018$ \cite{Hoferichter:2017olk} represents the form factor of nucleon and $\mu_n=\frac{m_n \mathfrak{m}_{\phi}}{m_n+\mathfrak{m}_{\phi}}$ 
stands for the reduced mass and $n$ stands for nucleon. Also note that the maximum direct search cross-section for $\phi$ is folded by the fraction of 
relic density that $\phi$ possess in the total DM relic density in a two component framework given by $\frac{\Omega_{\phi}}{\Omega_{\phi}+\Omega_{X}}$.
The expressions of $\lambda_{h_1\phi\phi}$ and $\lambda_{h_2\phi\phi}$ 
in terms of our model parameters are given by 
\bea
\lambda_{h_1\phi\phi}&=&-v\cos\theta\lambda_{\phi H}+\frac{m_X}{g_X}\sin\theta \lambda_{\phi S}\,, \nonumber\\
\lambda_{h_2\phi\phi}&=&-v\sin\theta\lambda_{\phi H}-\frac{m_X}{g_X}\cos\theta \lambda_{\phi S}. 
\eea

In our analysis, the $h_2$ mediation in the direct detection cross-section is suppressed by small mixing angle. Also, there will be some propagator suppression 
due to $h_2$ which is assumed heavier than the SM Higgs. It is to be noted that if $\sin\theta\sim 0$, ie, mixing is absent, Eq. \ref{eq:DD} boils down to the
typical scalar singlet direct detection cross-section, mediated by SM Higgs. The constraint on the mixing is propagated to constraining $\lambda_{HS}$, as per 
Eq. \ref{aEWSB_rel}, and importantly affects both WIMP and FIMP under abundance. The SI direct search limit from XENON1T is mentioned in \ref{sec:constraints}.

We further note that even if freeze-out occurs before EWSB, one can have direct search possibility as described above. Even the FIMP under abundant region
gets constrained by direct search bound due to the presence of $\lambda_{\phi S}$ in both the cases. Only when $s$ decay completes before 
EWSB, it does not mix with $h$ aEWSB and then FIMP has absolutely no connection to SM and no constraints from direct search.


\newpage
\bibliographystyle{JHEP}
\bibliography{WIMP-FIMP}
\end{document}